\documentclass[final,3p]{elsarticle}
\usepackage{amssymb}

 \usepackage{graphicx}          %
 \usepackage{amssymb}
 \usepackage{amsmath}
\usepackage{amsthm}
 \usepackage{tikz}

\newcommand{\tp}{\intercal}
\long\def\comment#1{}

\usepackage{stmaryrd}

\newtheorem{prop}{Proposition}
\newtheorem{cor}{Corollary}
\newtheorem*{assum}{Assumption}

\usepackage{algpseudocode}
\algrenewcommand\algorithmicindent{1.0em}

\newcommand{\pushcodeb}{\hspace{3.55em}\relax}

\DeclareMathOperator*{\argmax}{arg\,max}
\DeclareMathOperator*{\argmin}{arg\,min}

\usepackage{float}
\newfloat{algorithm}{t}{lop}
\floatname{algorithm}{Algorithm}

\def\imagetop#1{\vtop{\null\hbox{#1}}}

\makeatletter
\def\ps@pprintTitle{%
  \let\@oddhead\@empty
  \let\@evenhead\@empty
  \let\@oddfoot\@empty
  \let\@evenfoot\@oddfoot
}
\makeatother
\begin{document}

\begin{frontmatter}

\title{Traffic Predictive Control from Low-Rank Structure}
\tnotetext[t1]{This research was supported by National Science Foundation SBIR Phase II Award 1329477}

\author[First]{Samuel Coogan}
\ead{scoogan@ucla.edu}
\author[Second]{Christopher Flores}
\ead{chrisf@sensysnetworks.com}
\author[Third]{Pravin Varaiya}
\ead{varaiya@berkeley.edu}

\address[First]{Electrical Engineering Department, University of California, Los Angeles}
\address[Second]{Sensys Networks, Inc.}
\address[Third]{Department of Electrical Engineering and Computer Sciences, University of California, Berkeley}

\begin{abstract}
The operation of most signalized intersections is governed by predefined timing plans that are applied during specified times of the day. These plans are designed to accommodate average conditions and are unable to respond to large deviations in traffic flow.  We propose a control approach that adjusts time-of-day signaling plans based on a prediction of future traffic flow. The prediction algorithm identifies correlated, low rank structure in historical measurement data and predicts future traffic flow from real-time measurements by determining which structural trends are prominent in the measurements. From this prediction, the controller then determines the optimal time of day to apply new timing plans. We demonstrate the potential benefits of this approach using eight months of high resolution data collected at an intersection in Beaufort, South Carolina.
\end{abstract}

\end{frontmatter}

\section{Introduction}
The advent of ubiquitous traffic sensing provides unprecedented real-time, high-resolution data that elucidate historical trends and current traffic conditions. However, traditional signal timing approaches have yet to take full advantage of these data \cite{Kurzhanskiy:2015fj}.  Surveys of practitioners suggest that only sixty percent of the 300,000 signalized intersections in the United States are retimed at intervals less than five years \cite{NCHRP397}, and the National Transportation Operations Coalition has given a grade of ``C$-$'' to signal timing practices and a grade of ``F'' to traffic monitoring and data collection in the United States. These deficiencies contributed to the 6.9 billion hours of additional travel time caused by inefficient traffic management in 2015 \cite{Schrank:2015hs}.

Increased urbanization demands more efficient use of this transportation infrastructure, which in turn requires full use of measured data. At signalized intersections, standard actuated signal timing plans are designed to make limited use of real-time measurements to, \emph{e.g.}, extend green time for approaching vehicles or enable actuation phases to be skipped if no waiting vehicles are present \cite{Koonce:2008gd}. However, actuated traffic signal timing only accommodates modest deviations from the nominal traffic conditions and is thus unable to respond to systematic changes in the traffic flow.

This paper proposes a \emph{traffic predictive control} strategy for signalized intersections that predicts future traffic flow based on real-time measurements and adjusts the intersection's signal timing accordingly. The prediction algorithm first identifies trends in historical traffic flow and then uses real-time measurements to determine the degree to which these historical trends are exhibited by the current traffic conditions. For example, historical trends may indicate that increased flow in one commute direction during the morning correlates with increased flow in the opposite direction in the evening. Traffic predictive control identifies this relationship and uses it to adjust signal timing parameters in the afternoon using measurements of traffic flow in the morning.

The present work thus lies within the extensive literature on traffic prediction and forecasting. See, for example, \cite{Vlahogianni:2004ct} for a survey of the literature. 
The majority of the forecasting literature focuses on freeways rather than urban street traffic \cite{Vlahogianni:2004ct}. For example, \cite{Jabari:2012fv} and \cite{Jabari:2013dz} develop a stochastic traffic flow model for freeways, and it is shown that this model is amenable to Kalman filtering techniques to estimate traffic conditions. Kalman filtering is also used in \cite{Ojeda:2013ya} to estimate traffic flow as the result of a random walk biased by historical increments in measured flow over time. In \cite{Wu:2014fb},  $k$-means clustering is used to divide historical data, and an ARMAX prediction is computed for each possible cluster. The most likely cluster is used to provide a forecast of future traffic flow.

A large body of literature focuses on estimating flows throughout a network using flow measurements and conservation laws. For example, in \cite{Castillo:2008qc}, real-time link flow data is used to predict origin-destination and link flows by modeling a traffic network as a Gaussian Bayesian network, which provides conditional distributions and probability intervals for link flow. In \cite{Zhou:2007uo}, structural deviations from regular traffic patterns, modeled using polynomial trend filters, are used to estimate origin-destination flows. In contrast to these approaches, we focus on predicting traffic flow at a higher resolution for a single intersection.

A variety of model-based and statistical approaches exist for estimating travel times on arterial roads, which is closely associated with the problem of estimating traffic flows and volumes. A flow model is combined with GPS probe data in \cite{Hofleitner:2012oj} to predict arterial travel times using a Bayesian network learning approach. Each link is modeled as being in a state of congestion or undersaturation, and the Bayesian network models the transition between these states. A similar approach is considered in \cite{Jenelius:2013tg} for using GPS probe data to estimate travel time where additional explanatory variables such as speed limits and number of lanes per link are used to reduce the number of parameters in the model. In \cite{Du:2012hc}, an approach for short-term travel time estimation that fuses past and real-time data is presented. These data are weighted based on their quality, which incorporates properties of the data including sensor accuracy and delay. A queuing theory approach is used to provide probability distributions on queue lengths in traffic networks in \cite{Osorio:2011fu}, and the model accounts for finite-capacity queues and captures spatial correlations. While the above approaches focus on aggregate estimates for larger networks, the focus of this paper is on high-resolution estimates of traffic flow for all movements at a single intersection.

In \cite{Guardiola:2014ij}, aggregate daily traffic flow patterns are studied, and a functional principal component decomposition is used reduce the dimensionality of the data. This approach is similar to the analysis proposed in Section \ref{sec:princ-comp-analys} of the present paper, however, \cite{Guardiola:2014ij} focuses on identifying changes in daily traffic patterns for long term traffic monitoring, whereas, in the present paper, we use a principal component decomposition to lay the foundation for our traffic prediction approach. Moreover, \cite{Guardiola:2014ij} focuses on freeway networks and not signalized networks.

When prediction is applied to signalized intersections for adaptive control, prediction horizons are short, \emph{e.g.}, seconds or minutes \cite{Mirchandani:2001zp}. In contrast, the present work focuses on longer prediction horizons on the order of hours for use in conjunction with more traditional traffic signal timing methods such as pre-defined timing plans.

In this paper, we develop a principal-component based prediction scheme for signalized intersections using the \emph{projection to latent structures (PLS)} algorithm \cite{Rosipal:2006kx}. Abstractly, this algorithm decomposes two sets of data to find low-rank, correlated structure between the data sets. For example, in the case of traffic, suppose the objective is to predict traffic flow from 2pm to 8pm using traffic flow measurements up to time 10am. First, historical measurements of traffic flow up to 10am and of traffic flow between 2pm and 8pm are collected. Next, the PLS algorithm decomposes the data into a set of pairs of \emph{latent structures} which provide low-rank approximations of the data sets with the additional requirement that each pair of latent structures is highly correlated. Then, given real-time measurements of traffic flow for a particular day up to 10am, future traffic from 2pm to 8pm is predicted by computing weights for the latent structures.

An important property of this approach is that the proposed traffic predictive control builds on existing standard practices for traffic signal timing. In particular, we consider the common practice of signal timing based on time-of-day plans which are preprogrammed to apply during certain periods of the day \cite[Chapter 5]{Koonce:2008gd}. Each plan is designed to accommodate a certain level of traffic flow at the intersection.  Traditionally, this level of traffic flow is determined based on averaged historical measurements; often, these measurements span only a limited time window over several days and may have been collected years ago. Critically, averaged historical flow is unable to capture anomalous traffic patterns. In \cite{Abbas:2006zh}, a genetic algorithm selects from among a set of predefined plans based on current conditions.

The primary contribution of this paper is a traffic predictive control scheme that uses a prediction of future traffic flow to adjust the time periods for which the time-of-day plans are active and, additionally, suggests predicted levels of traffic around which the timing plans should be designed. Our approach thus does not depend on the exact algorithm that is used to determine timing plans (\emph{i.e.}, green splits) from traffic flow, however, in our case study we employ a delay minimization policy. By ensuring that the proposed control approach is well-aligned with existing practices,  the proposed controller integrates well with existing traffic control hardware which universally accommodate time-of-day timing plans and are often capable of remote changes to these plans. Additionally, practitioners familiar with standard practices are likely to be more receptive to the proposed traffic predictive control.

This paper is organized as follows: Section \ref{sec:preliminaries} describes the problem setup, available data, and details of the case study. Section \ref{sec:princ-comp-analys} analyzes structural trends in the traffic flow data using a principal component analysis, which establishes the foundation for the prediction algorithm presented in Section \ref{sec:traff-pred-from}. Section \ref{sec:traff-pred-contr} proposes a traffic predictive control scheme that uses predictions of future traffic flow to adjust signal timing plans. Section \ref{sec:conclusions} provides concluding remarks, future directions of research, and plans for implementation.

\section{Preliminaries}
\label{sec:preliminaries}
We first characterize the requirements and assumptions generally and then specialize to a specific test site as our case study throughout the paper.
\subsection{Available Data}

We consider a single traffic intersection consisting of a set of $M$ \emph{turn movements} indexed $1$ through $M$. For example, a prototypical intersection consists of four approaches, each approach consisting of a left turn movement, a right turn movement, and a through movement so that $M=12$, as in the case below.

 The flow rate of vehicles along each turn movement is measured and recorded once per interval of time $\Delta$. Thus, $T\triangleq (24\text{ hours})/\Delta$ measurements of the flow are made per movement per day. We assume measurements are available for a total of $D$ days.

For day $d\in\{1,\ldots,D\}$ and movement $m\in\{1,\ldots,M\}$, let $x^d_m(t)$ denote the flow rate of vehicles executing the $m$-th turn movement on day $d$ during  time interval $t\in\{1,\ldots,T\}$ in vehicles per hour (vph). From this notation, we aggregate the measurements into vectors and matrices as follows: 
\begin{alignat}{2}
  \label{eq:4}
  x^d_m&=\begin{bmatrix}x^d_m(1)&x^d_m(2)&\cdots&x^d_m(T)\end{bmatrix}^{\tp},&&\qquad d=1,\ldots, D,\quad m=1,\ldots,M\\
x^d&=\begin{bmatrix}(x^d_1)^{\tp}&(x^d_2)^{\tp}&\cdots&(x^d_M)^{\tp}\end{bmatrix}&&\qquad d=1,\ldots, D
\end{alignat}
where $(\cdot)^{\tp}$ denotes vector transpose. That is, $x^d_m\in \mathbb{R}^{T}$ is the vector of flow measurements along the $m$-th movement on day $d$ and $x^d\in\mathbb{R}^{TM}$ is the vector of flow measurements along all $M$ movements on day $d$.

We define the aggregated data measurement matrix as
\begin{align}
\label{eq:1}
  X=
  \begin{bmatrix}
    (x^1)^{\tp}\\
\vdots\\
(x^D)^{\tp}
  \end{bmatrix}\in \mathbb{R}^{D\times (TM)}.
\end{align}
Typically, $TM\gg D$ so that $X$ is a wide matrix. Table \ref{tab:notation} contains a summary of the notation.

We consider measured turn movements to represent the exogenous demand on the system from the external environment. In particular, it is assumed that measured turn movements are negligibly affected by the choice of control action:
\begin{assum}
\label{assum:exo}
  Turn movement flows originate from an exogenous process and, in particular, are not influenced by the choice of control actions at the intersection.
\end{assum}
For this assumption to be reasonable, we must have the sampling period $\Delta$ long enough so that the impact of the control signal is negligible (\emph{e.g.}, if $\Delta$ is 5 minutes and the cycle time of the signal actuation at the intersection is 2 minutes, then a measurement $x^d_m(t)$ may include between two and three periods of actuation for movement $m$. Thus, $\Delta$ is too short since $x^d_m$ will exhibit undesired oscillations caused by the control signal). Empirical evidence suggests that $\Delta\approx 15$ minutes is reasonable for minimizing such effects.
Additionally, implicit in Assumption \ref{assum:exo} is that vehicles do not reroute in response to changes in signal actuation. The test site presented below is a large intersection for which few alternative routes exist. Thus, the assumption that vehicles do not reroute is reasonable for our study.%

\begin{table}
  \centering
  \begin{tabular}{r |p{4in} |l r}
&& \multicolumn{2}{l}{Value (val) or}\\
&& \multicolumn{2}{l}{Dimension (dim)}\\
    Notation& Meaning&\multicolumn{2}{l}{for Case Study}\\
\hline \hline
 $M$&Number of turn movements at the intersection &$12$&(val)\\
$\Delta$&Interval of time between subsequent measurement of turn movement flows&$15$ min& (val)\\
$D$&Number of days&$132$& (val)\\
 $T$&Number of measurements per movement per day& $96$ &(val)\\
$x^d_m(t)$&Flow rate of the $m$-th turn movement on day $d$ at time $t$ in vehicles per hour&$1$ &(dim)\\
$x^d_m$&Vector of $T$ measurements of the flow rate of the $m$-th turn movement on day $d$&$96$ &(dim)\\
$x^d$& Vector of $(TM)$ measurements of the flow rate on all movements on day $d$&$1152$ &(dim)\\
$X$& Matrix of dimension $D\times (TM)$ containing measures of the flow rate on all movements for all days&$132\times 1152$& (dim)
  \end{tabular}
  \caption{Summary of notation.}
  \label{tab:notation}
\end{table}

\subsection{Case Study Test Site}

\label{sec:case-study-test}
\begin{figure}
  \centering
  \begin{tabular}{c c}
    \includegraphics[height=2.5in]{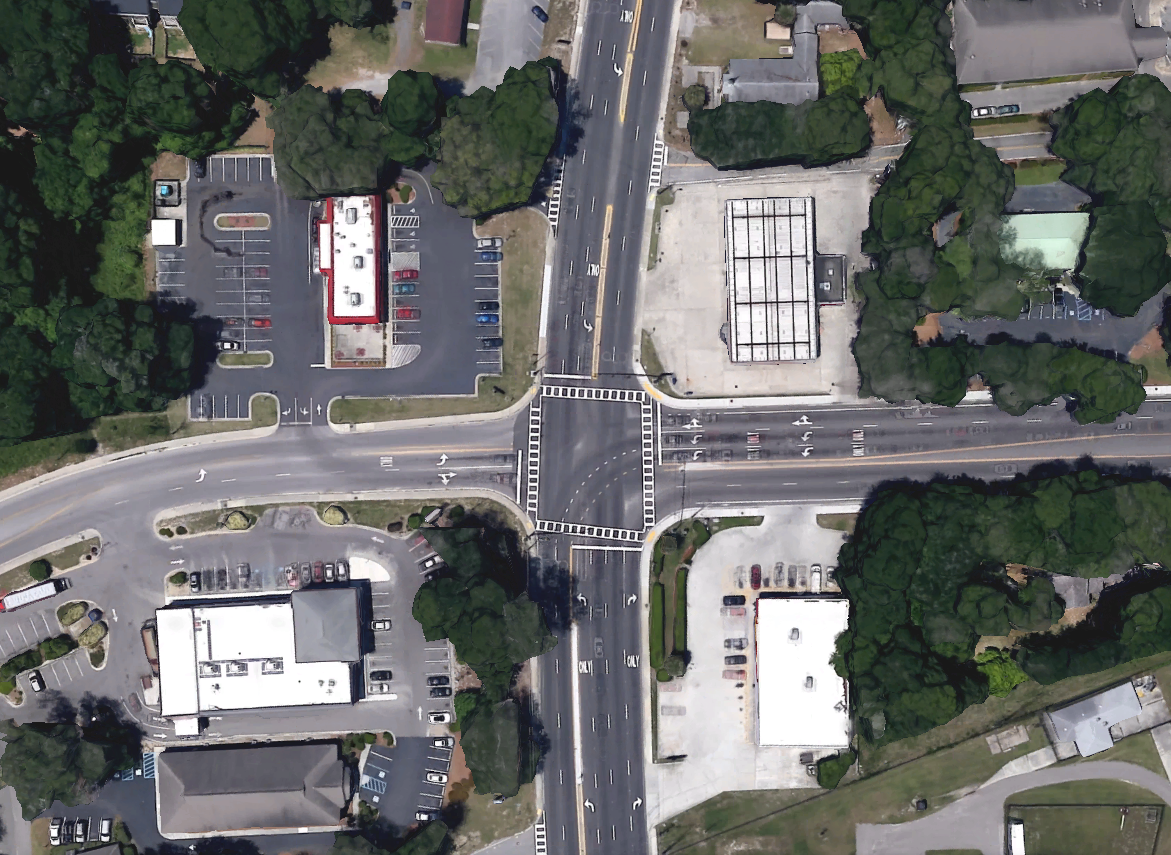}&   \includegraphics[height=2.5in, clip=true, trim=2in 0in 2in 0in]{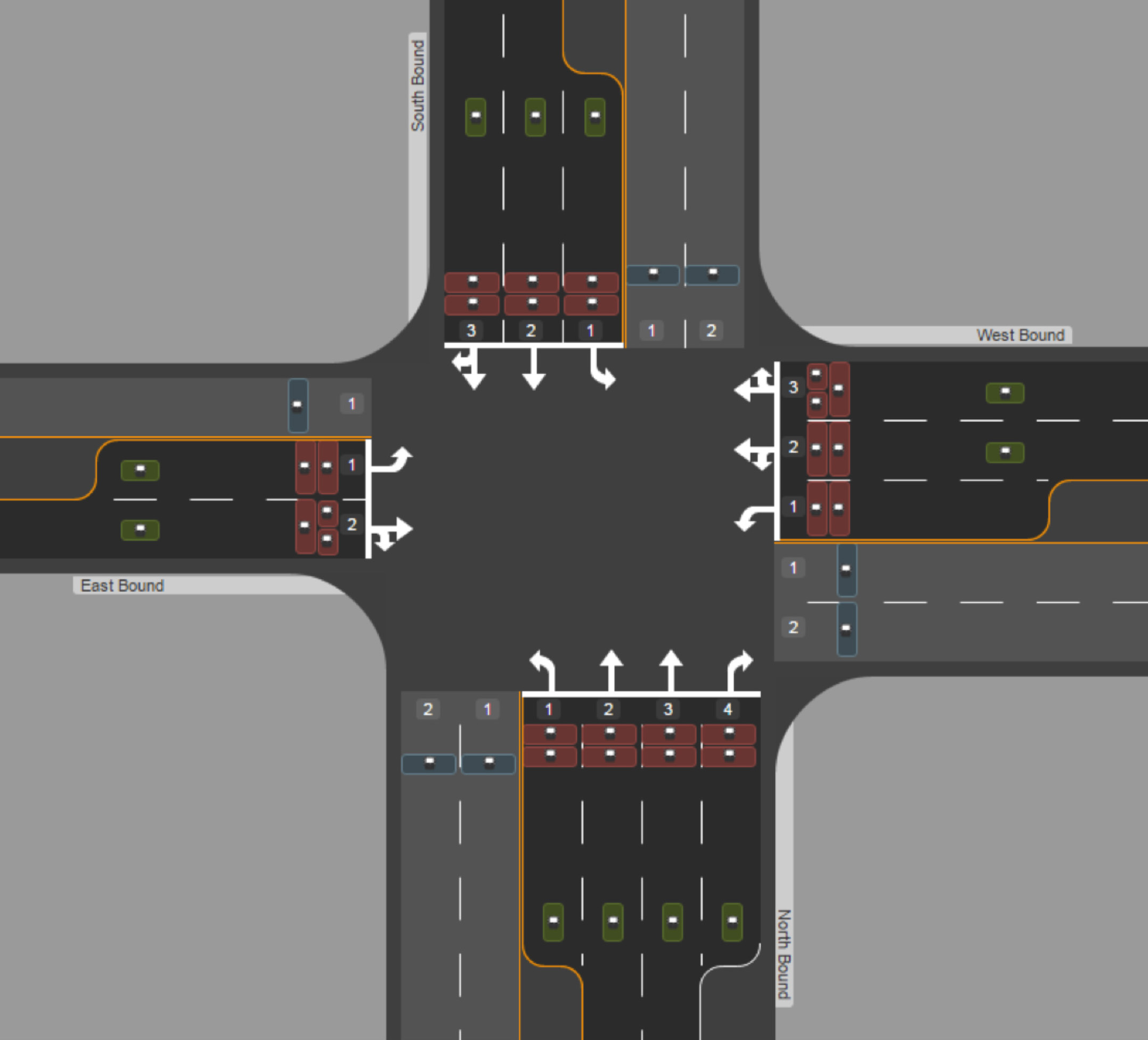}\\
(a)&(b)
  \end{tabular}
  \caption{Test site in Beaufort, South Carolina. (a) Image of the test site which consists of one intersection with four approaches. Each approach consists of a left turn movement, a right turn movement, and a through turn movement. (b) Schematic depiction of the lane configuration and placement of sensors at the intersection.  Stopbar sensors are indicated with red blocks, and departure lane sensors are indicated with blue blocks, and advance sensors placed upstream are indicated with green block. The sensors are manufactured by Sensys Networks, Inc. and their measurements are fused to determine the turn movement of each vehicle that transits the intersection.}
  \label{fig:beaufort}
\end{figure}
In this paper, we focus on a test site in Beaufort, South Carolina consisting of one intersection with four approaches. Each approach consists of a left turn movement, a right turn movement, and a through turn movement for a total of $M=12$ movements. On the left of Figure \ref{fig:beaufort} is a satellite image of the intersection and on the right of Figure \ref{fig:beaufort} is a schematic depiction of the lane configuration and sensor placements at the intersection. A total of 44 magnetometer sensors provide real-time measurements of the movement of vehicles through the intersection; the sensors are manufactured by Sensys Networks, Inc. \cite{Haoui:2008qf}. By measuring changes in magnetic field, the sensors are able to detect the presence of each vehicle at the intersection. Detection events from sensors on approaching lanes with subsequent detection events on departure lanes together with the signal phase are then used to determine the turn movement of each vehicle. These data are aggregated every $\Delta=15$ minutes to provide a measurement of the number of vehicles executing each turn movement at the intersection. These data are reported in vehicles per hour. There are therefore a total of $T=96$ measurements of flow per movement per day. This high-resolution data acquisition system provides a rich dataset for managing traffic at intersections \cite{Muralidharan:2016dp}.

The intersection consists of four \emph{approaches} and four \emph{departures}. A physically adjacent approach/departure pair is referred to as a \emph{leg} of the intersection. As is labeled in Figure \ref{fig:beaufort}b, the left leg is the Eastbound (EB) leg, the right leg is the Westbound (WB) leg, the top leg is the Southbound (SB) leg, and the bottom leg is the Northbound leg (NB). Each approach consists of a left turn (LT) movement, a through (T) movement, and a right turn (RT) movement. We sometimes use, \emph{e.g.}, the notation ``$m=\text{EB LT}$'' to indicate the movement index corresponding to the Eastbound left turn movement.

\begin{figure}
  \centering
  \begin{tabular}{@{}l@{} @{}l@{}}
    \includegraphics[height=1.6in]{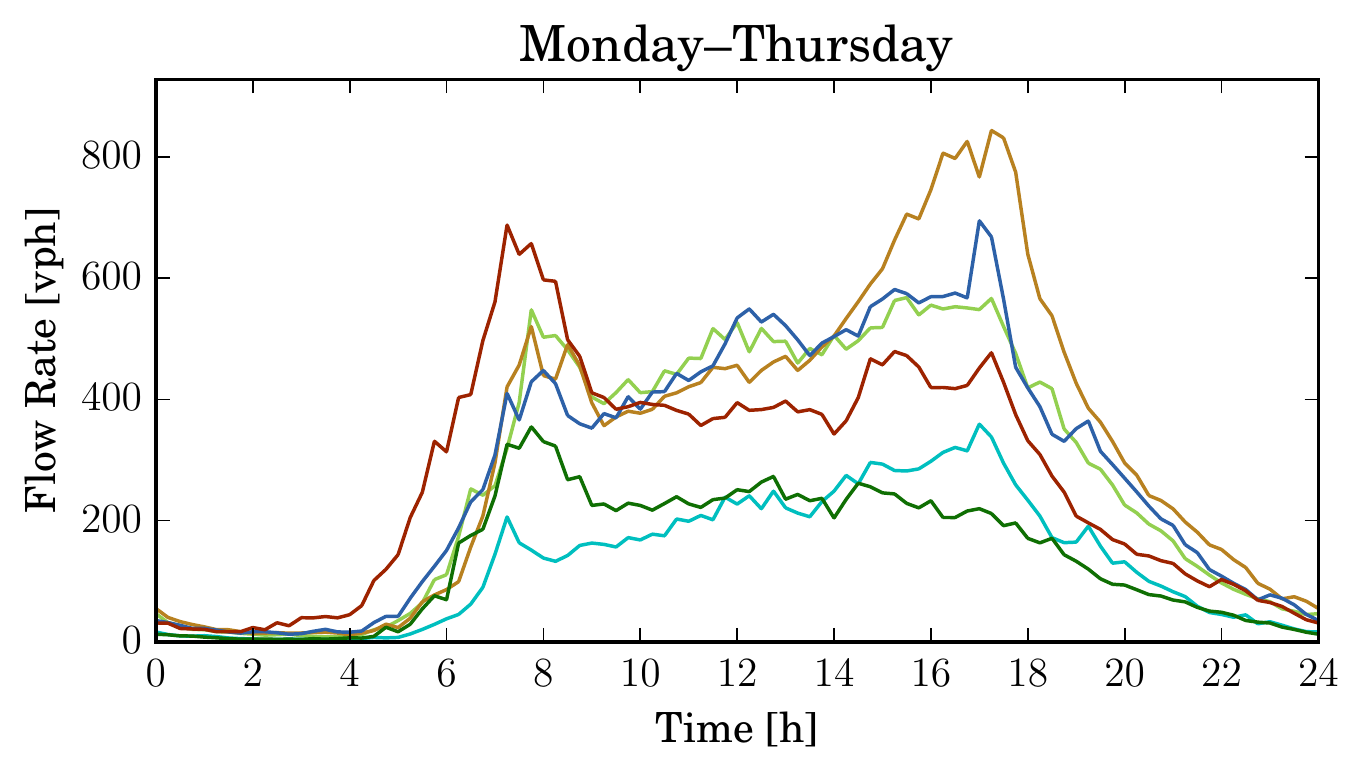}&    \includegraphics[height=1.6in]{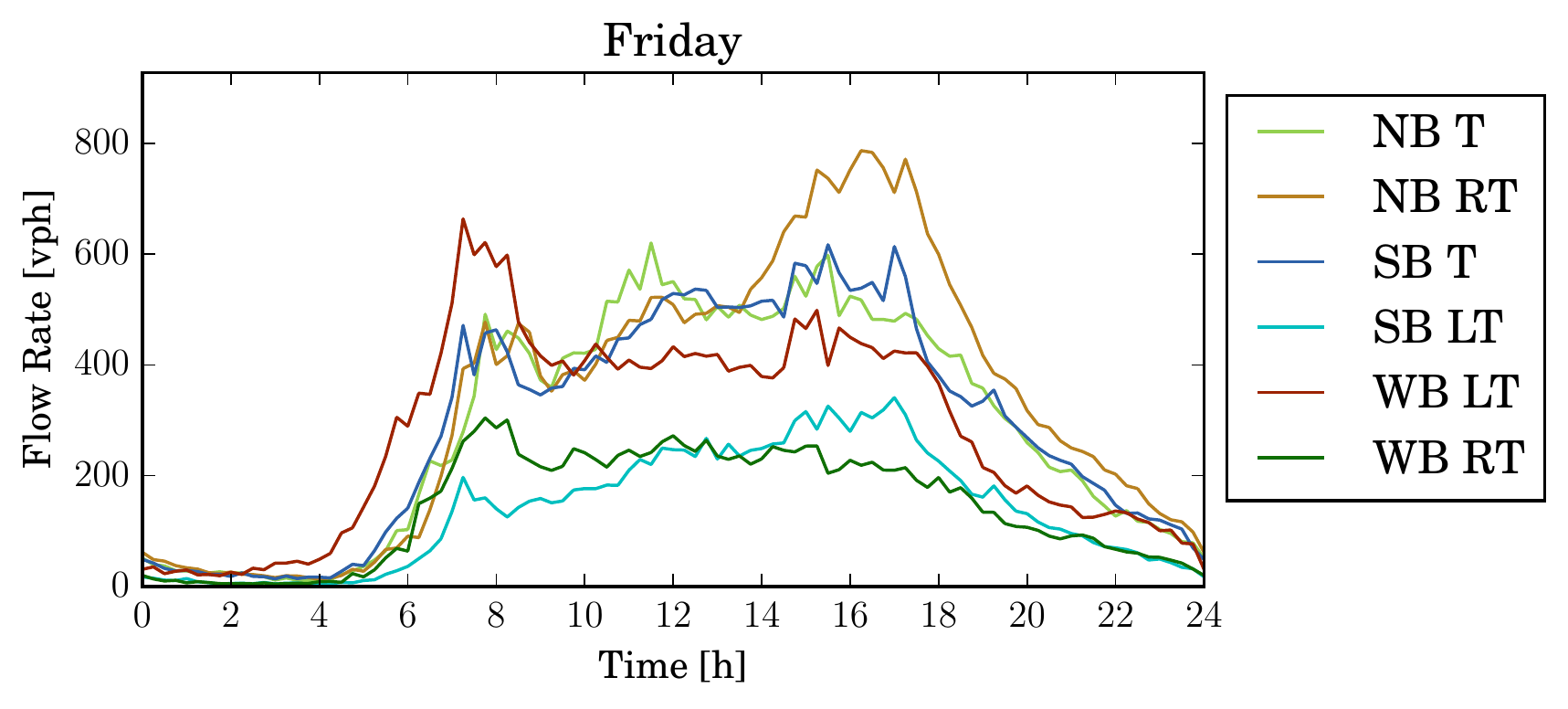}\\
    \includegraphics[height=1.6in]{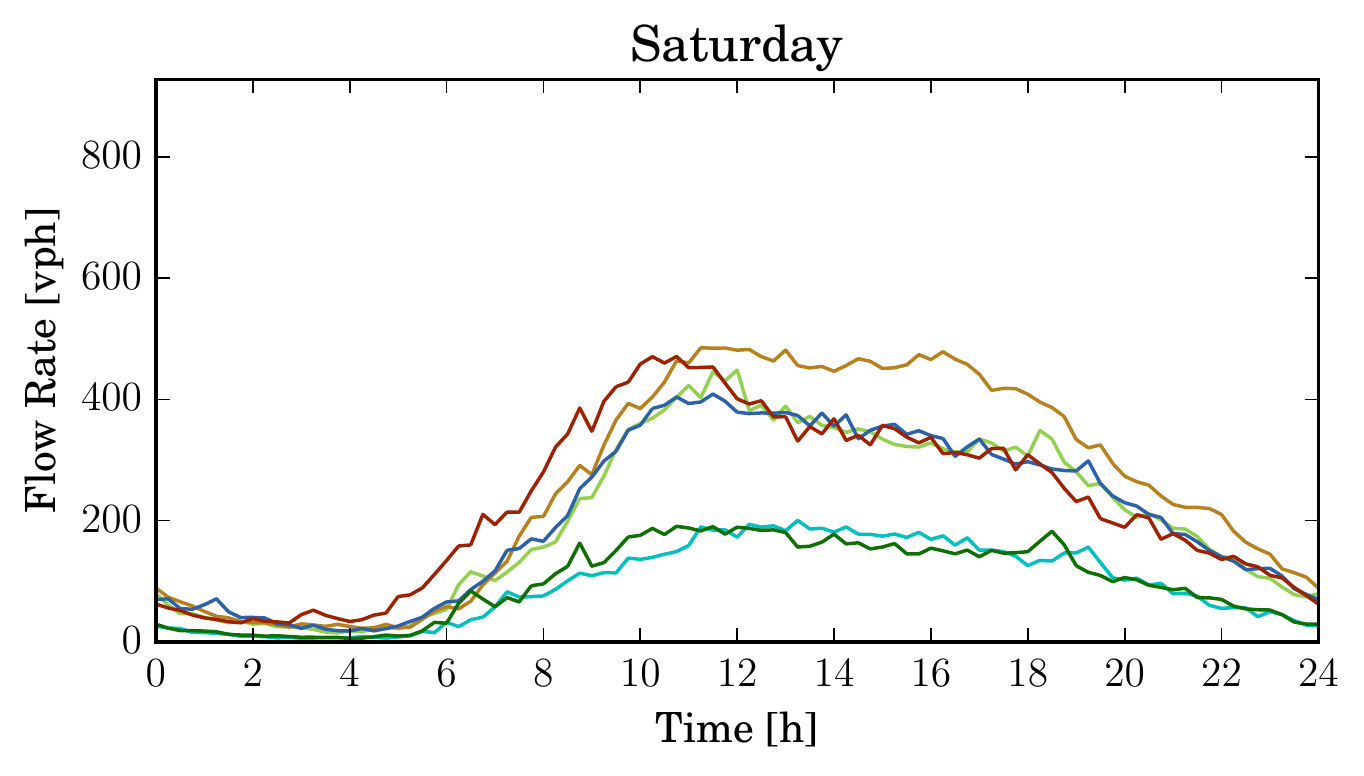}&    \includegraphics[height=1.6in]{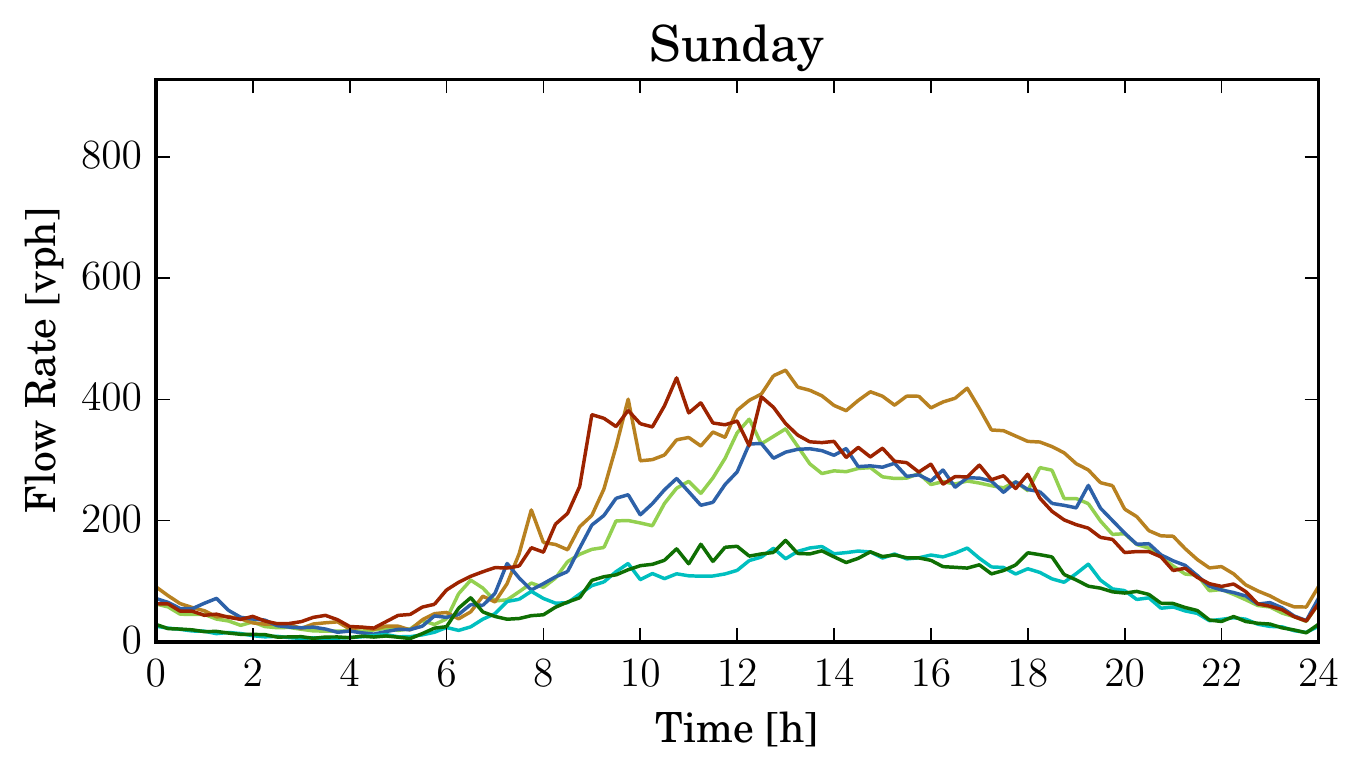}
  \end{tabular}
  \caption{Average flow rates over a 7 month period from December 2014 to July 2015 for Monday--Thursday (132 days), Friday (32 days), Saturday (34 days), and Sunday (34 days). Traffic originating from and bound for the West (\emph{i.e.}, EB LT, EB T, EB RT, NB LT, SB RT, and WB T movements) are not shown since traffic volumes for these movements are much lower than the remaining movements.  Eleven days in this period contain missing measurements and are omitted from the analysis.}
  \label{fig:DOW}
\end{figure}

Traffic at the case study intersection throughout the week has similar profiles each Monday through Thursday, and a different profile on Fridays, on Saturdays, and on Sundays. Figure \ref{fig:DOW} shows the average flow at the intersection for each of these groups for data spanning December 2014 to July 2015. Eleven days from this period are omitted due to missing measurements on these days. To ensure the figures are legible, the plots do not include traffic flows originating from or bound for the leftmost (that is, EB) leg (\emph{i.e.}, movements WB T, NB LT, SB RT, and all EB movements) because these movements contribute much lower flow than the remaining movements (approximately 50--100 vph during peak periods).

The grouping depicted in Figure \ref{fig:DOW} is verified via standard $k$-means clustering for which, when sufficiently many clusters are computed, the clusters tend to contain days from only one of these four groups, but for more than four clusters, no meaningful division of the Monday--Thursday group is discernible.

\section{Principal Components of Traffic Flow}
\label{sec:princ-comp-analys}
In this section, we consider a low-rank decomposition of the measured traffic flow. This decomposition is obtained via a principal component (PC) analysis of the data, and we will see that this simple approach reveals much about the data.  Furthermore, a PC analysis establishes the foundation for the PLS-based traffic prediction strategy that is the focus of Section~\ref{sec:traff-pred-from} and is the main contribution of this paper.
\subsection{Computation of Principal Components}
Recall our data matrix $X$ constructed in \eqref{eq:1} from the vectors $x^d$, $d=1,\ldots, D$ containing the turn movement flow measurements for each movement over the course of each day $d$. Throughout the remainder of the paper, we focus exclusively on data from December 2014 to July 2015, Monday--Thursday for a total of $D=132$ days.
Define
\begin{align}
  \label{eq:2}
  \bar{x}_m=\frac{1}{D}\sum_{d=1}^Dx_m^d\in\mathbb{R}^{T}
\end{align}
to be the mean measured flow along movement $m$ over the course of a day, and define
\begin{align}
  \label{eq:3}
  \bar{x}=\begin{bmatrix}\bar{x}_1^{\tp}&\ldots &\bar{x}_M^{\tp}\end{bmatrix}^{\tp}.
\end{align}

\begin{figure}
  \centering
  \begin{tabular}{@{}c@{}  @{}c@{}  @{}c@{}}
    \includegraphics[width=.33\textwidth, clip=true, trim=0in .8in 0in 0in]{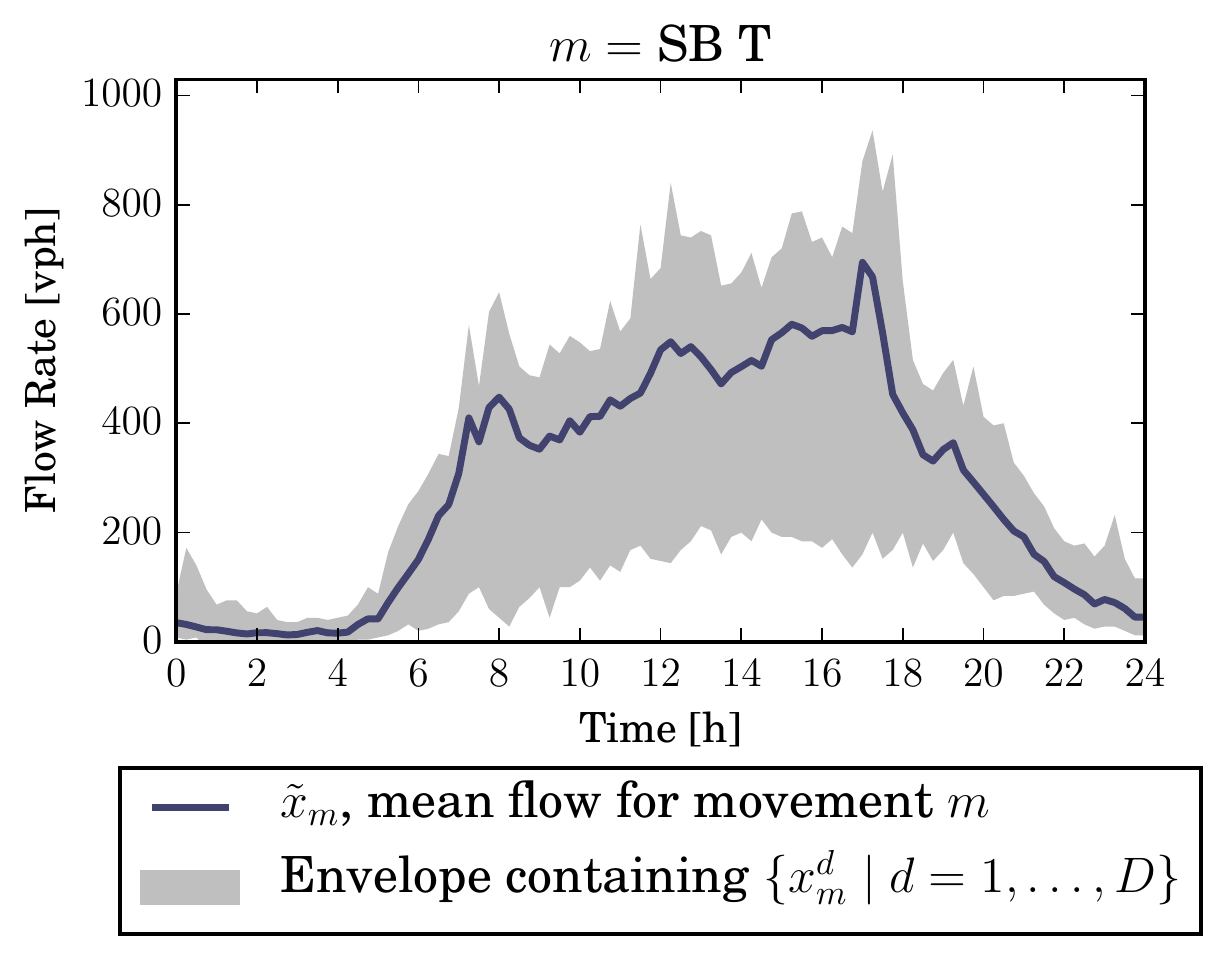}&
    \includegraphics[width=.33\textwidth, clip=true, trim=0in .8in 0in 0in]{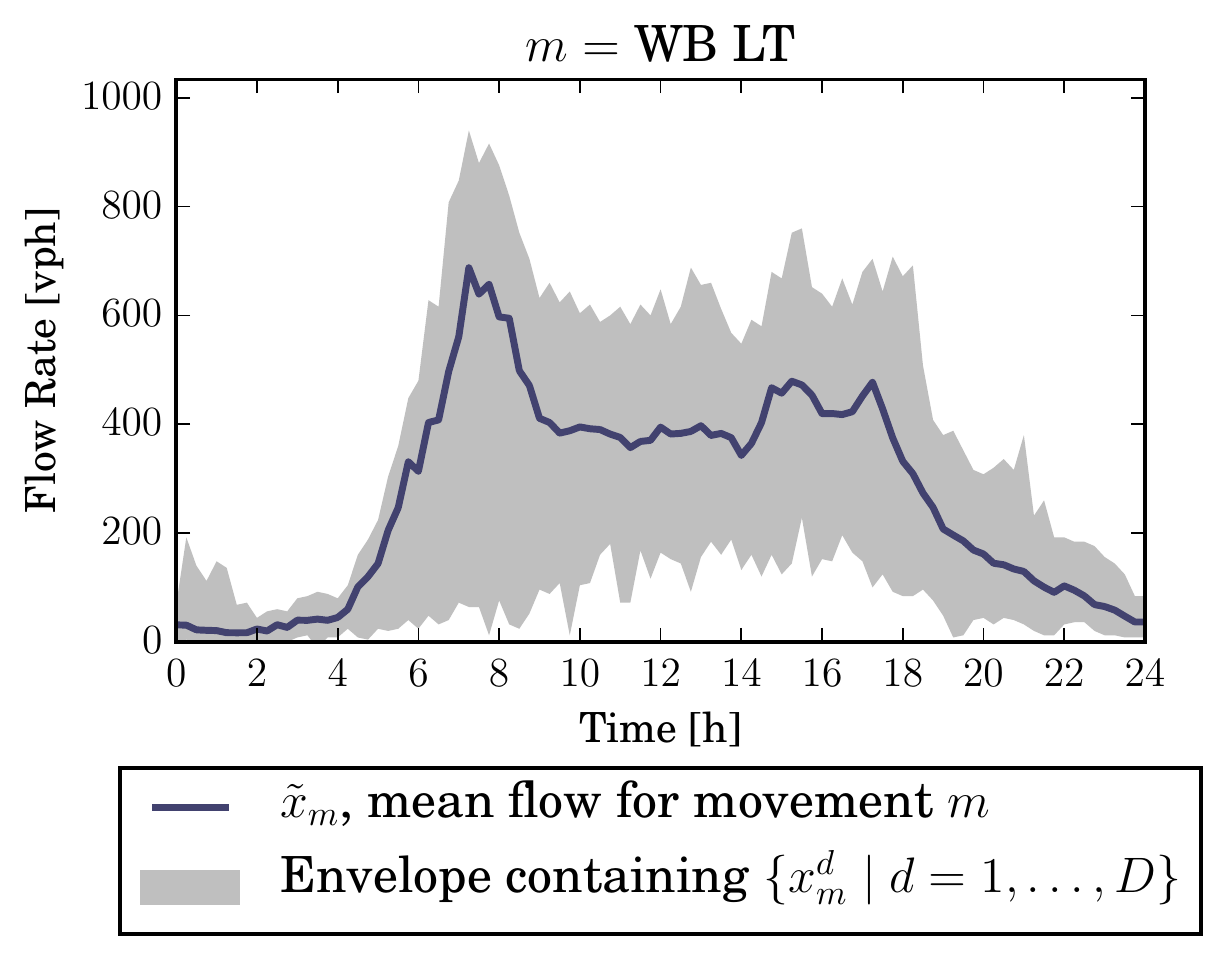}&
    \includegraphics[width=.33\textwidth, clip=true, trim=0in .8in 0in 0in]{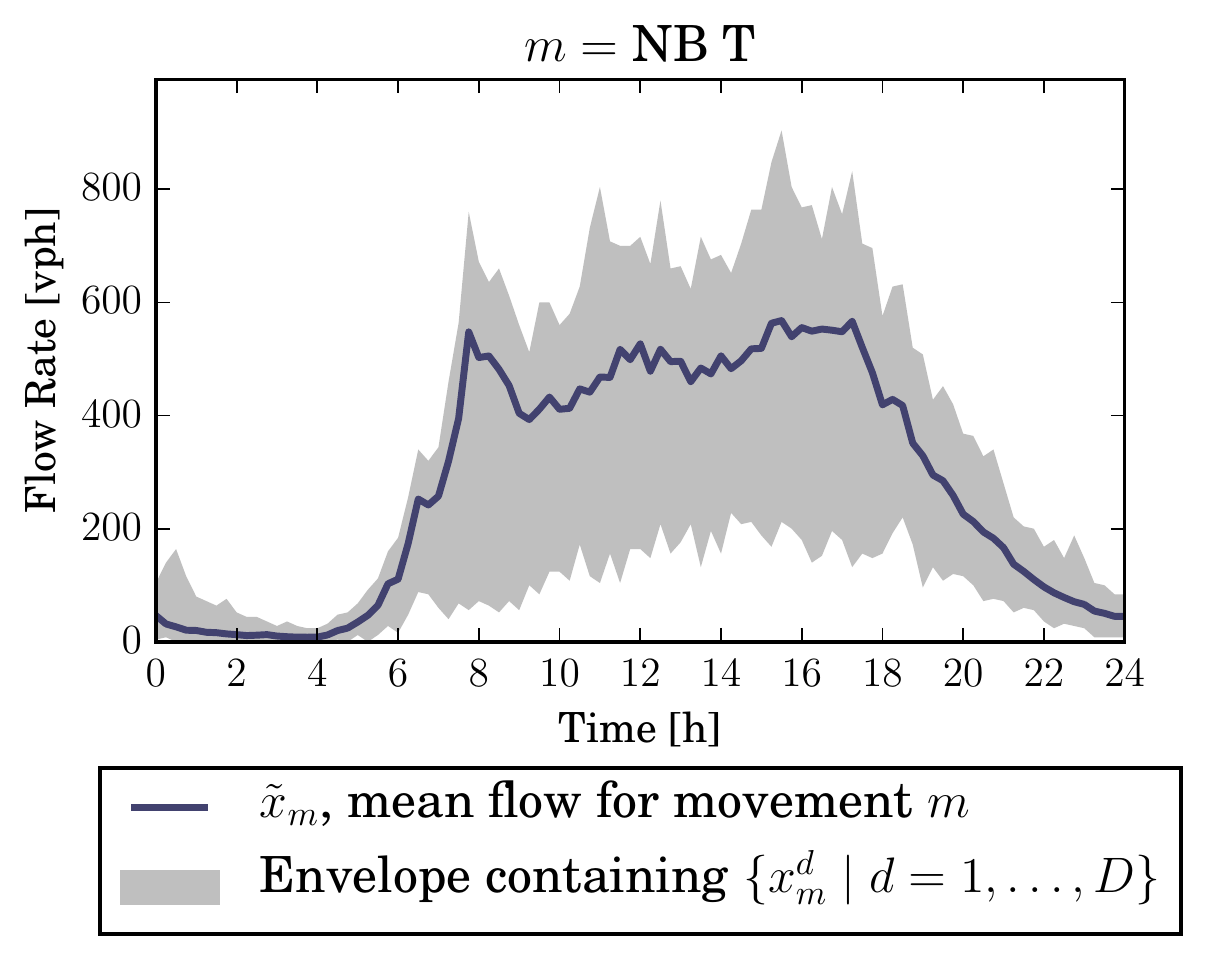}\\
  \imagetop{\includegraphics[width=.33\textwidth, clip=true, trim=0in .8in 0in 0in]{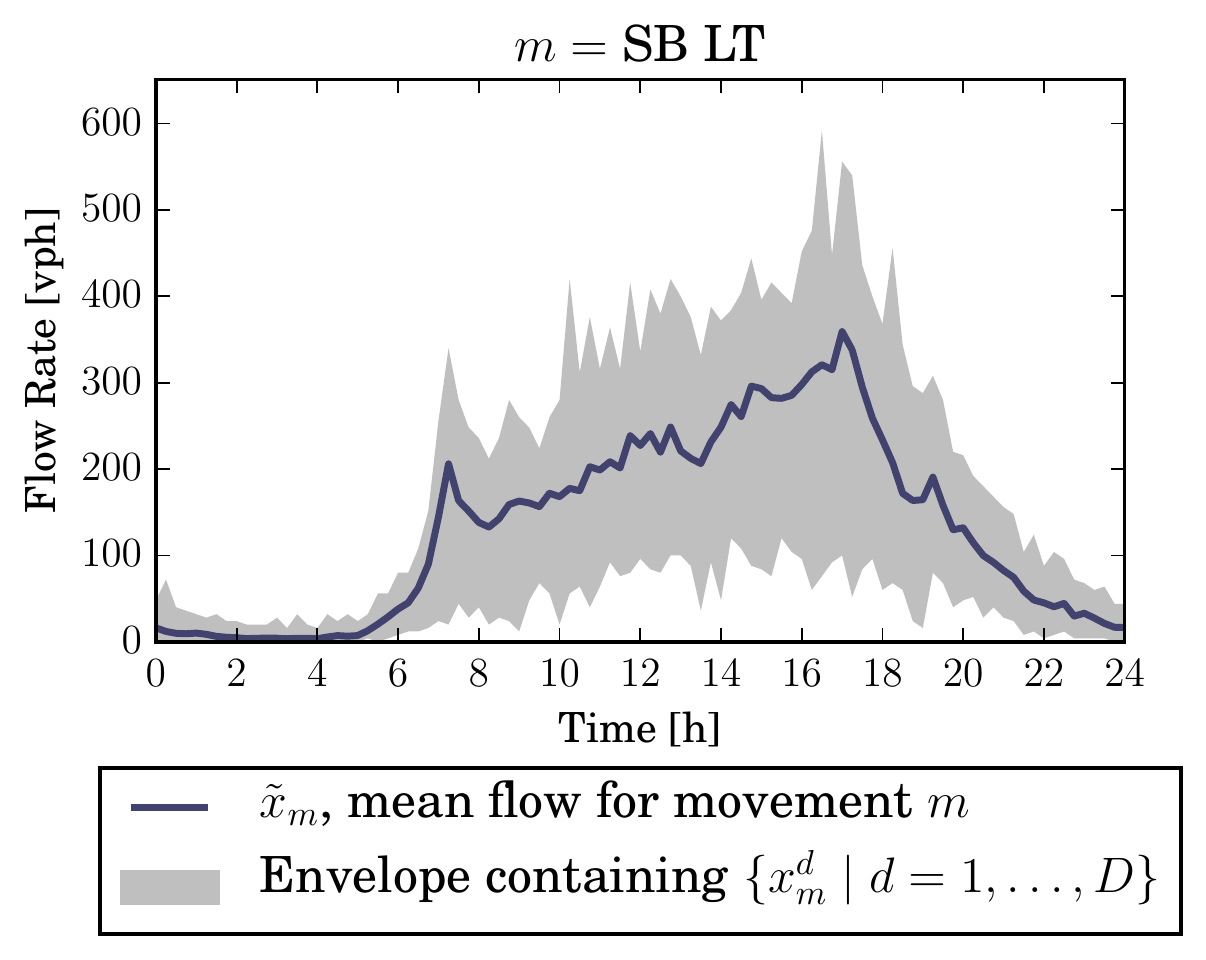}}&
  \imagetop{\includegraphics[width=.33\textwidth, clip=true, trim=0in 0in 0in 0in]{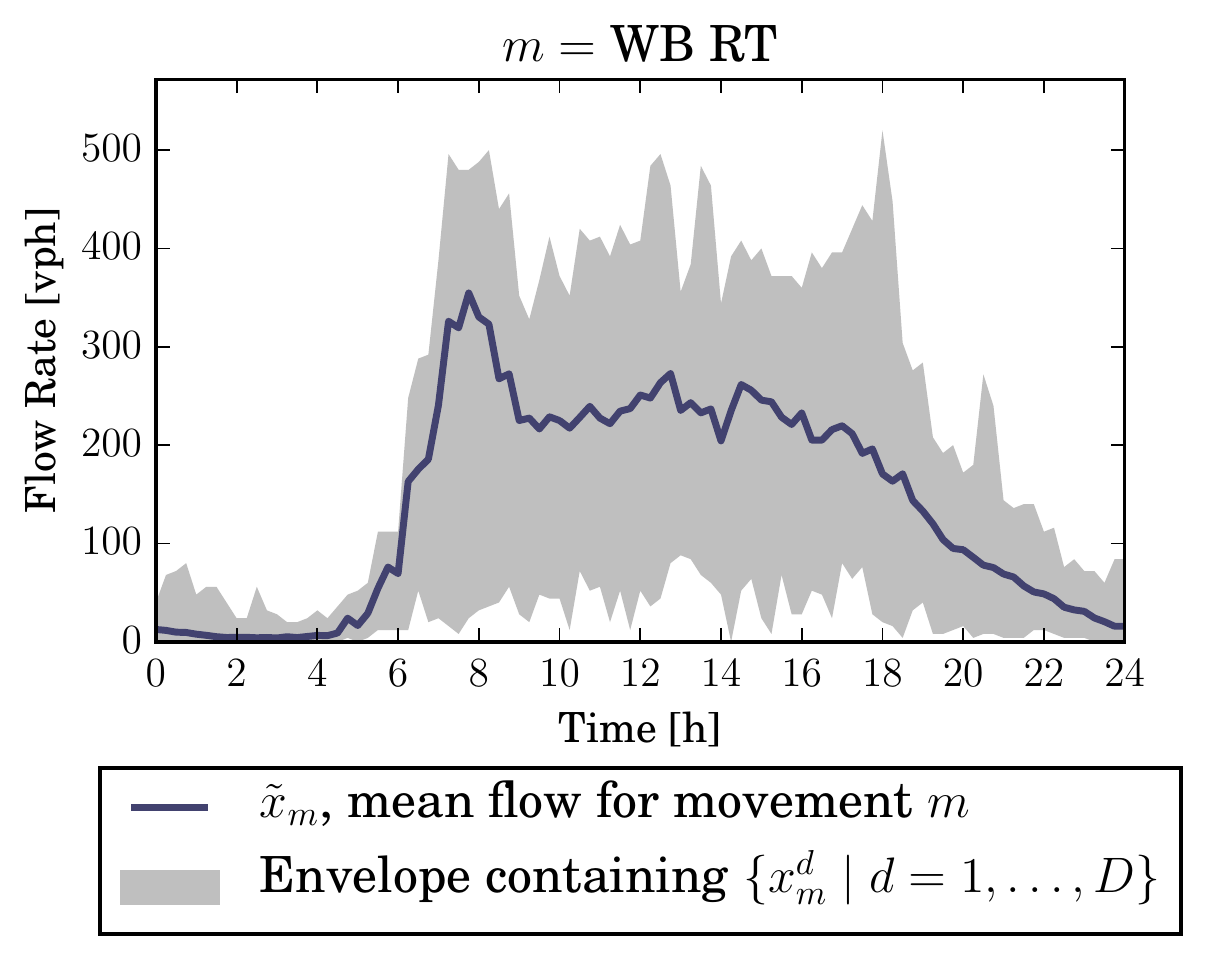}}& 
  \imagetop{\includegraphics[width=.33\textwidth, clip=true, trim=0in .8in 0in 0in]{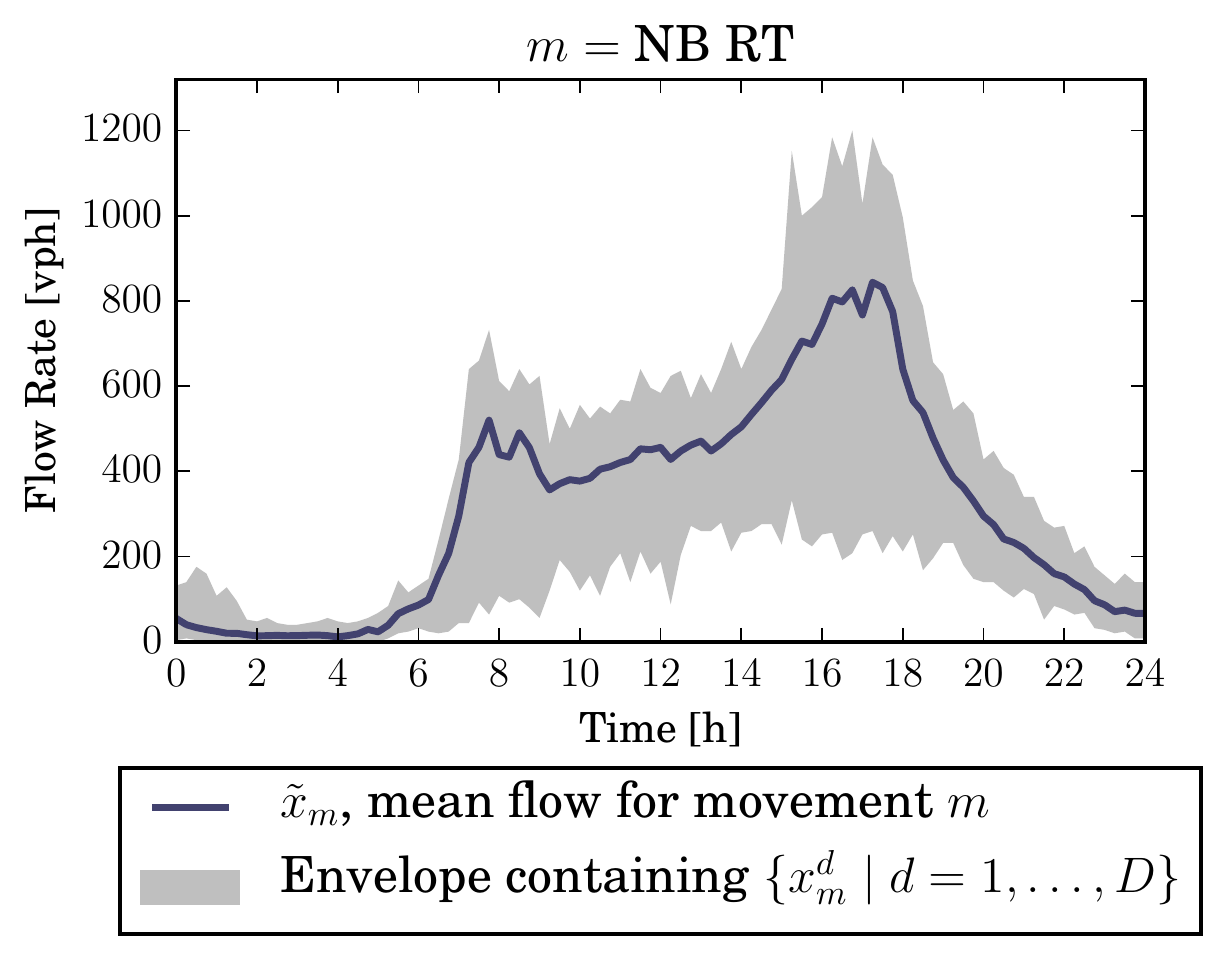}}
  \end{tabular}
  \caption{Plots of mean flow for six turn movements along with the envelope containing all measured data for all days Monday--Thursday. }
  \label{fig:env}
\end{figure}

The top-left image of Figure \ref{fig:DOW} plots $\bar{x}_m$ for the six indicated movements. Figure \ref{fig:env} shows separately the mean measured flow for each of these six movements and additionally displays a shaded region that represents the envelope containing $x^d_m$ for all $d=1,\ldots,D$. Clearly, there is a large variation in measured flow rate around the mean. Our first objective is to find a low-rank decomposition of the data to characterize this variation. Specifically, given the rank parameter $N\geq 1$, we wish to find a  collection of $N$ principal components $q^1,q^2,\ldots, q^N$ with each $q^i\in\mathbb{R}^{TM}$ and for each $d=1,\ldots,D$ a vector of weights $w(d)\in \mathbb{R}^N$ with
\begin{equation}
  \label{eq:8}
    w(d)= \begin{bmatrix}w^1(d)&w^2(d)&\ldots&w^N(d)\end{bmatrix}^{\tp}
\end{equation}
such that
\begin{align}
  \label{eq:5}
  x^d\approx \bar{x}+\sum_{i=1}^Nw^i(d)q^i,
\end{align}
that is, each of the 132 daily  mean-centered measurement vectors $x^d-\bar{x}$ is approximately represented by a linear combination of the principal components. If \eqref{eq:5} holds, then we may effectively replace $x^d$ by its weight vector $w(d)$.
As will be evident below, much of the day-to-day variation can be captured by a few (three to five) principal components.

To make our search for principal components precise, let $\tilde{x}^d=x^d-\bar{x}$ for all $d=1,\ldots, D$ and define
\begin{align}
  \label{eq:9}
  \tilde{X}=
  \begin{bmatrix}
(\tilde{x}^1)^{\tp}\\
\vdots\\
(\tilde{x}^D)^{\tp}
  \end{bmatrix} = X-\mathbf{1}_{D}\bar{x}^{\tp}
\end{align}
where $\mathbf{1}_n$ denotes the all-ones vector of length $n$, and let
\begin{alignat}{2}
  \label{eq:10}
  Q&=
  \begin{bmatrix}
    q^1&\cdots&q^N
  \end{bmatrix}&&\in\mathbb{R}^{(TM)\times N}\\
W&=
  \begin{bmatrix}
    w^{\tp}(1)\\
\vdots\\
w^{\tp}(D)
  \end{bmatrix}&&\in\mathbb{R}^{D\times N}
\end{alignat}
where $w^{\tp}(d)$ denotes the transpose of the weight vector $w(d)$.

We reformulate \eqref{eq:5} and specifically seek $Q$ and $W$ to minimize
\begin{align}
  \label{eq:11}
  ||\tilde{X}-WQ^{\tp}||_F
\end{align}
where $||\cdot||_F$ denotes the Frobenius matrix norm. It is well known that a pair $(W,Q)$ minimizing \eqref{eq:11} is obtained via the singular value decomposition of $\tilde{X}$. This decomposition results in $q^1,\ldots,q^N$ that are orthonormal and are assumed ordered with respect to the (descending) order of the singular values of $\tilde{X}$. In this way, $q^1$ is the principal component that lies in the direction maximizing the explained variance of the data, \emph{i.e.}, maximizes $\sum_{d=1}^D (q^1)^{\tp}\tilde{x}^d$, $q^2$ is the principal component that lies in the direction maximizing the explained variance subject to the constraint that $(q^1)^{\tp}q^2=0$, \emph{etc.} It is standard to assume without loss of generality that each $q^i$ is of unit norm; here, we multiply each $q^i$ by a factor of 100 and therefore divide each $w^i(d)$ by a factor of 100 to make the plots more intuitive and legible.

\subsection{Case Study}
\label{sec:case-study-principal}

\begin{figure}
  \centering
  \begin{tabular}{l l}
    \includegraphics[height=1.6in]{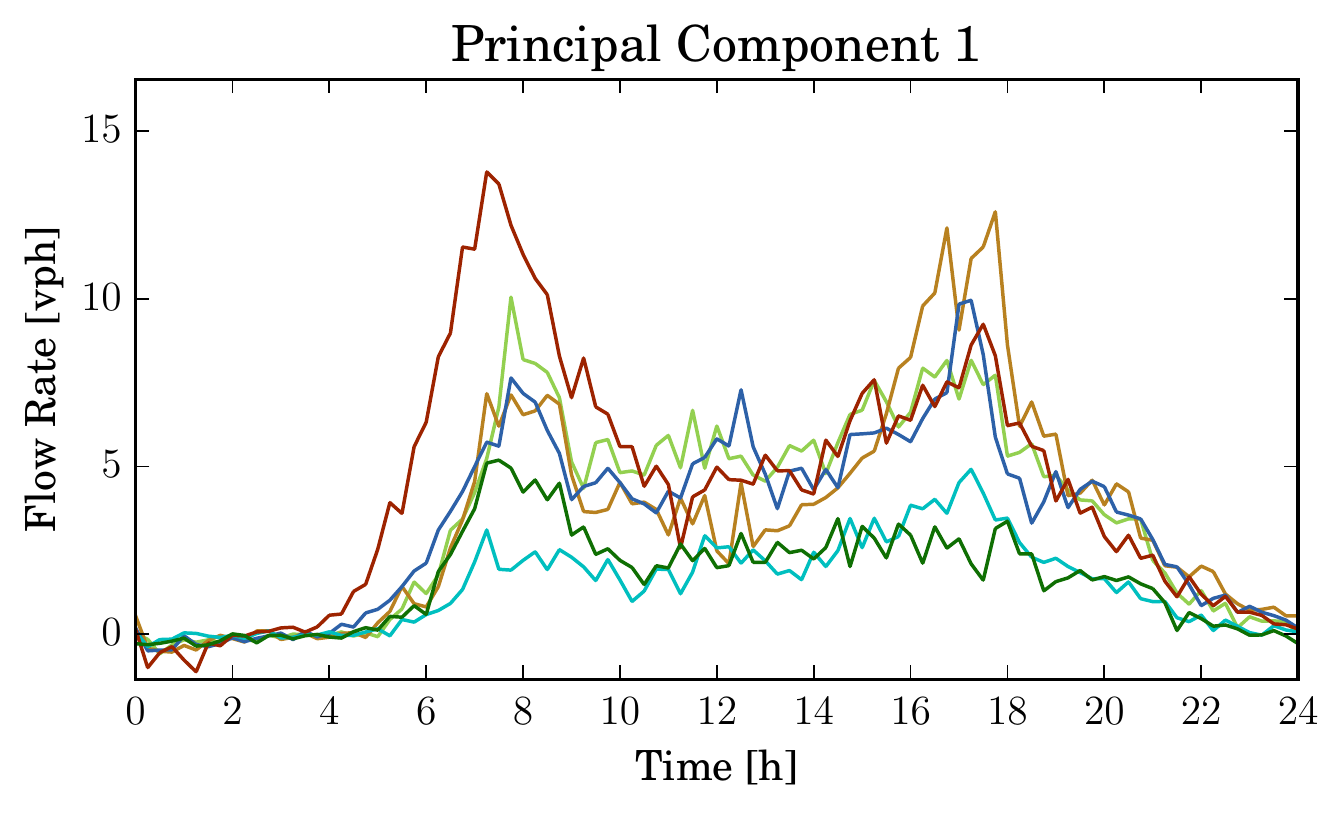}&    \includegraphics[height=1.6in]{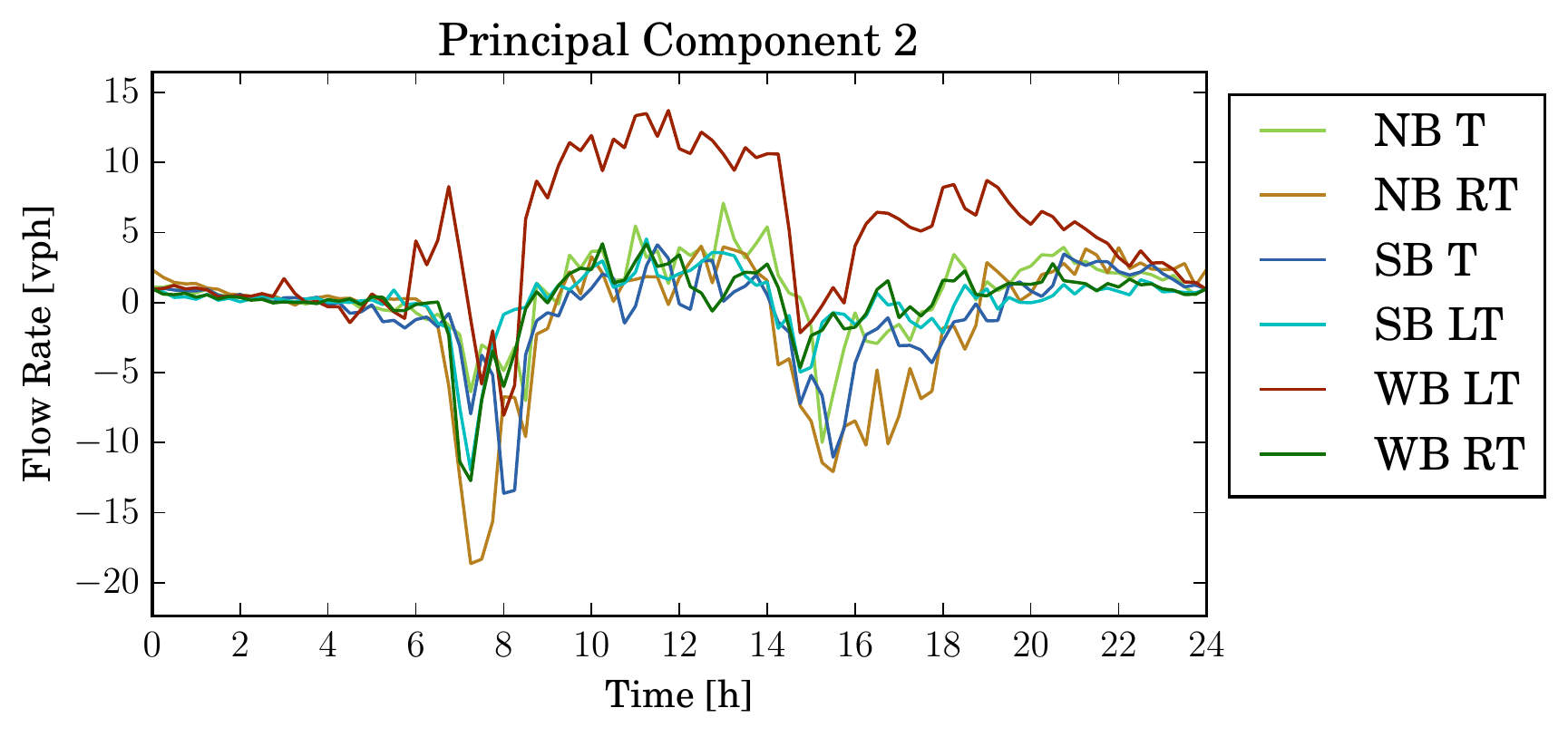}\\
    \includegraphics[height=1.6in]{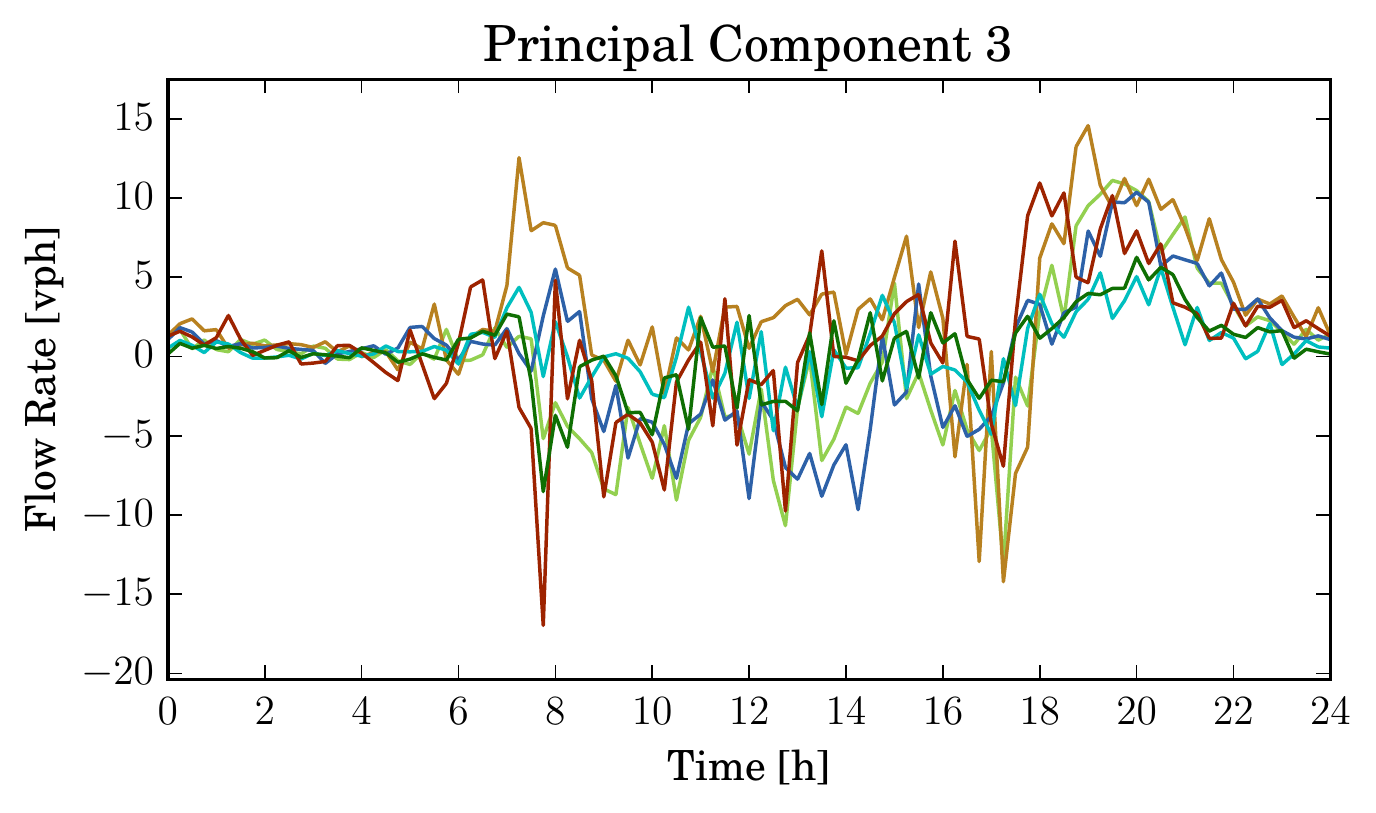}&    \includegraphics[height=1.6in]{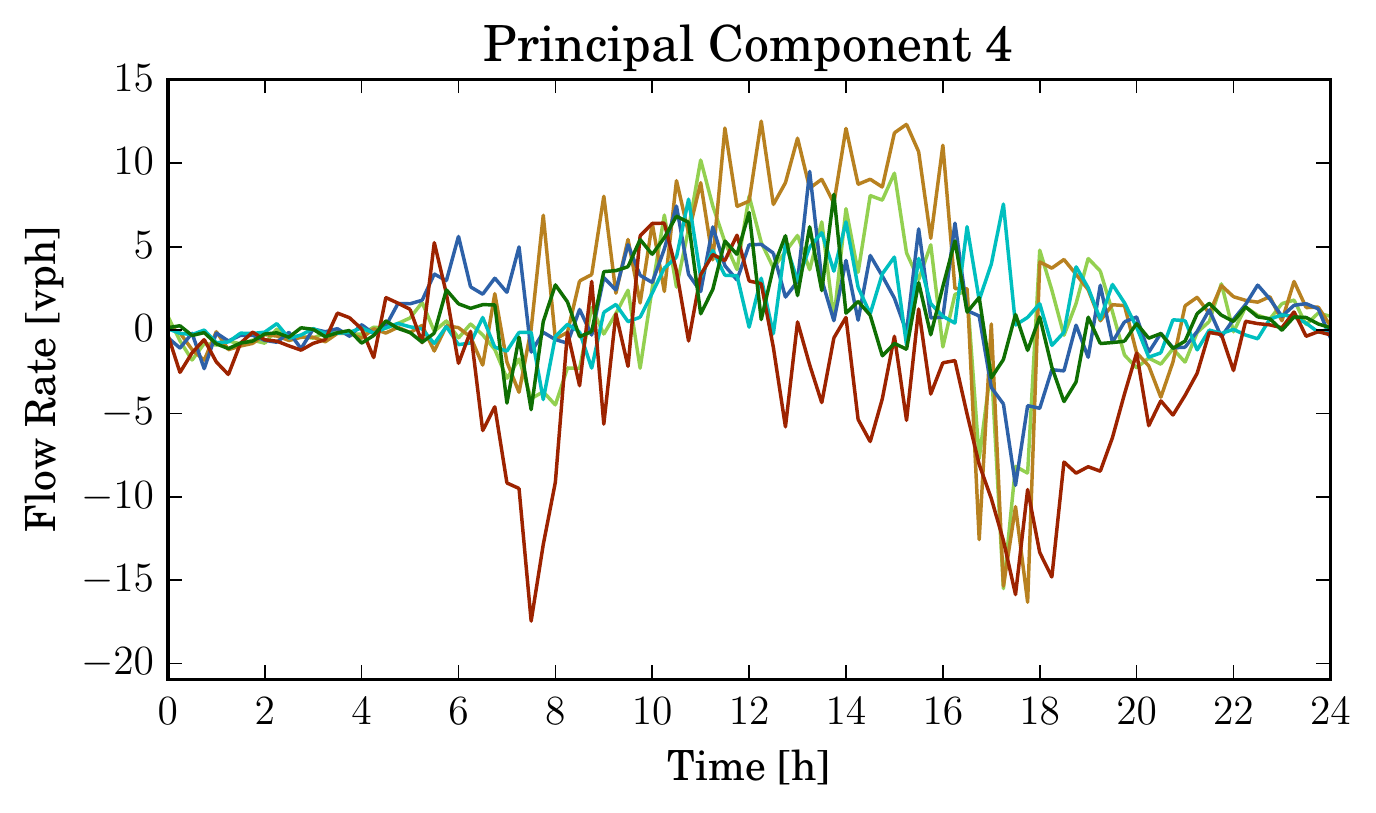}
  \end{tabular}
  \caption{The first four principal components of the traffic data. Each component is $TM$-dimensional but is plotted as $M$ traces of length $T$ in the figures, each trace corresponding to one turn movement over 24 hours.}
  \label{fig:pca}
\end{figure}

\begin{figure}
  \centering
\includegraphics[width=.5\textwidth]{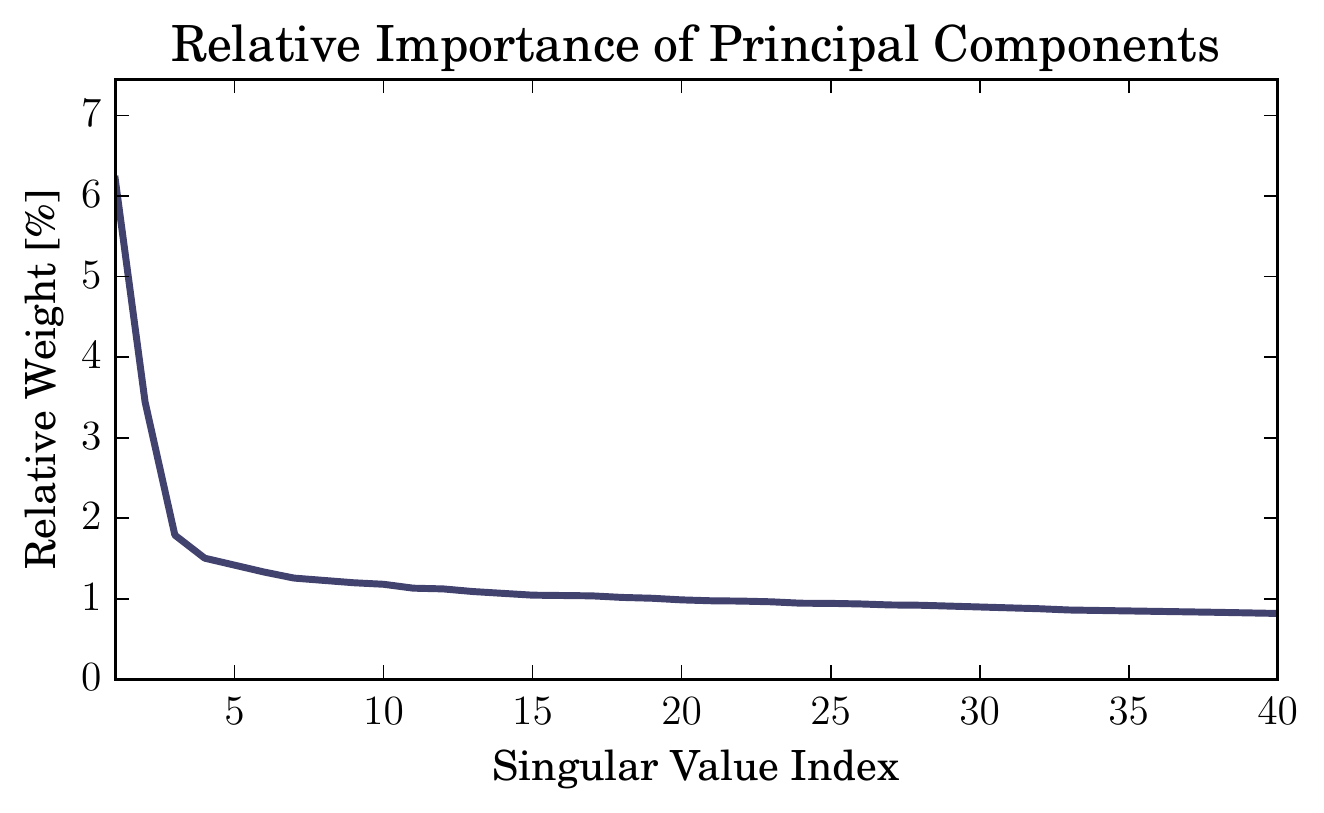}
  \caption{Relative weight of the singular values obtained from a singular value decomposition of $\tilde{X}$. The singular values are ordered by magnitude and the relative weight is a percent of the sum of all singular values. The plot indicates that a significant portion of the variation in traffic flow is explained by the first few principal components of the decomposition.}
  \label{fig:variance}
\end{figure}

In Figure \ref{fig:pca}, we plot the first four principal components of our dataset. Each $q^i$ is $TM$-dimensional, however, for ease of comprehension, each principal component is plotted as $M$ traces of length $T$, each trace corresponding to one turn movement.    From the figure, we observe identifiable trends in each of the principal components, which are discussed next. We also note that the principal components exhibit some high-frequency oscillation, especially in the AM peak period. This oscillation is generally of lower magnitude than the main trends in the components and would likely be ameliorated with more data.

Figure \ref{fig:variance} plots the singular values of $\tilde{X}$ normalized by the sum of all singular values. From the plot, we see that a significant portion of the variation in the traffic flow is explained by the first few principal components of the decomposition.

\begin{figure}
  \centering
\begin{tabular}{c c c}
  \includegraphics[width=.48\textwidth]{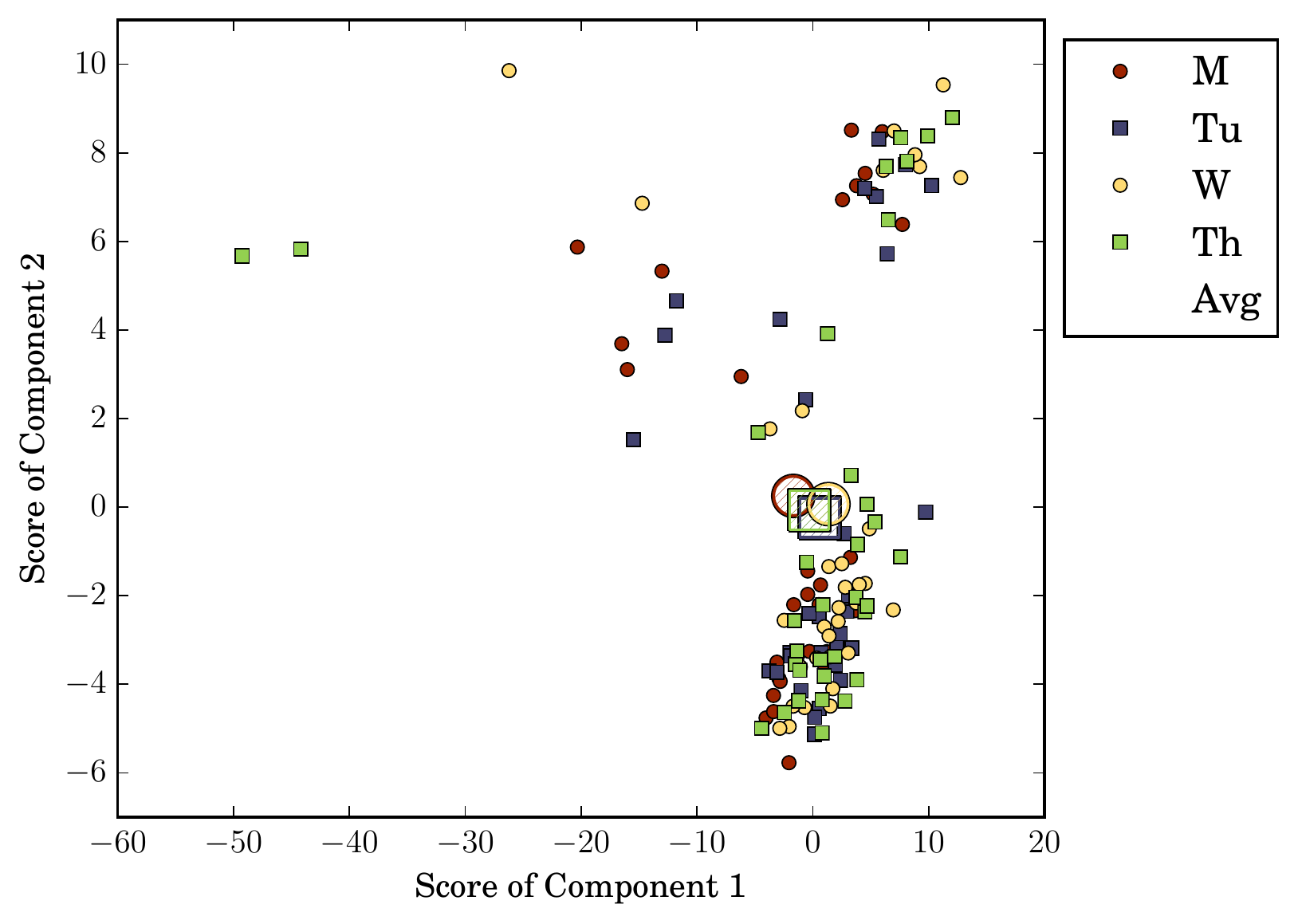}& \includegraphics[width=.48\textwidth]{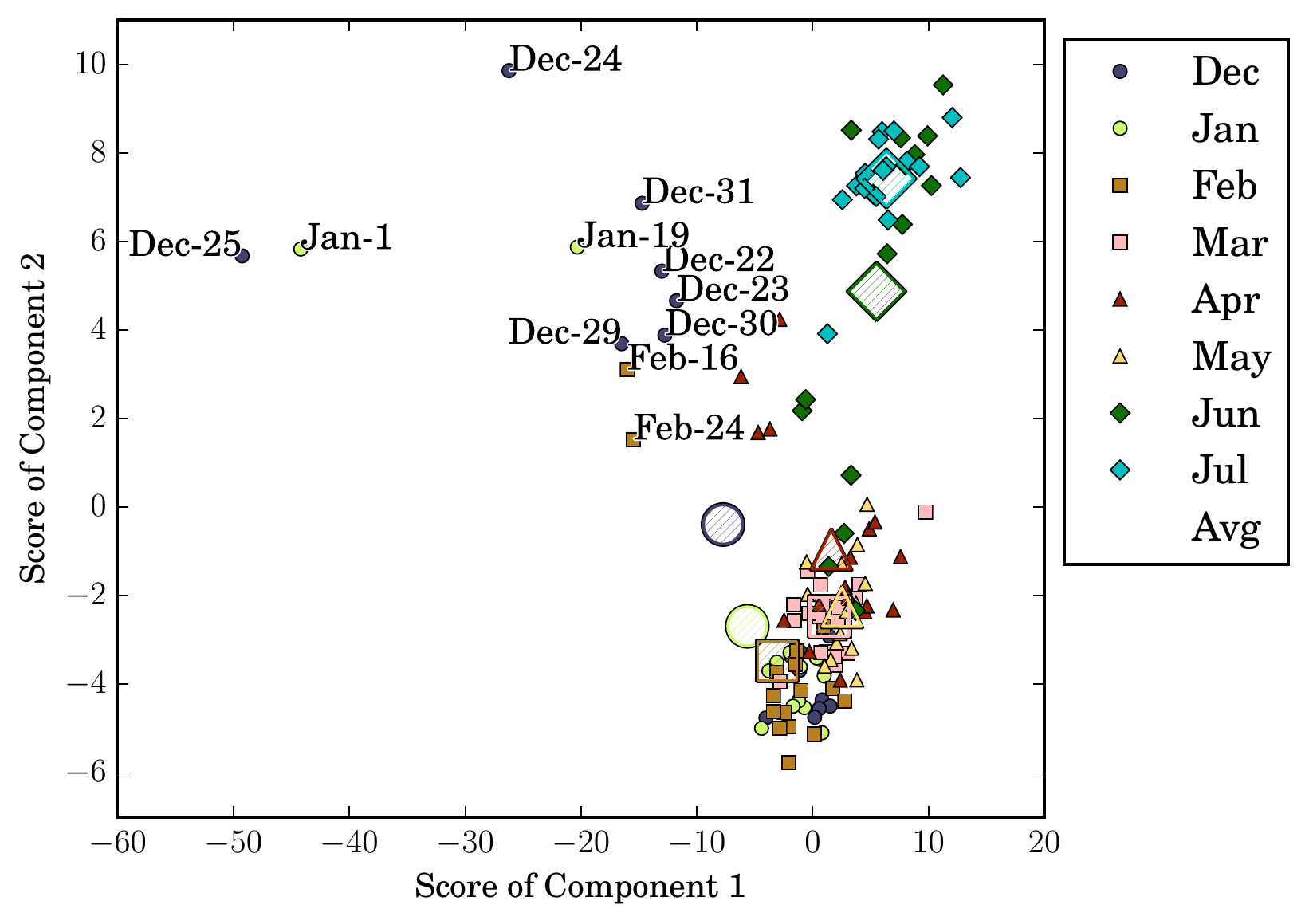}\\
(a)&(b)\\
\includegraphics[width=.48\textwidth]{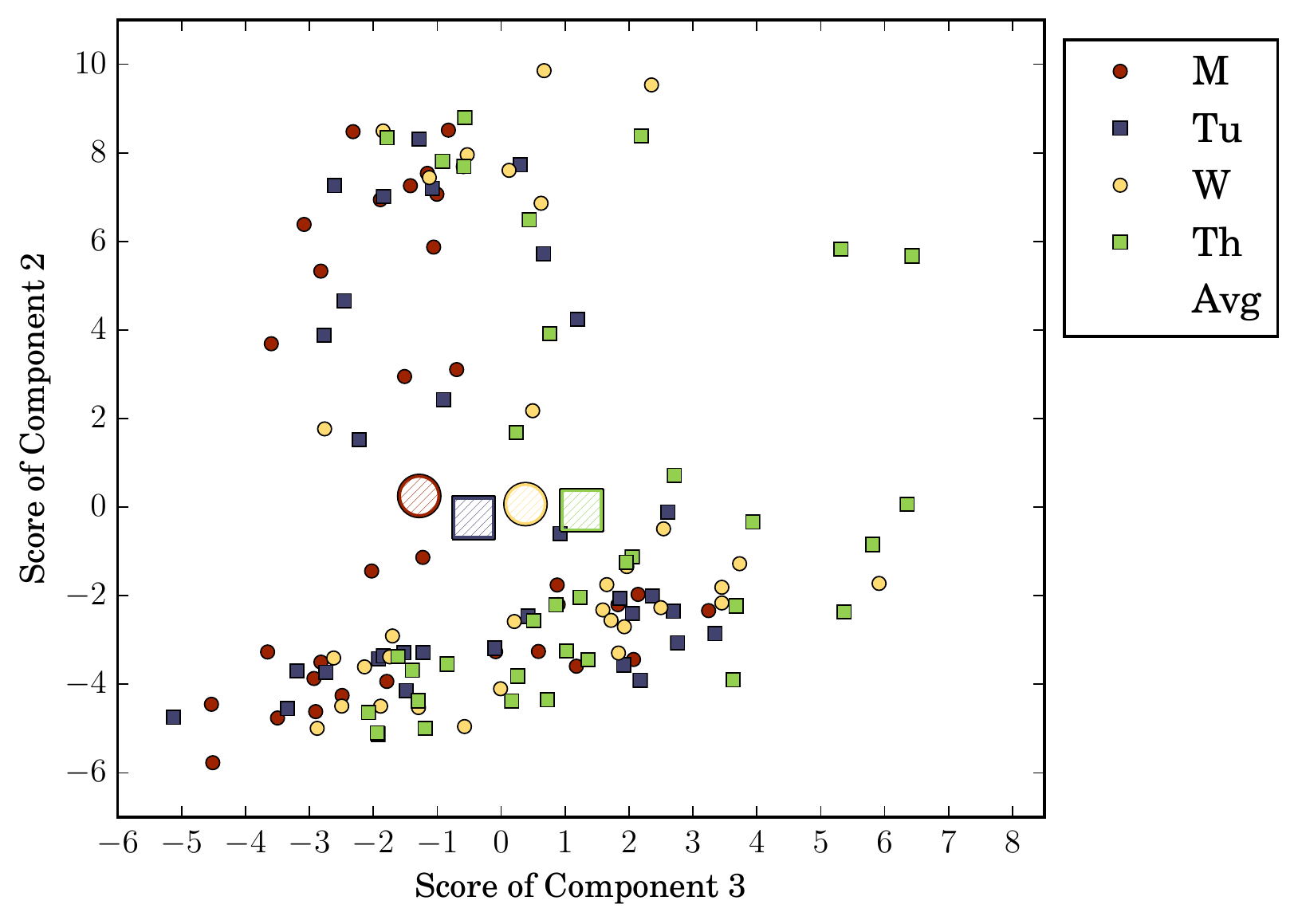}&\includegraphics[width=.48\textwidth]{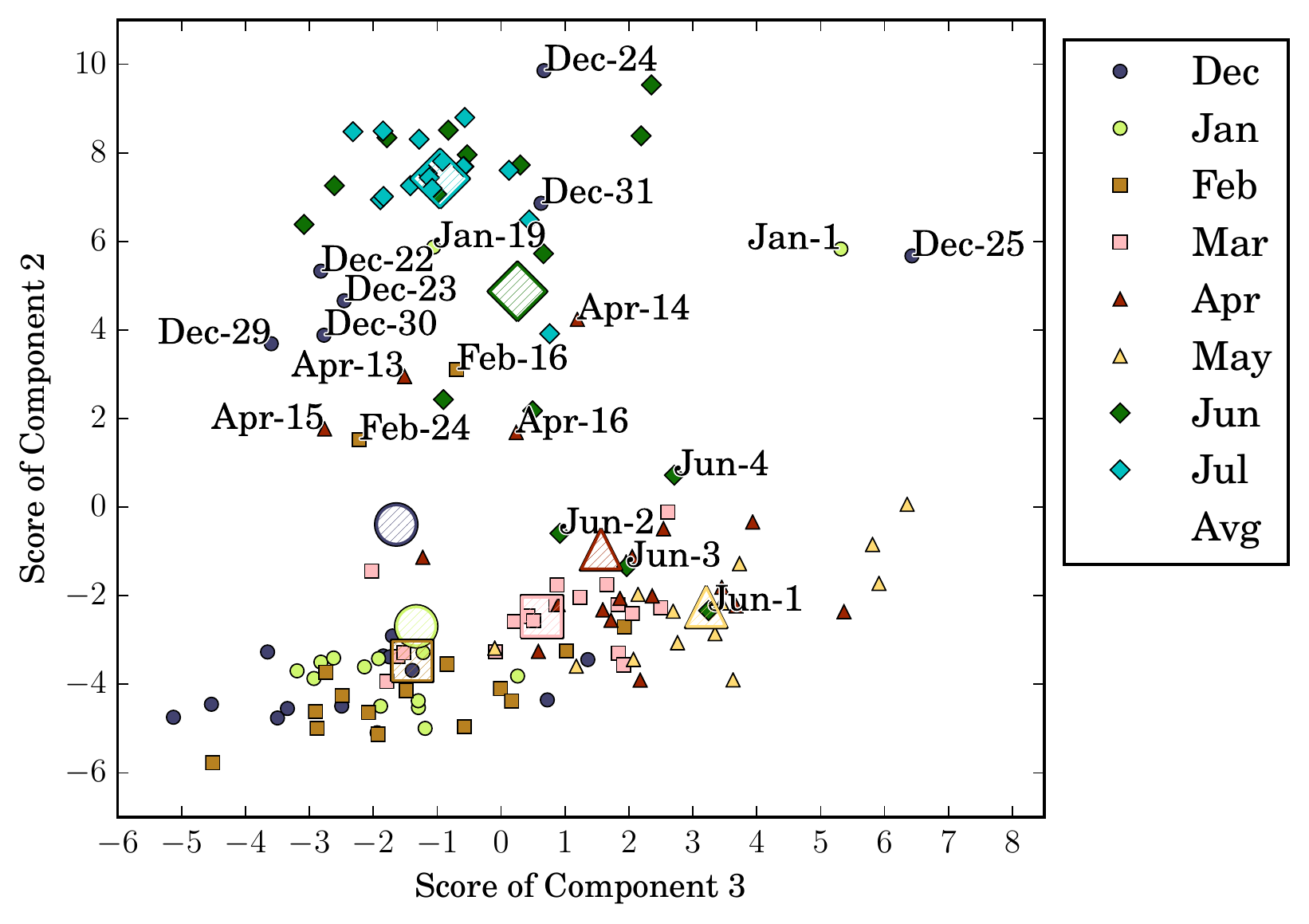}\\
(c)&(d)
\end{tabular}
\caption{Scatter plots of weights for pairs of principal components. Each day is indicated by a small marker and the centroid average for each label category is indicated with a large hatched marker. (a) The weight of component 1 versus the weight of component 2 for each day, labeled by day of week. The centroid points for all four days are close, supporting the notion that no further clustering by day-of-week is necessary. (b) The weight of component 1 versus the weight of component 2, labeled by month. A large negative weight for component 1 indicates days when work/business traffic is low. (c) The weight of component 2 versus the weight of component 3 labeled by day. There is a relationship between day of week and the weight of component 3 but it is small compared to the spread of weights. (d) The weight of component 2 versus the weight of component 3 labeled by month. Positive weights for component 3 indicate days when school is not in session.}
  \label{fig:score}
\end{figure}

In Figure \ref{fig:score}, we present scatter plots of the weights for pairs of components. In Figure \ref{fig:score}a we plot the weight of component 1 versus the weight of component 2 for each day where the data is labeled by the day of the week. In Figure \ref{fig:score}b we again plot the weight of component 1 versus the weight of component 2 but now label the data by month. Similarly, Figure \ref{fig:score}c and Figure \ref{fig:score}d plot the weight of component 3 versus the weight of component 2 and label the data by day and by month, respectively. Each marker corresponds to one day and the centroid average for each label category is indicated with a large hatched marker.

We now discuss some clear trends that emerge from these plots. In Figure \ref{fig:score}b, it is apparent that a large negative weight for component 1 indicates a nonbusiness day; all of the labeled days are on or near holidays in the U.S. except February 24 for which there were school and business closings due to weather conditions, \emph{i.e.}, it was a snow day. This observation is congruous with the plot of principal component 1 in Figure \ref{fig:pca} which corresponds to overall higher traffic volume, especially during the morning and afternoon periods. Thus, a negative score for component 1 indicates lower traffic during these periods. Furthermore, the morning and afternoon peaks for component 1 in Figure \ref{fig:pca} correspond to reversed commute directions, \emph{i.e.}, the morning peak for WB LT corresponds with the afternoon peak of NB RT, likewise for the NB T peak in the morning and the SB T peak in the afternoon.

Approximately one mile to the north of the intersection is a school. In Figure \ref{fig:score}d, we see that a positive score for component 2 indicates days when school is not in session, which includes the entire months of June and July except June 1--4, which are labeled.  We additionally label dates when school is not in session due to holidays or weather conditions; April 13--16 is  the Spring Break holiday. Again, this observation is congruous with the plot of principal component 2 in Figure \ref{fig:pca}, which exhibits two particularly telling features. First, this component has narrow spikes around 7:30 and 15:30, corresponding to the school session. Second, the WB LT movement clearly differs from the others; this is the only movement that is not going to or coming from the north (\emph{i.e.}, Southbound leg) where the school is located.

Figure \ref{fig:score}c indicates that the weight of component 3 generally increases from Monday to Thursday, although the increase is modest compared to the spread of the weights for this component, and the centroids in Figure \ref{fig:score}a are closely clustered. Thus these plots generally support our decision to cluster Monday--Thursday together. %

Lastly, we observe longer term trends in the data. In Figure \ref{fig:score}d, the score of component three generally increases from December to May, the months when school is in session. This may correspond with seasonal variations; as the plot of principal component 3 in Figure \ref{fig:pca} suggests, a larger score for component 3 indicates more traffic in the evening. The days get longer from December to May, which may explain higher evening traffic. In Figure \ref{fig:score}b, there is also an increase in the score of component 1 per month for all months December to July, although this trend is somewhat obscured by the axis scaling. This suggests that for the entire data set, there is a general increase in business-related traffic, which may be seasonal or may reflect improving economic conditions from December 2014 to July 2015.

\subsection{Discussion}

By its nature, a PC-based approach offers a more nuanced interpretation of the data than cluster-based approaches, as we saw above, since it indicates the \emph{degree} to which a set of measurements exhibits a particular component. For example, rather than simply identifying that traffic patterns are different in the Winter season than in the Spring season (which clustering may reveal), we are able to quantify this difference. Furthermore, a PC-based approach retains the ability to identify categorical differences in the data. For example, our analysis indicates that traffic patterns for days when school is in session are qualitatively different than when school is not in session.%
 
We also note that in a PC-based approach  the components frequently offer an interpretation or explanation for the observed trends. For example, we could associate each of the observed trends in the scatter plots of Figure \ref{fig:score} with an intuitive hypothesis regarding the variation in traffic movement flows based on the principal components in Figure \ref{fig:pca}.  We note that the trends observed above are unique for the available dataset. This means that the principal components and corresponding interpretations are unlikely to transfer to other datasets obtained from other traffic intersections or networks. An interesting direction for future research is to study the transferability of prediction and learning algorithms to other datasets.

\section{Traffic Prediction from Low-Rank Structure}
\label{sec:traff-pred-from}
Suppose it were possible to obtain an estimate of the weight vector $w(d)$ for some day $d\in D$ by using measurements only up to some time $T^*<T$, that is, from the vector
\begin{align}
  \label{eq:6}
  z^d&=\begin{bmatrix}(z^d_1)^{\tp}&\ldots&(z^d_M)^{\tp}\end{bmatrix}^{\tp} \in \mathbb{R}^{T^*M}
\end{align}
where
\begin{align}
  \label{eq:7}
  z^d_m=\begin{bmatrix}x^d_m(1)&\ldots&x^d_m(T^*)\end{bmatrix}^{\tp}\in\mathbb{R}^{T^*}\quad m=1,\ldots,M.
\end{align}
Then it would be possible to \emph{predict} traffic flow for times $t>T^*$ by constructing an estimate of $x^d$ as $\hat{x}^d=Q\hat{w}(d)$ (see \eqref{eq:5}) where $\hat{x}^d$ is our estimate of the traffic flow for the entire day and $\hat{w}(d)$ is our estimate of the weight vector $w(d)$.

A naive approach to estimating $w(d)$ using measurements up to time $T^*$ is to project the vector $z^d$ onto the corresponding truncation of each $q^i$. However, this approach is unreliable because the PC-based decomposition uses data for the whole day and does not consider that we wish to make a prediction after time $T^*$. For example, consider principal component 3 in Figure \ref{fig:pca} which, as argued in Section \ref{sec:case-study-principal}, corresponds with increased/decreased traffic in the evening. This component lies largely in the direction corresponding to flow measurements later in the day and thus it is difficult to predict $w^3(d)$ given $z^d$ for $T^*$ early in the day. 

To overcome this limitation, we propose computing a set of components that simultaneously explains the variation in the measured data \textit{before} time $T^*$ and  correlates with a second set of components that explains the variation in the traffic flow data measured \textit{after} time $T^*$. To make this precise, we define the following vector of flow measurements after time $T^*$, complementary to $z^d$ defined in \eqref{eq:6}--\eqref{eq:7}:
\begin{align}
  \label{eq:12}
  y^d&=\begin{bmatrix}(y^d_1)^{\tp}&\ldots&(y^d_M)^{\tp}\end{bmatrix}^{\tp}\in  \mathbb{R}^{(T-T^*)M}
\end{align}
where
\begin{align}
  \label{eq:13}
 y^d_m=\begin{bmatrix}x^d_m(T^*+1)&\ldots&x^d_m(T)\end{bmatrix}^{\tp}\in  \mathbb{R}^{(T-T^*)},\quad m=1,\ldots,M.
\end{align}
Let
\begin{alignat}{2}
  \label{eq:14}
  \bar{z}&=\frac{1}{D}\sum_{d=1}^Dz^d,\qquad&  \bar{y}&=\frac{1}{D}\sum_{d=1}^Dy^d
\end{alignat}
and define
\begin{alignat}{2}
  \label{eq:15}
  Z&=
  \begin{bmatrix}
    (z^1)^{\tp}\\
\vdots\\
(z^D)^{\tp}
  \end{bmatrix}
,\qquad&
Y&=  \begin{bmatrix}
    (y^1)^{\tp}\\
\vdots\\
(y^D)^{\tp}
  \end{bmatrix}\\
\label{eq:15-2}\tilde{Z}&=Z-\mathbf{1}_D\bar{z}^{\tp},\qquad&
\tilde{Y}&=Y-\mathbf{1}_D\bar{y}^{\tp}.
\end{alignat}

The matrix $Z$ collects all flow measurements up to time $T^*$ and the matrix $Y$ collects all flow measurements after time $T^*$ onward to the final time $T$. The matrices $\tilde{Z}$ and $\tilde{Y}$ are mean-centered versions of $Z$ and $Y$. Note that $Z$ and $Y$ partition $X$, and, similarly, $\tilde{Z}$ and $\tilde{Y}$ partition $\tilde{X}$.

Our objective is to find a collection of \emph{predictor} components $p^1,p^2,\ldots,p^N$ with each $p^i\in\mathbb{R}^{T^*M}$; a collection of \emph{predicted} components $c^1,c^2,\ldots,c^N$ with each $c^i\in\mathbb{R}^{(T-T^*)M}$; and, for each $d=1,\ldots,D$, a vector of weights $\omega(d)\in\mathbb{R}^N$ with
\begin{align}
  \label{eq:16}
  \omega(d)=\begin{bmatrix}\omega^1(d)&\omega^2(d)&\ldots&\omega^N(d)\end{bmatrix}^{\tp}
\end{align}
such that 
\begin{align}
  \label{eq:17}
  z^d&\approx \bar{z} +\sum_{i=1}^N\omega^i(d)p^i\\
  \label{eq:17-2}  y^d&\approx \bar{y} +\sum_{i=1}^N\omega^i(d)c^i.
\end{align}
Let
\begin{align}
  \label{eq:19}
  \omega^i=\begin{bmatrix}\omega^i(1)&\omega^i(2)&\ldots&\omega^i(D)\end{bmatrix}^{\tp}\in\mathbb{R}^D \quad \text{for all }i=1,\ldots,N.
\end{align}

The use of \eqref{eq:17}--\eqref{eq:17-2} for prediction is immediately apparent: if we are able to determine the weights $\omega(d)$ using only $z^d$, that is, measurements up to time $T^*$, then we are able to predict traffic flow after time $T^*$ using the same weights and the collection of predicted components $c^i$, $i=1,\ldots, N$. We formalize this prediction procedure in Section \ref{sec:pred-from-latent}.

\subsection{The Projection to Latent Structures Algorithm}
\label{sec:proj-latent-struct}
To compute the collection of predictor and predicted components, we use a statistical technique called \emph{projection to latent structures (PLS)}, also known as \emph{partial least squares}. Given two sets of measured variables (\emph{e.g.}, measured traffic flow over two time intervals) $Z$ and $Y$, the PLS algorithm identifies low-rank approximations of both sets as in \eqref{eq:17}--\eqref{eq:17-2} in such a way that the low-rank components are highly correlated. The low-rank approximations are then used for prediction. While the PLS algorithm is a general statistical tool, it has particularly been developed in the domain of chemometrics beginning with the work of Wold \emph{et. al.} \cite{Wold:1984fy} and has since been applied to a wide range of chemical data analysis problems. References \cite{Geladi:1986fq}, \cite{Manne:1987xw}, and \cite{Hoskuldsson:1988by} provide early overviews of the PLS algorithm with emphasis on applications to chemometrics, and references \cite{Rosipal:2006kx} and \cite{Abdi:2010dn} provide more recent tutorials on the PLS algorithm.

The PLS technique is iterative; it first computes a pair of components $(p^1,c^1)$ from the data matrices $\tilde{Z}$ and $\tilde{Y}$, then the algorithm deflates the data matrices by removing the contribution of this pair of components. Next, a new pair $(p^2,c^2)$ is computed from the updated data matrices, \emph{etc.} To determine the first pair $(p^1,c^1)$, we solve the following optimization problem:
\begin{alignat}{2}
  \label{eq:18}
  (r^*,s^*)=\argmax_{r,s}&\qquad &&\hspace*{-25pt} (\tilde{Z}r)^{\tp}(\tilde{Y}s)\\
\text{such that}& &|| r ||_2&=1\\
&&||s||_2&=1.
\end{alignat}
The interpretation of \eqref{eq:18} is as follows: we wish to find directions $r^*\in\mathbb{R}^{T^*M}$ and $s^*\in\mathbb{R}^{(T-T^*)M}$ to maximize the empirical covariance of the score vectors $u=\tilde{Z}r^*\in\mathbb{R}^D$ and $v=\tilde{Y}s^*\in\mathbb{R}^D$. These score vectors contain the projection of each day's data $z^d$ and $y^d$ onto the directions $r^*$ and $s^*$. We define the first score component vector
\begin{align}
  \label{eq:20}
  \omega^1\triangleq \frac{1}{||\tilde{Z}r^*||_2}\tilde{Z}r^*.
\end{align}

Note that $r^*$ and $s^*$ are the first left and right, respectively, singular vectors of $\tilde{Z}^{\tp}\tilde{Y}$. To obtain $p^1$ we project $\tilde{Z}$ onto $\omega^1$, and, similarly, to obtain $c^1$ we project $\tilde{Y}$ onto $\omega^1$:
\begin{align}
  \label{eq:21}
  p^1=\tilde{Z}^{\tp}\omega^1\\
  c^1=\tilde{Y}^{\tp}\omega^1.
\end{align}
Note that our treatment of the data matrices $\tilde{Z}$ and $\tilde{Y}$ is asymmetric, that is, $\omega^1$ is obtained from $\tilde{Z}$ and $r^*$. This asymmetry is because we ultimately wish to use the scores and components for \emph{prediction}. We next {deflate} the data matrices $\tilde{Z}$ and $\tilde{Y}$ using the first score vector $\omega^1$ and the computed components:
\begin{align}
  \label{eq:22}
  \tilde{Z}^+=\tilde{Z}-\omega^1p^1\\
  \tilde{Y}^+=\tilde{Y}-\omega^1p^1.
\end{align}
Above, $\tilde{Z}^+$ and $\tilde{Y}^+$ are the updated data matrices. To compute $\omega^2$, $p^2$, and $c^2$ we repeat the above procedure, replacing $\tilde{Z}$ and $\tilde{Y}$ with their updated versions $\tilde{Z}^+$ and $\tilde{Y}^+$. We repeat this process until we obtain $N$ score vectors and components where $N$ is a design parameter. We gather the computed scores and components:
\begin{align}
  \label{eq:23}
  \Omega&=\begin{bmatrix}\omega^1&\ldots&\omega^N\end{bmatrix}\in\mathbb{R}^{D\times N}\\
P&=\begin{bmatrix}p^1&\ldots&p^N\end{bmatrix}\in\mathbb{R}^{(T^*M)\times N}\\
  \label{eq:23-3} C&=\begin{bmatrix}c^1&\ldots&c^N\end{bmatrix}\in\mathbb{R}^{((T-T^*)M)\times N}.
\end{align}
We then have
 $ \tilde{Z}\approx \Omega P^{\tp}$ and  $ \tilde{Y}\approx \Omega C^{\tp}.$

Above, to compute $r^*$ and $s^*$, one approach is to first multiply $\tilde{Z}^{\tp}$ and $\tilde{Y}$ and then compute the singular value decomposition of the product, and this operation is performed for each iteration. For traffic flow measurements, it is typically the case that $D\ll \min\{T^*M,(T-T^*)M\}$, thus $\tilde{Z}^{\tp}\tilde{Y}$ is a large matrix of dimension $T^*M\times (T-T^*)M$ and computing its singular value decomposition is computationally expensive. There exists efficient implementations of the PLS algorithm that use the \emph{kernel} matrices  $\tilde{Z}\tilde{Z}^{\tp}$ and $\tilde{Y}\tilde{Y}^{\tp}$ to compute the predictor and prediction components by finding the principal eigenvector and eigenvalue pair of a (much smaller) $D\times D$ matrix instead \cite{Rannar:1994ee}. %
The PLS algorithm is described by \eqref{eq:18}--\eqref{eq:23-3}. %

\subsection{Prediction From Latent Structures}
\label{sec:pred-from-latent}
Once the predictor and predicted components have been computed, we are able to construction a prediction of future traffic flow given sample measurements. To this end, let 
\begin{align}
  \label{eq:24}
  z^s\in\mathbb{R}^{T^*M} 
\end{align}
denote measured traffic flow for the $M$ turn movements up to time $T^*$. Our objective is to use $z^s$ along with the predictor and predicted components calculated from historical data to predict traffic flow for our sample day for time periods after $T^*$.   To do so, we first calculate $\hat{\omega}$, a vector of scores for the sample day. As above, let $\bar{z}$ and $\bar{y}$ denote means of historical traffic flow up to time $T^*$ and after $T^*$, respectively. The score vector is computed as
\begin{align}
  \label{eq:25}
  \hat{\omega}=\big((z^s-\bar{z})^{\tp}(P^{\tp})^\dagger\big)^{\tp}\in\mathbb{R}^{N}
\end{align}
where $(P^{\tp})^\dagger$ denotes the Moore-Penrose pseudoinverse of $P^{\tp}$. Then
\begin{align}
  \label{eq:26}
  \hat{y}^s&=\hat{\omega}^{\tp}C^{\tp} +\bar{y}\in\mathbb{R}^{(T-T^*)M}
\end{align}
is the prediction of traffic flow after time $T^*$.
Combining \eqref{eq:25} and \eqref{eq:26}, we obtain one succinct equation for prediction:
\begin{align}
  \label{eq:27}
  \hat{y}^s=(z^s-\bar{z})^{\tp} (P^{\tp})^\dagger C^{\tp}+\bar{y}.
\end{align}
Notice that after $P^{\tp}$ and $C^{\tp}$ are computed via the PLS algorithm  and the pseudoinverse of $P^{\tp}$ is computed and stored, predicting traffic flow given a sample vector of measurements requires only matrix multiply operations and is thus computationally easy.

\subsection{Extensions and Case Study}
We now discuss a few immediate extensions to the PLS prediction algorithm presented above that we utilize in Section \ref{sec:traff-pred-contr}. Firstly, we assume above that our objective is to predict flow measurements for all time periods beyond time $T^*$ up to the end of the day, $T$. However, our objective may be to predict traffic flow for a shorter horizon or for a period not immediately subsequent to the measurement period. For example, we may wish to predict traffic during the evening from observed traffic flow in the morning. We can accommodate this scenario by appropriately truncating the matrices $Z$ and $Y$ above. 

Secondly, we assume above that the interval of time between subsequent flow measurements, $\Delta$, is the same for the measured data as well as the prediction. However, it is straightforward to accommodate differences in these intervals, or even nonuniform intervals so long as the measurement times are the same for each day. For example, we may wish to use flow measurements at 15 minutes intervals up to time $T^*$ to predict traffic flow at one hour intervals subsequently. 

\begin{figure}
  \centering
  \begin{tabular}{l l}
    \includegraphics[height=1.6in]{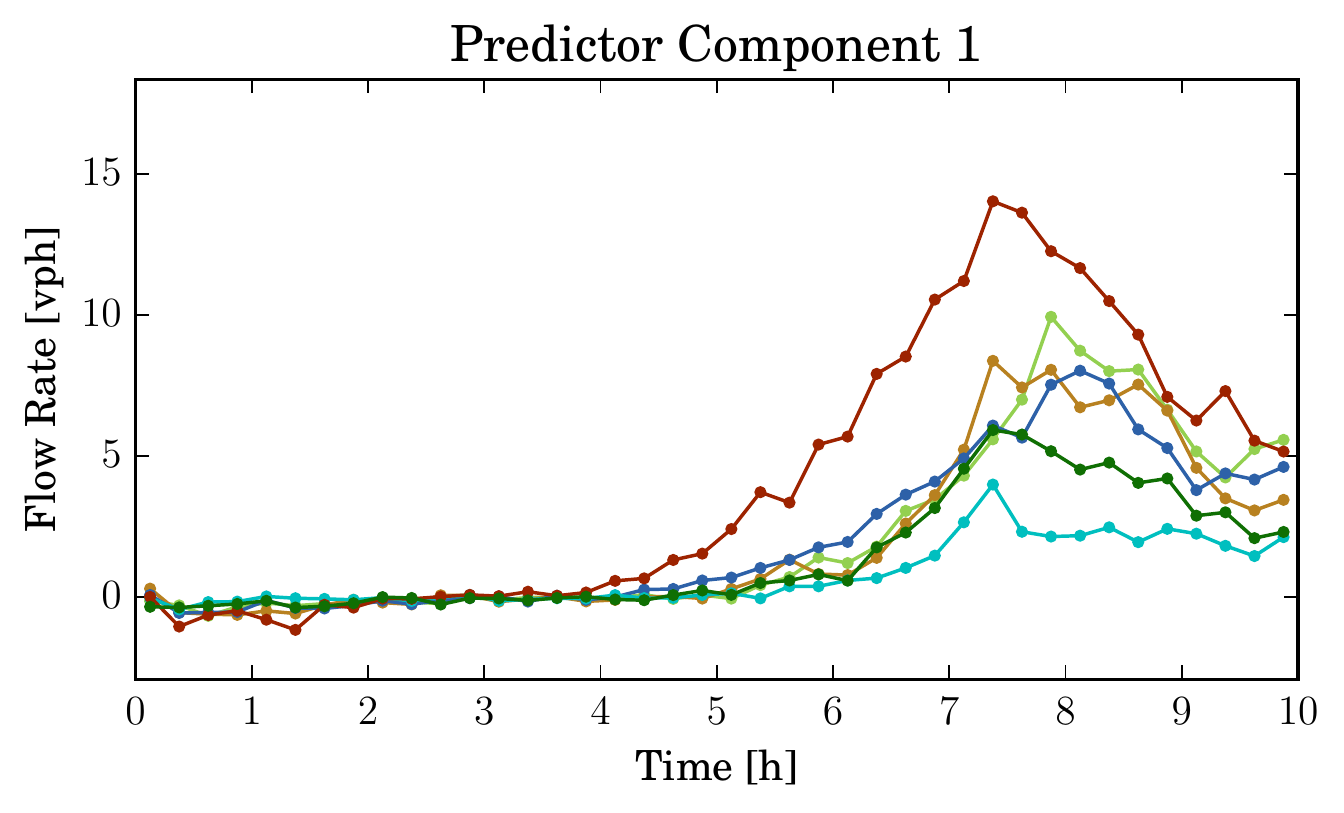}&    \includegraphics[height=1.6in]{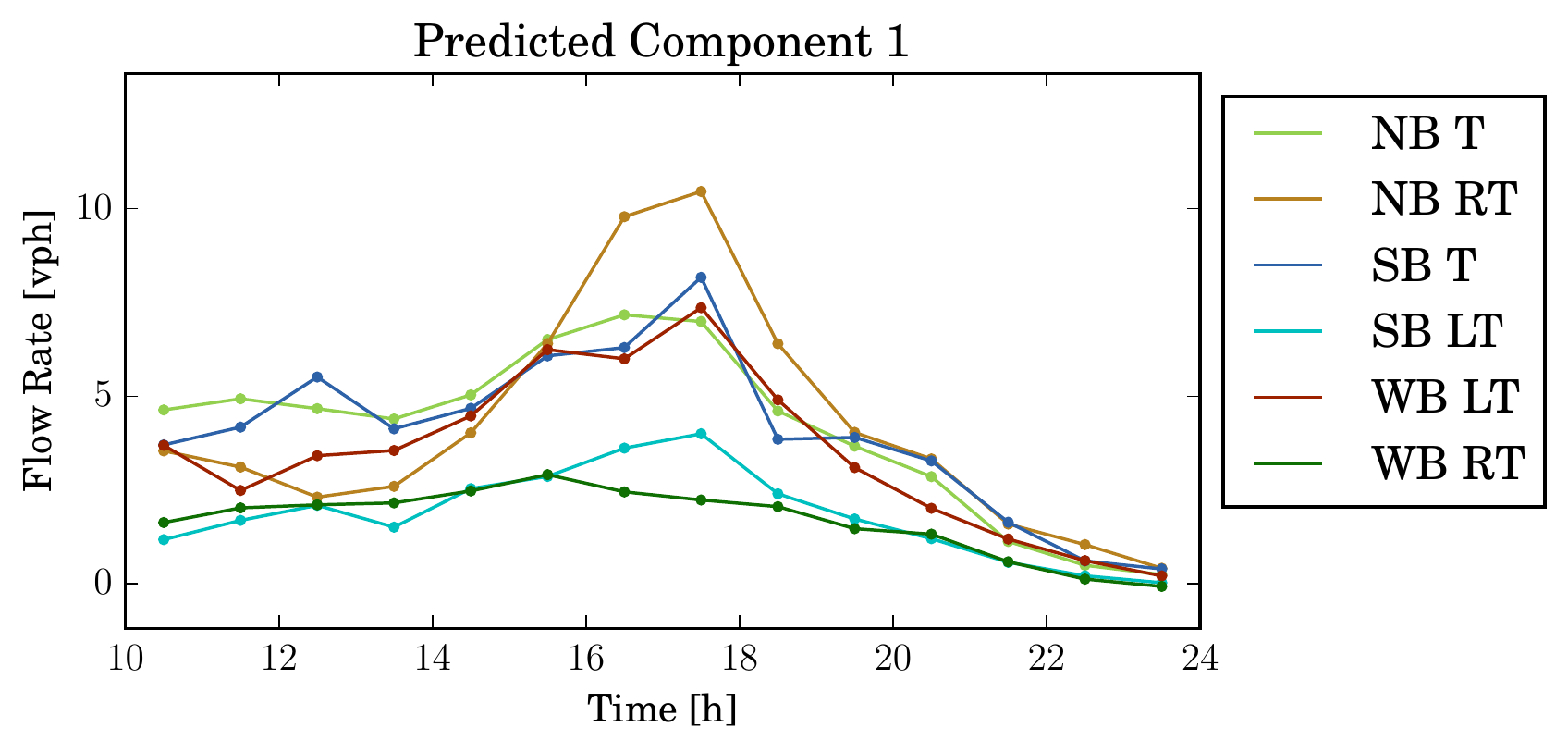}\\
    \includegraphics[height=1.6in]{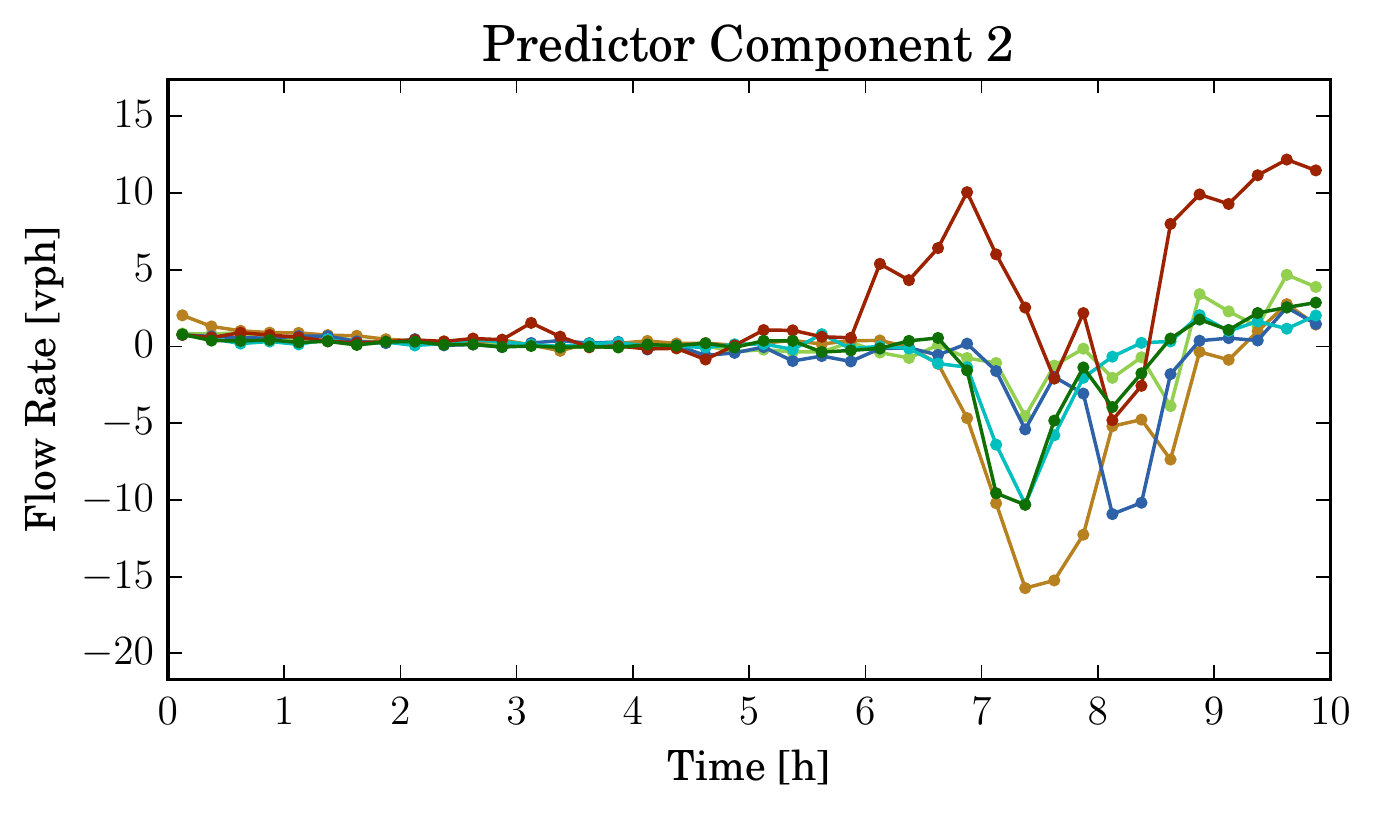}&    \includegraphics[height=1.6in]{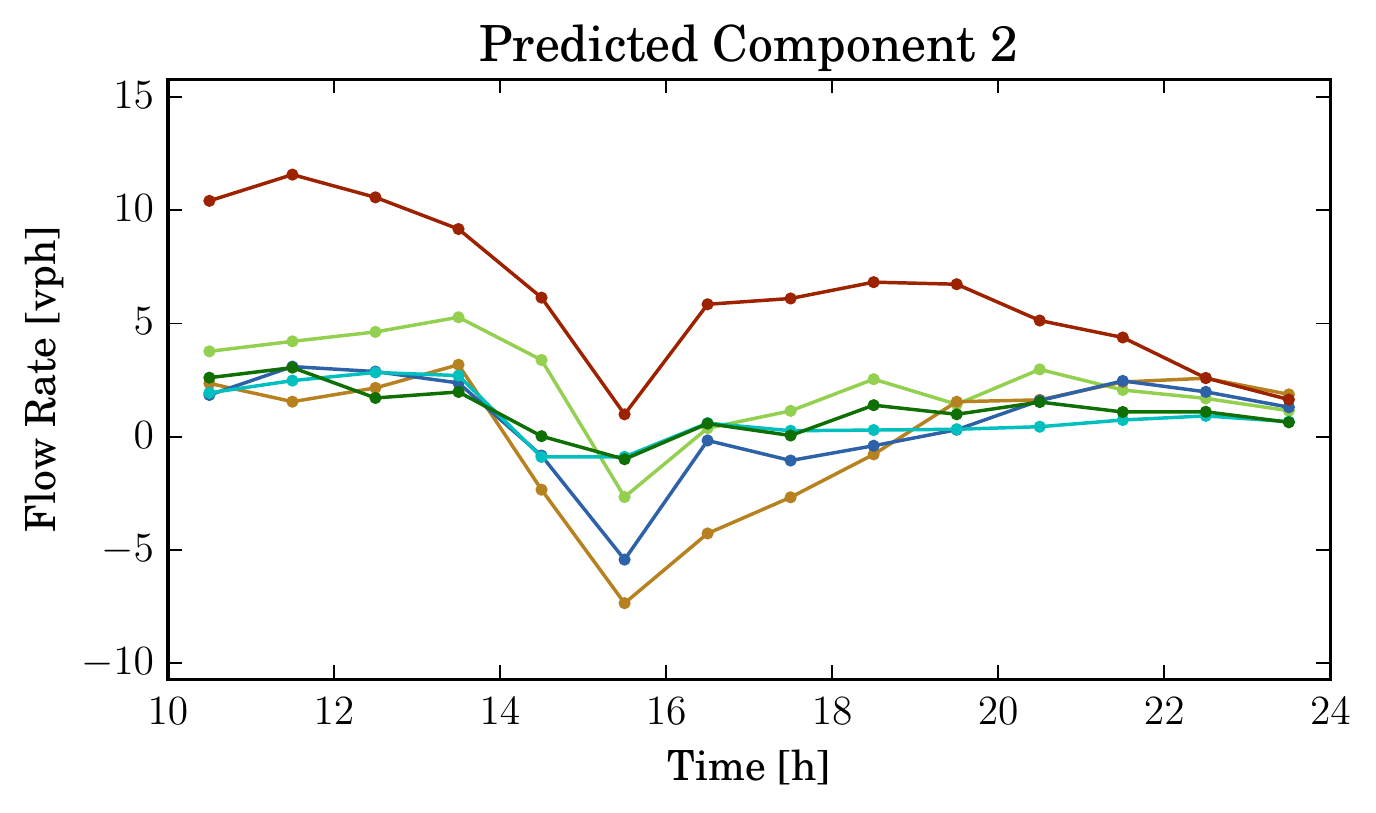}\\
    \includegraphics[height=1.6in]{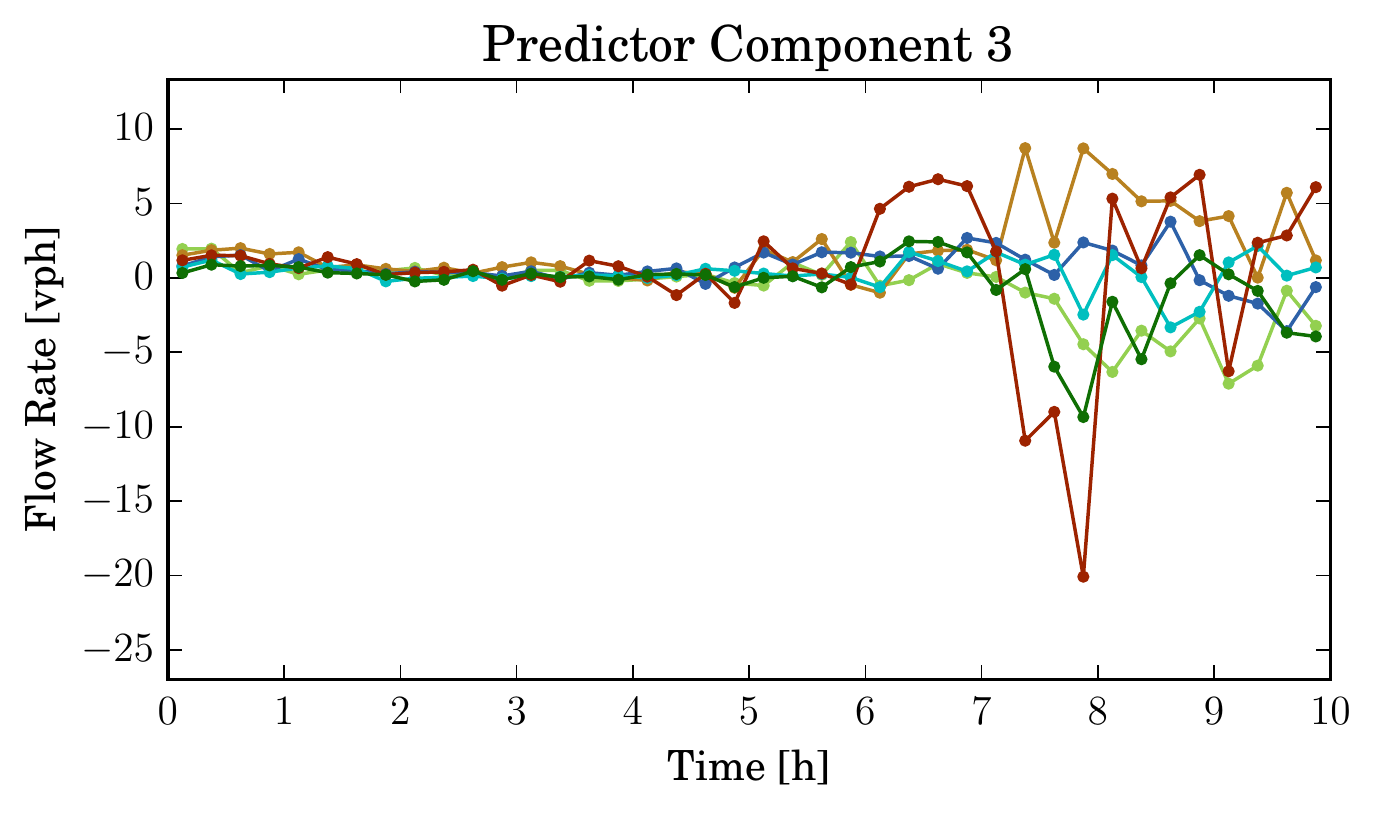}&    \includegraphics[height=1.6in]{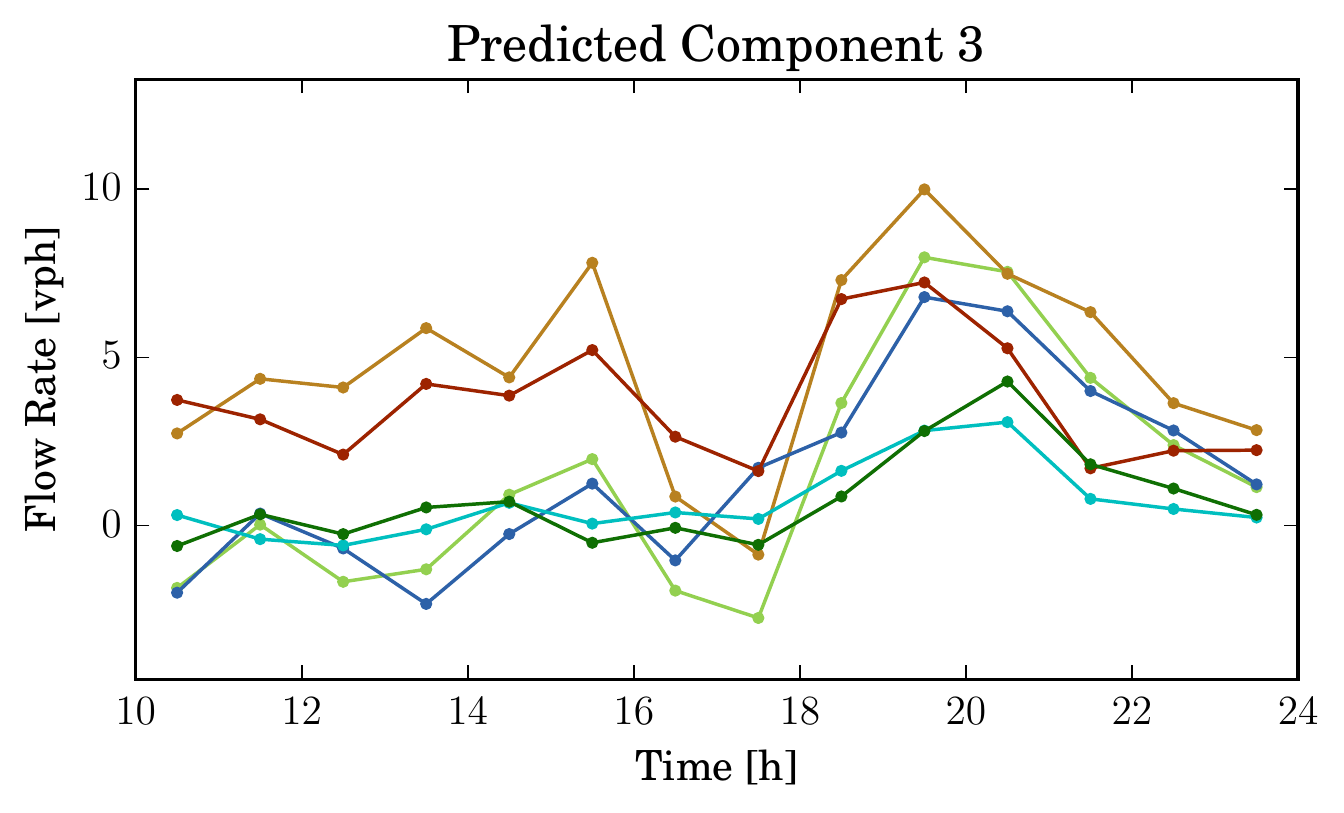}
  \end{tabular}  
  \caption{Projection to latent structures algorithm. We take $T^*$ corresponding to 10:00 and consider the case for which flow measurements are available in 15-minute increments up to 10:00 and available in 1-hour increments thereafter. The left column of plots contains the first three predictor components, the right column of plots contains the corresponding prediction components.  }
  \label{fig:PLS}
\end{figure}

To demonstrate the PLS prediction algorithm as well as these extensions, we return to our case study. We let $T^*$ correspond to 10:00 and consider the case when flow measurements are available up to $T^*$ in 15-minute increments and available thereafter in 1-hour increments. Figure \ref{fig:PLS} shows the first three predictor and predicted components generated from the dataset. To emphasize the different measurement intervals, individual data points are indicated with a solid marker point.  Comparing Figure \ref{fig:PLS} to Figure \ref{fig:pca}, we see that each predictor/predicted component pair of Figure \ref{fig:PLS} resembles the corresponding principal component of Figure \ref{fig:pca}. The differences reflect that the predictor and predicted components are calculated to maximize their correlation.

\begin{figure}
  \centering
  \begin{tabular}{@{}l@{}   @{}l@{}   @{}l@{}}
    \multicolumn{3}{c}{Southbound Through Movement}\hspace{.6in}\\
    {\includegraphics[height=1.55in]{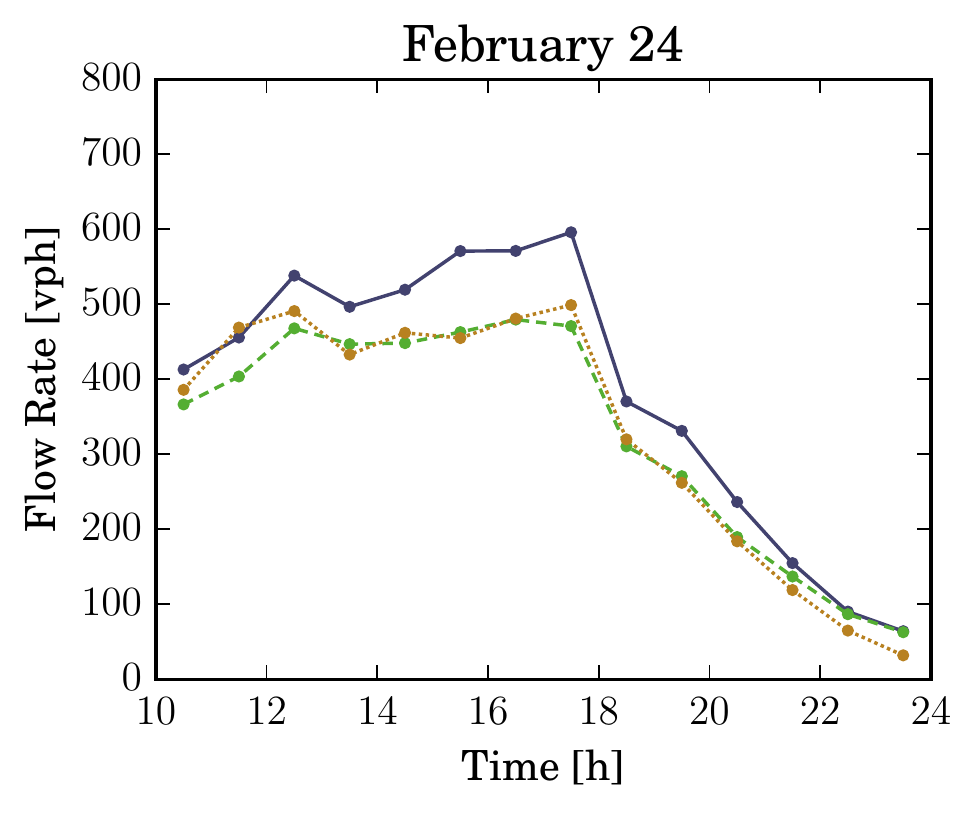}} &    \includegraphics[height=1.55in]{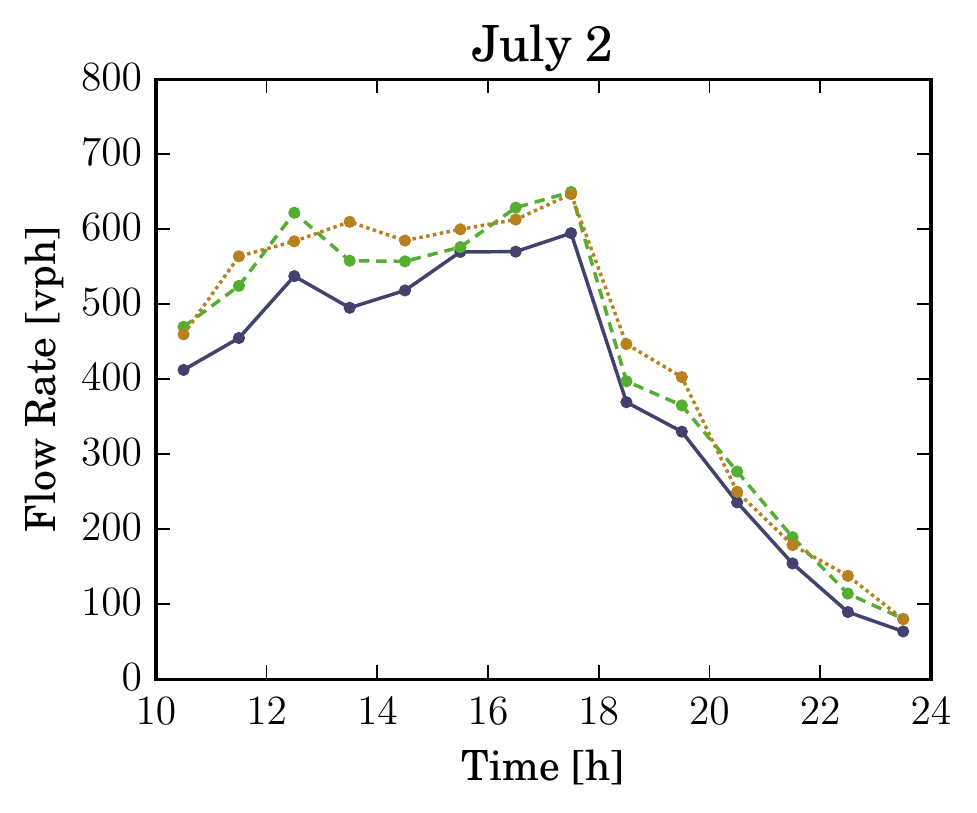}&    \includegraphics[height=1.55in]{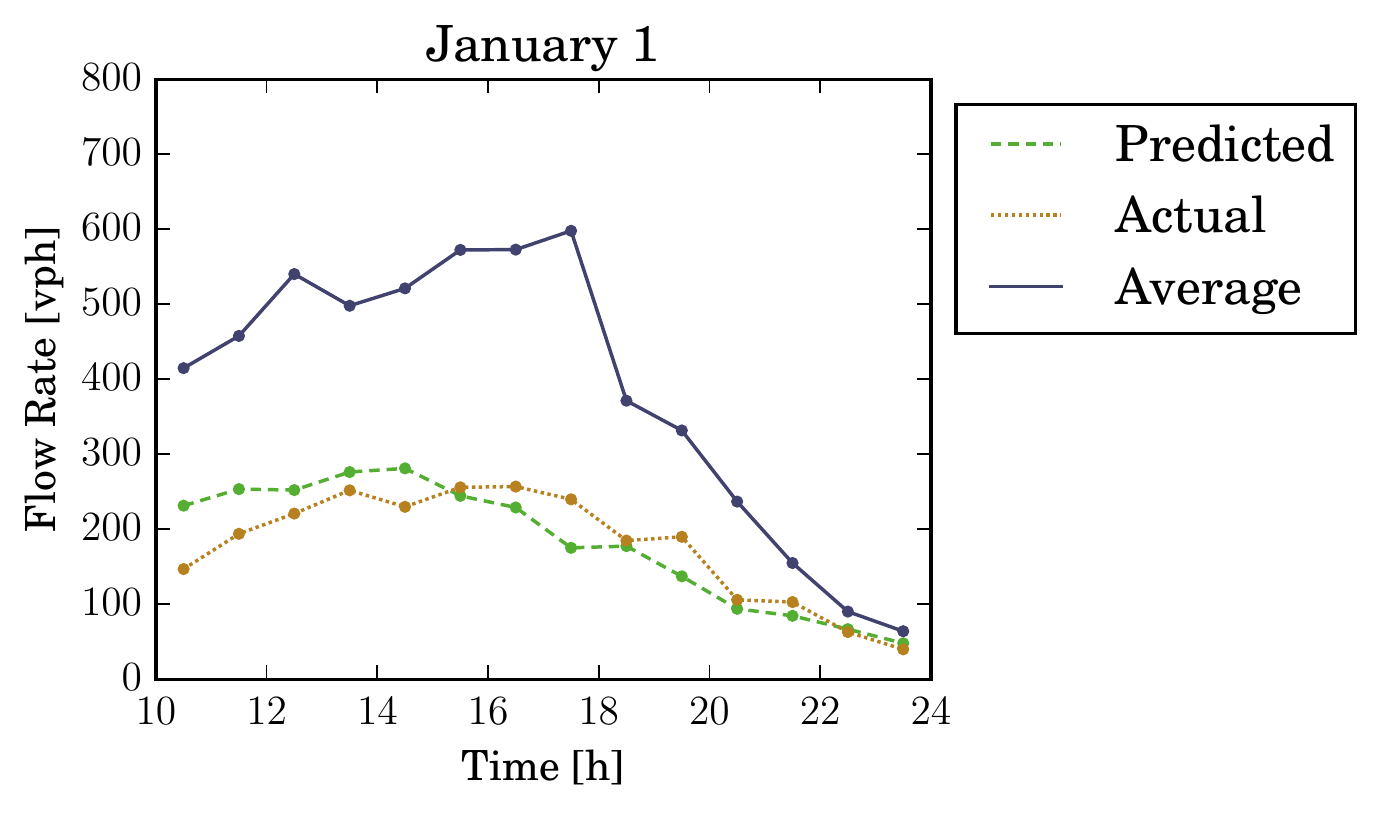}\\
    \multicolumn{3}{c}{Northbound Right Turn Movement}\hspace{.6in}\\
    \includegraphics[height=1.55in]{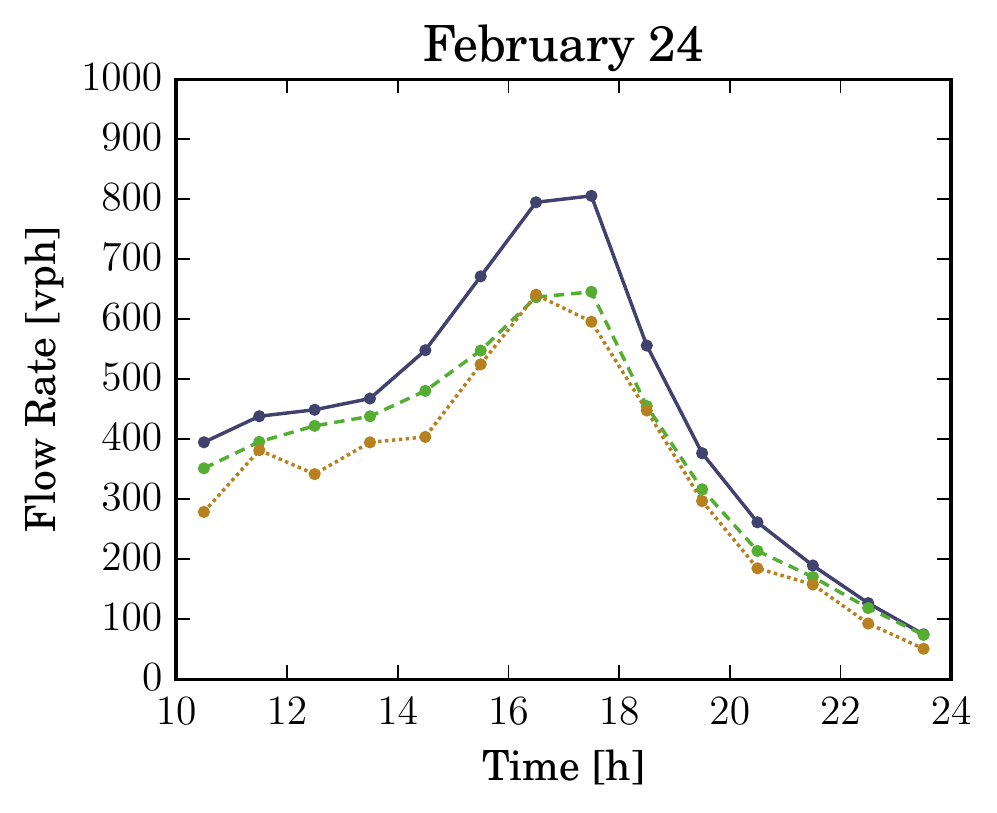}&    \includegraphics[height=1.55in]{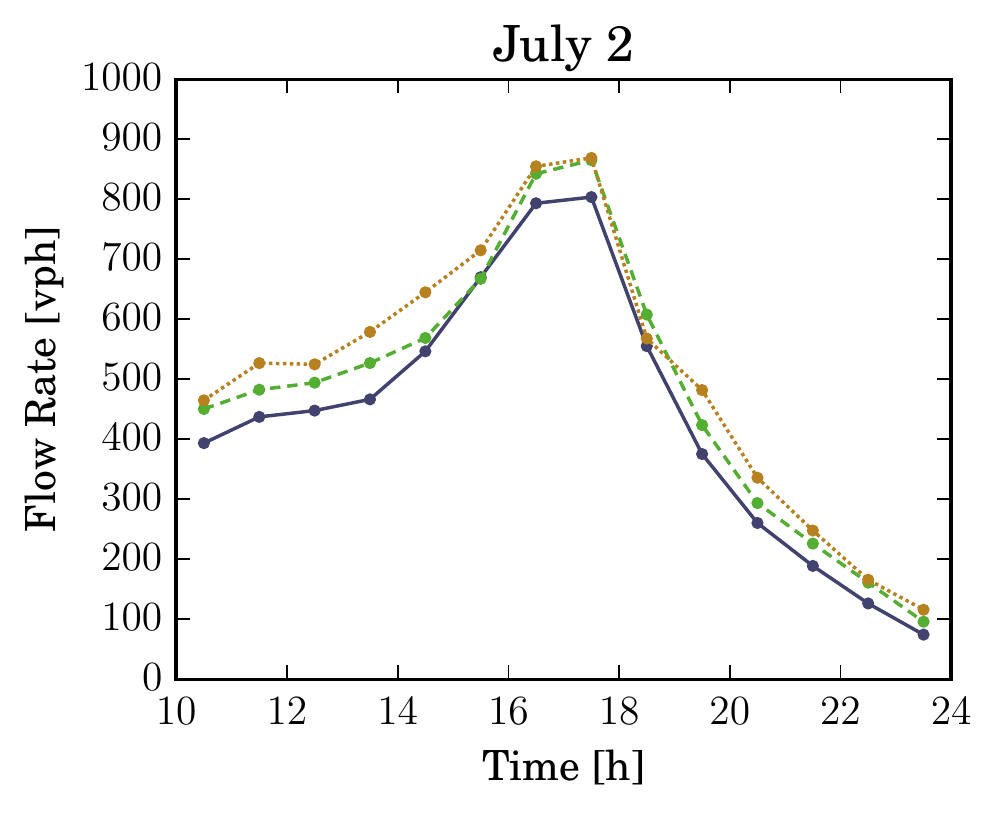}&    \includegraphics[height=1.55in]{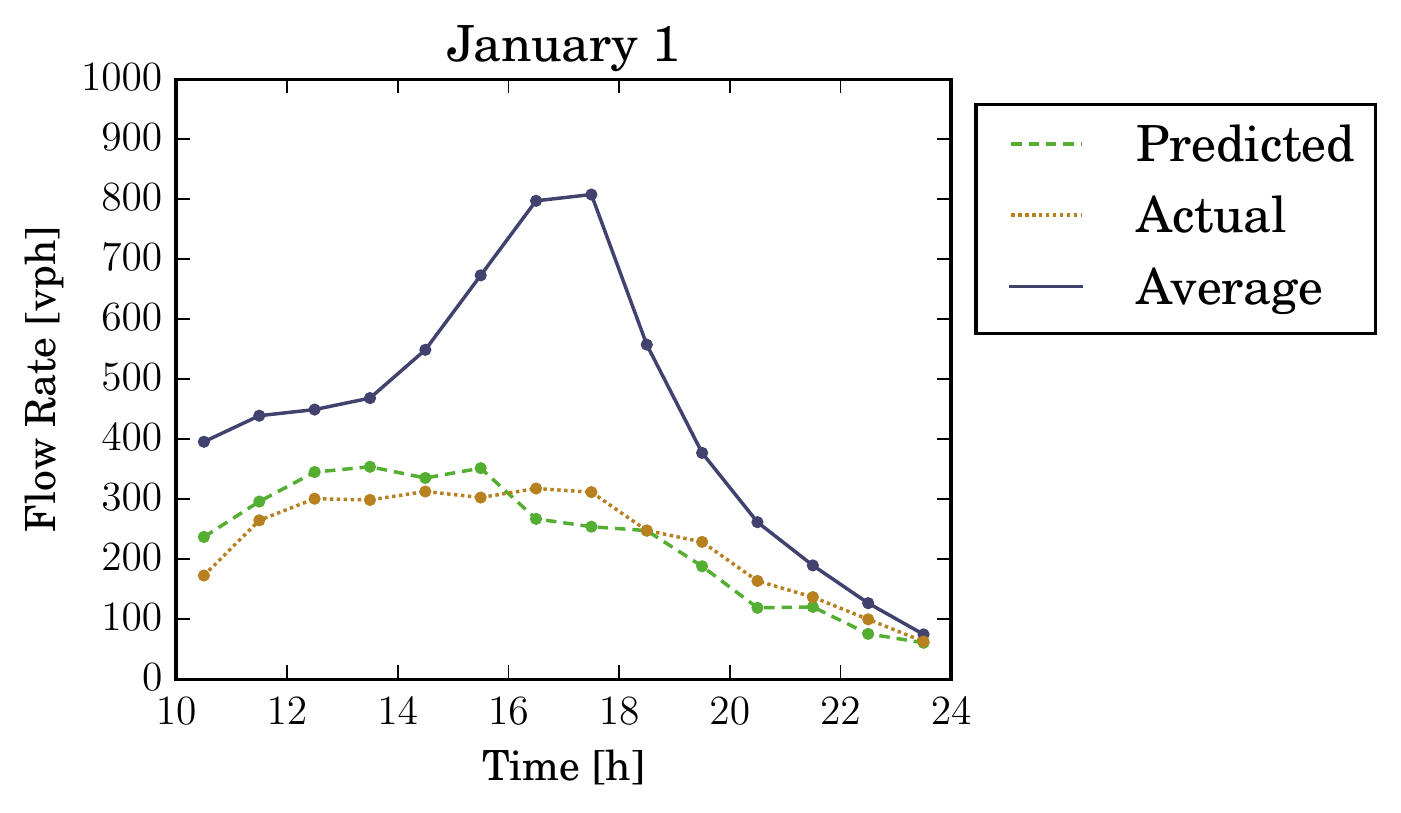}\\
  \end{tabular}  
  \caption{Example prediction results on three unusual days for two movements. On February 24, the algorithm correctly predicts below average flow caused by winter weather and closed schools. On July 2, the algorithm correctly predicts above average flow caused by the subsequent holiday weekend. The algorithm additionally predicts the well below average traffic on the holiday January 1 which deviates substantially from the average flow.}
  \label{fig:pred_ex}
\end{figure}

Figure \ref{fig:pred_ex} illustrates the use of the PLS algorithm for predicting traffic flow. We consider three unusual  days from our data set: February 24, which, as described above, experienced winter weather resulting in school closures; July 2, which is a Thursday and preceded a long holiday weekend in the U.S.; and January 1, which is a significant holiday and resulted in dramatically different traffic patterns. As a demonstration, we only plot results for two of the twelve movements; the first (respectively, second) row of Figure \ref{fig:pred_ex} plots flow measurements and predictions for the SB through movement (respectively, NB right turn movement). 

The blue trace shows the average flow over the fourteen one-hour periods from 10:00 to midnight, the gold trace shows the actual flow on the given days, and the green trace shows the flow as predicted using \eqref{eq:27} with four predictor/predicted component pairs (three of which are shown in Figure \ref{fig:PLS}). To compute the prediction, we employ leave-one-out cross validation whereby, for each unusual day, the predictor/predicted components are computed from the dataset excluding the sample day of interest. From the plots, we see that the PLS algorithm correctly predicts the below-average flow on February 24 and the above-average flow on July 2 where the algorithm nearly exactly predicts the peak flow for both movements between 17:00 and 18:00. In addition, the algorithm adeptly predicts the well below-average flow on January 1 for which traffic conditions greatly differ from the norm. Thus, while the PLS algorithm is ultimately a linear prediction scheme as is apparent in \eqref{eq:27}, the technique ably accommodates significant variation in the traffic flow.

To quantitatively assess the quality of the prediction, we compute a prediction for each of 132 days in the dataset using leave-one-out cross validation and compute the one-norm distance from the prediction to the actual flow measurements. That is, for each day $d=1,\ldots,D$, we compute the prediction error $E_\text{pred}^d$ as
\begin{align}
  \label{eq:28}
  E_\text{pred}^d=||y^d-\hat{y}^d||_1=\sum_{m=1}^M\sum_{t=T^*+1}^{\tp} |y_m^d(t)-\hat{y}_m^d(t)|
\end{align}
where $\hat{y}^d$ is the predicted traffic flow on day $d$ computed as in Section \ref{sec:pred-from-latent}.  We use the one-norm as our distance metric because it corresponds to the absolute total difference of the number of vehicles that are measured versus predicted at the intersection along each movement. Likewise, we define the baseline error to be the difference between the average flow measurements and the actual flow measurements\footnote{Since we employ a cross validation scheme, $\bar{y}$ will change slightly for each day because we remove a different set of measurements from the dataset in each case.}:
\begin{align}
  \label{eq:29}
  E^d_\text{base}=||y^d-\bar{y}||_1.
\end{align}

\begin{figure}
\begin{center}
\begin{tabular}{r}
  \includegraphics[width=.8\textwidth]{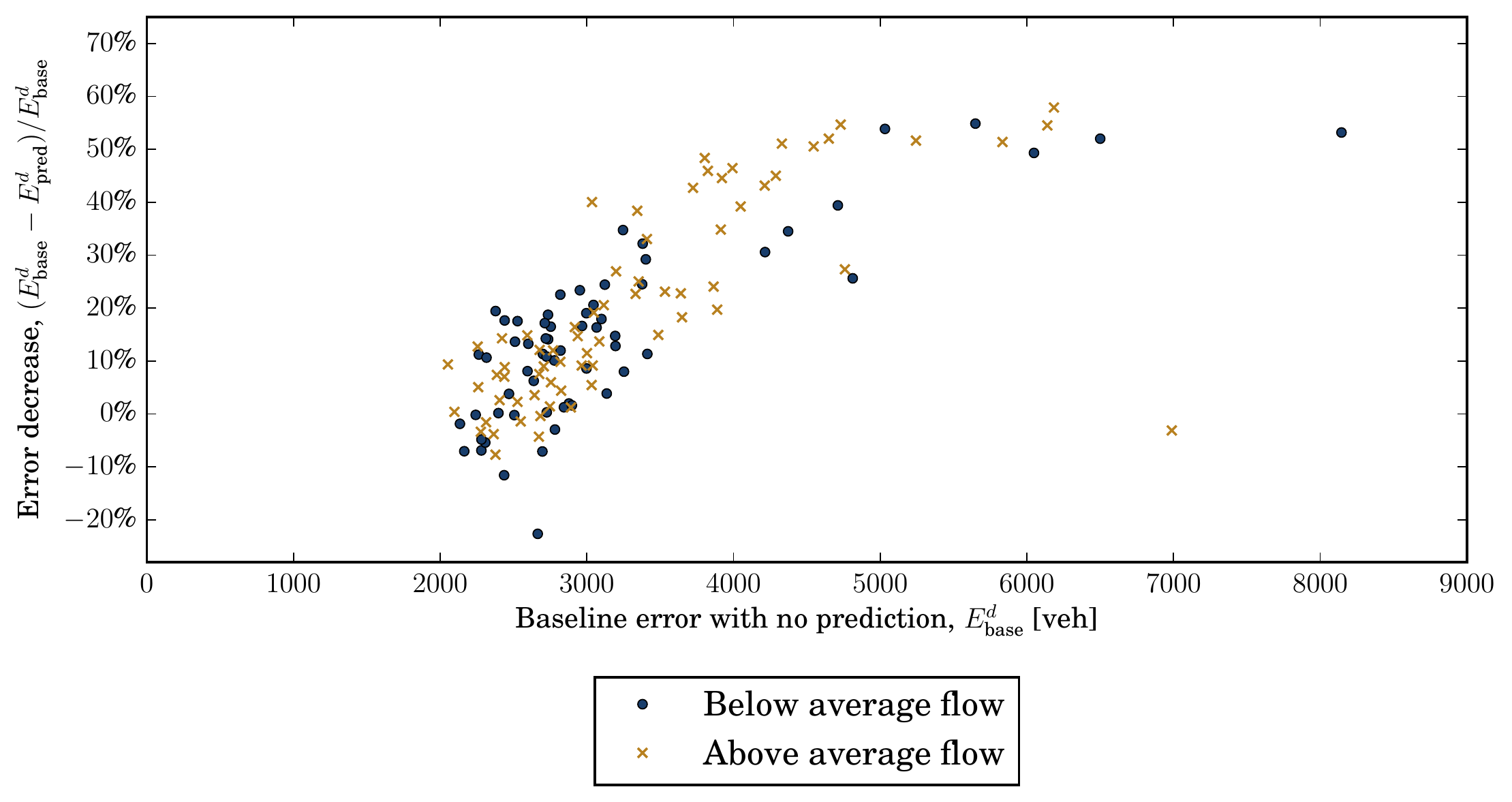}
\end{tabular}
\end{center}
\caption{Scatter plot of prediction error. The plot shows baseline error $E_\text{base}^d$ versus the normalized error decrease  $(E_\text{base}^d-E_\text{pred}^d)/E_\text{base}^d$ for each day $d=1,\ldots,D$. Days with below average (respectively, above average) flow are indicated with circles (respectively, crosses). For clarity, the plot excludes December 25 and January 1 which experience error decreases exceeding 60\% with baseline errors exceeding $14\,000$.}
  \label{fig:scatter}
\end{figure}

Figure \ref{fig:scatter} is a scatter plot of the normalized error decrease $(E_\text{base}^d-E_\text{pred}^d)/E_\text{base}^d$ versus the baseline error with no prediction $E_{\text{base}}^d$,  for each day $d=1,\ldots, D$. Excluded are December 24 and January 1 which both have baseline errors exceeding 14\,000 with prediction errors less than 5\,500, an error decrease of over 60\%. %
Nearly all of the points (113 out of 132) correspond to positive error decreases indicating that the prediction almost always is an improvement over the baseline average. The circles (respectively, crosses) are those days for which the total flow volume throughout the day is below (respectively, above) average. We see that the PLS algorithm performs well in both cases. For most days, the improvement in error is approximately 10\% to 20\%. However, for those days for which the baseline error is high, the improvement is higher and can be as high as 50\% to 60\%, and this holds for both those days for which traffic is below average and those days for which traffic is above average. Thus for days that differ significantly  from the average, the PLS prediction is especially accurate.

There is one particularly anomalous point for which the baseline error is high, approximately 7\,000, but the normalized error decrease is not significant and is in fact slightly negative. It appears that, on this day, total traffic volume was only moderately higher than normal but traffic flow was significantly above average for some movements and below average for other movements, although the cause is unclear. While the PLS algorithm was unable to predict this difference, we observe that the PLS algorithm did not result in a prediction error that was significantly worse than the baseline error.

\section{Traffic Predictive Control}
\label{sec:traff-pred-contr}
How do we utilize traffic flow predictions based on low-rank structure to improve traffic control? We first review a common technique for programming traffic signal controllers and describe an approach for optimizing this technique using only the average flow measurements. We will then use traffic predictions from the PLS algorithm to improve upon this approach.
A primary goal is to ensure the algorithm is easily applied to existing hardware and well-aligned with conventional traffic control practices.

\subsection{Time-of-Day Signal Operation}

A common approach to traffic control supported by nearly all traffic control hardware is \emph{time-of-day (TOD)} scheduling \cite[Chapter 5]{Koonce:2008gd} whereby a pre-defined control plan is applied during specified periods of the day. For example, a pre-defined plan may be specified for 6:00 to 9:00, and a different plan may be specified for 9:00 to 13:00, \emph{etc.}  Typically, the TOD periods are obtained via limited data collection at the intersection or simply by prior experience. 

Once the TOD periods have been selected, timing parameters are designed for each period and applied to the controller, constituting a TOD plan. Typical timing parameters include the cycle time of the intersection and green time allocations for the turn movements. A TOD plan may be completely fixed or it may be \emph{actuated} in which case the presence or absence of queued vehicles  extends green time allocations by a certain amount or leads to early termination of a phase actuation. 

The timing parameters for a given TOD period are determined by expected turn movement flows at the intersection. Since the TOD plan is applied at fixed periods and the parameters are fixed, the  plan is  designed around a set of nominal turn movement flows, for example, average turn movement flows over the TOD period.  Methods for determining desirable timing parameters given nominal turn movement flows have a long history going back at least to the seminal work of \cite{Webster:1958cs} and is not the focus of this paper. Instead, we assume that what is required for signal timing is only an estimation of the turn movement flows during a TOD period, for which any signal timing optimization scheme may be employed. This approach is reasonable if traffic flow does not deviate excessively from these estimated flows during a TOD period. In the following section, we suggest an algorithm for identifying TOD periods that minimize such deviations.

\subsection{Optimal Time-of-Day Segmentation}
\label{sec:optimal-time-day}
We focus on the problem of determining TOD periods and representative turn movement flows for each period. In the sequel, we will consider adjusting the TOD periods and representative turn movement flows based on predictions using the PLS algorithm. %

To this end, we now present an algorithm for optimally segmenting the 24-hour day into TOD periods. The intuitive idea is to identify segmentation times that minimize the variability in turn movement flows within each TOD period so that a particular fixed TOD plan works well throughout the period. For example, turn movement flows may remain relatively steady through the morning period but change substantially in the afternoon, requiring a different TOD plan. Increasing the number of TOD periods reduces variability in any given period, but there are practical limitations to the number of TOD periods that may employed at an intersection. For example, it typically takes several cycles totaling up to ten minutes to fully switch from one TOD plan to another; during this intermediate time, mobility at the intersection may be reduced. Furthermore, excessive changes to timing plans may confuse drivers. Traffic intersections commonly employ up to seven TOD plans throughout the day.

Approaches for TOD segmentation proposed in the literature include randomized clustering algorithms \cite{Smith:2002xi, Guo:2014zt}, heuristic genetic algorithms \cite{Park:2003mq}, and simulation based algorithms \cite{Brian-Park:2004ta}. To improve performance, the paper \cite{Guo:2014zt} suggests explicitly incorporating time as a variable in determining the clusters. Here, we consider an optimal segmentation approach suggested for generic data sets in \cite{Auger:1989jt} and adapt it to the context of traffic flow measurements. Like \cite{Guo:2014zt}, this approach explicitly accounts for time when identifying clusters. However, our approach is not based on $k$-means clustering and is able to identify optimal clusters in computational time quadratic in the number of time steps. In contrast, $k$-means clustering requires exponential computational time, although efficient heuristics exist.

Suppose we wish to obtain $S$ TOD periods for some $S>1$. We consider the equivalent problem of choosing $S-1$ segmentation times $\tau_1, \tau_2, \ldots, \tau_{S-1}$ with each $\tau_i\in\{1,2,\ldots, T-1\}$ satisfying
\begin{align}
  \label{eq:30}
  \tau_1<\tau_2< \ldots< \tau_{S-1}
\end{align}
so that
\begin{align}
  \label{eq:31}
  \{\tau_{i-1}+1,\tau_{i-1}+2,\ldots,\tau_{i}\}
\end{align}
defines the $i$th TOD period for $i\in\{1,\ldots,S\}$, with $\tau_0:=0$ and $\tau_{S}:=T$.

To assess the quality of a given segmentation $(\tau_1,\ldots,\tau_{S-1})$, we consider a vector of turn movement flows $x\in\mathbb{R}^{TM}$ throughout the day and define the \emph{cost} of a time segment $(t_a,t_a+1,\ldots,t_b)$ given these flows for some $t_a\leq t_b$ as follows:
\begin{align}
  \label{eq:33}
\text{Cost}&=\min_{\mu\in\mathbb{R}^{P}} F(t_a,t_b,x,\mu)
\end{align}
where $\mu\in\mathbb{R}^P$ is a parameter vector of dimension $P\geq 1$ and $F(t_a,t_b,x,\mu)$ is a positive function that is convex in $\mu$. We call $F(t_a,t_b,x,\mu)$ the \emph{fit} of $x$ with the parameter vector $\mu$ on time segment $(t_a,t_a+1,\ldots,t_b)$. In this way, $F(\tau_{i-1}+1,\tau_i,x,\mu)$ is the fit of $x$ with $\mu$ on TOD period $i$. For $t_b<t_a$, we define the fit $F$ to be $0$.  To make this concrete, in this paper, we assume $P=M$ so that $\mu\in\mathbb{R}^M$ and take
\begin{align}
  \label{eq:32}
   F(t_a,t_b,x,\mu)&=\sum_{t=t_a}^{t_b}\Phi(x(t),\mu)
\end{align}
for $\Phi:\mathbb{R}^{M}\times\mathbb{R}^M\to\mathbb{R}^M_{\geq 0}$ a positive function that is convex in its second argument where $\mathbb{R}^M_{\geq 0}=\{x\in\mathbb{R}^M\mid x_m\geq 0 \text{ for all } m=1,\ldots,M\}$.

For period $i$, let $\mu_i\in\mathbb{R}^{M}$ denote the minimizer of \eqref{eq:33}  with $t_a:=\tau_{i-1}+1$ and $t_b:=\tau_i$. We interpret $\mu_i$ as a vector of flows that suitably represents the turn movement flows throughout the $i$th period and is used to determine the signal timing parameters. For example, if
\begin{align}
  \label{eq:48}
  \Phi(x,\mu)=(x-\mu)^{\tp}(x-\mu),
\end{align}
then the minimizing $\mu_i$ is given by $\mu_{i}=\frac{1}{\tau_i-\tau_{i-1}}\sum_{t=\tau_{i-1}+1}^{\tau_i}x(t)$, that is, $\mu_{i}$ is the average flow for the movements during period $i$.

Choosing $\Phi(x,\mu)$ as in \eqref{eq:48} penalizes the difference between the turn movement flow and the corresponding value in the parameter vector. This choice equally penalizes turn movement flows that are above and below the parameter vector flow. However, it is often case that flows which exceed the parameter vector result in worse performance degradation at the intersection than flows which are less than the parameter vector. For this reason, we propose an asymmetric function
\begin{align}
  \label{eq:34}
\Phi(x(t),\mu)&=\sum_{m=1}^M\varphi(x_m(t),\mu_m)\\
  \varphi(a,b)&=
  \begin{cases}
    C(a-b)^2&\text{if }a>b\\
(a-b)^2&\text{else}
  \end{cases}
\quad m=1,\ldots,M \label{eq:46}
\end{align}
for some choice $C\geq 1$ where $\mu_m$ denotes the $m$th entry of $\mu$. For $C=1$, we recover \eqref{eq:48}.

Finally, we choose segmentation times to minimize the sum of costs for all TOD periods:
\begin{align}
  \label{eq:35}
  (\tau_1,\ldots,\tau_{S-1})=\argmin_{t_1,\ldots,t_{S-1}}\sum_{i=1}^S\min_{\mu}F(\tau_{i-1}+1,\tau_i,x,\mu).
\end{align}

Convexity of $\Phi$ ensures that \eqref{eq:33} is a convex optimization problem. Assuming that solving \eqref{eq:33} requires computational time of $\Gamma(M)$, then \eqref{eq:35} is solved in time $O(M^2\Gamma(M))$ \cite{Auger:1989jt}. We note that our segmentation approach prevents oscillations in implemented timing plans since TOD periods are required to be contiguous intervals of time, in contrast to some clustering-based segmentation algorithms in the literature that do not account for contiguity in the TOD periods and for which the implemented signal timing plans have been observed to oscillate.

\begin{figure}
  \centering
  \begin{tabular}{@{}l@{} @{}l@{}}
    \includegraphics[height=1.6in]{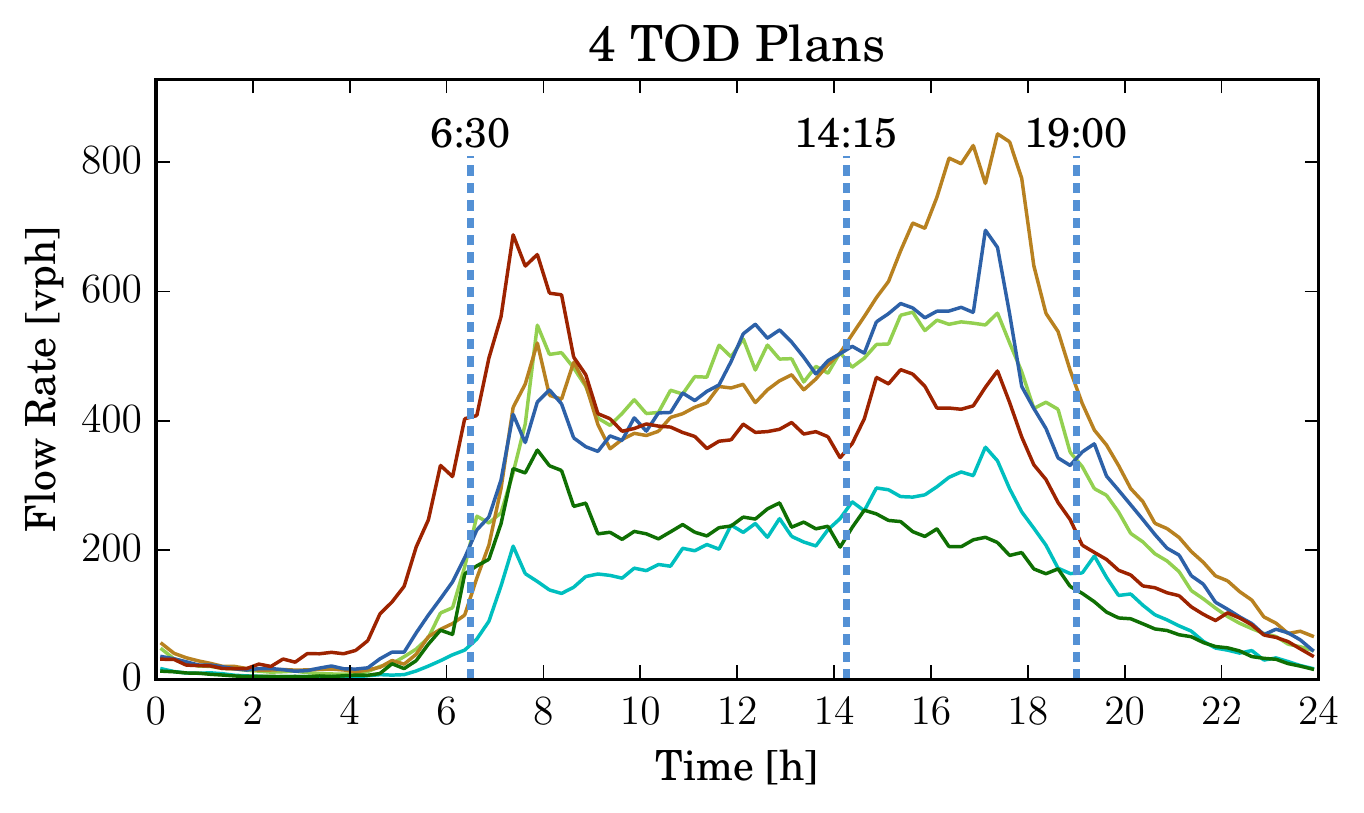}&    \includegraphics[height=1.6in]{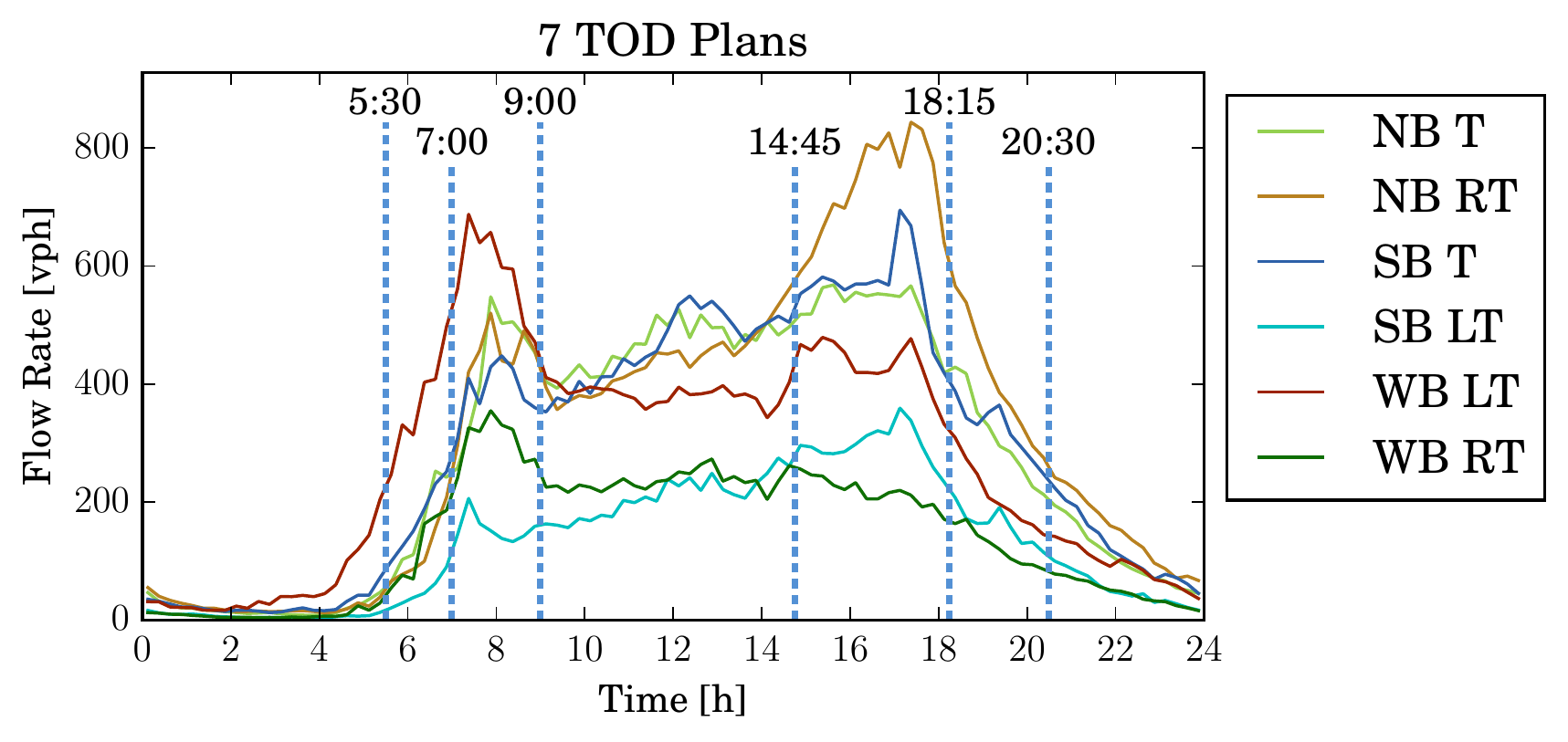}
  \end{tabular}
  \caption{Optimal segmentation of average traffic flow over the day. (a) Optimal segmentation allowing for four time-of-day periods. (b) Optimal segmentation allowing for seven time-of-day periods.}
  \label{fig:seg}
\end{figure}

We return to our case study dataset and apply the TOD segmentation algorithm above where we take $\varphi$ as defined in \eqref{eq:46} with $C=2$. We first take $x$  to be the mean flow across the dataset as defined in \eqref{eq:2}--\eqref{eq:3}, that is, we first consider $x:=\bar{x}$ in \eqref{eq:33}--\eqref{eq:35}. This constitutes the nominal TOD periods. In Figure \ref{fig:seg}, we plot the results of the optimal segmentation problem \eqref{eq:35} with $S=4$ and $S=7$ for this nominal case. In general, we see that the segmentation algorithm chooses segmentation times corresponding to when turn movement flows are changing rapidly, resulting in intuitive divisions of the day. For example, with seven TOD periods, we clearly obtain the morning peak period from 7:00 to 9:00 and an afternoon peak period from 14:45 to 18:15.

\subsection{Predictive Time-of-Day Plans}
\label{sec:predictive-time-day}
The segmentation algorithm presented in Section \ref{sec:optimal-time-day} where we use the mean flow $\bar{x}$ is a method for optimally determining nominal TOD periods, an important component of conventional traffic control. Thus this is a useful innovation in itself that can be immediately implemented with existing hardware, but it does not leverage the traffic prediction algorithm developed in Section \ref{sec:traff-pred-from}. Here we propose a method for extending this idea to allow real-time adjustments to TOD periods and TOD plans based on predicted traffic flow. The key idea  is to enable limited and intuitive modifications to the standard TOD scheduling paradigm so that these methods can be implemented with modest modification to existing hardware, and traffic engineers are able to easily envision the benefits of traffic predictive control and therefore more easily adopt this approach.

To this end, we assume a set of nominal segmentation times $\tau_1,\ldots,\tau_{S-1}$ have been chosen according to \eqref{eq:35}, defining a set of $S$ TOD periods. Additionally, a set of parameter vectors $\mu_1,\ldots,\mu_S$ are obtained for each TOD period as the minimizers of \eqref{eq:33}. These TOD periods and parameter vectors constitute the nominal intersection timing plans. 

We now suggest the following intuitive predictive traffic control scheme: throughout the day, online measurements are used to predict future traffic flow. When the current time approaches a nominal segmentation time, the predicted traffic flow is used to decide if the intersection controller should switch to a new TOD plan earlier or later than the nominal segmentation time.

To formalize this idea,  let $x^s\in\mathbb{R}^{TM}$ denote the traffic flow for a particular sample day and suppose that the current time is $t$ so that
\begin{align}
  \label{eq:36}
  x^\text{meas}:=\begin{bmatrix}x^s(1)&x^s(2)&\ldots x^s(t)\end{bmatrix}\in\mathbb{R}^{tM}
\end{align}
is the vector of currently available time measurements. We let 
\begin{align}
  \label{eq:42}
\textsc{SegmentWindow}(\tau)\subseteq \{1,2,\ldots,T\}
\end{align}
denote the window of times around a nominal segmentation time $\tau$ for which it is acceptable to switch early or late to a new TOD period. That is, the signal controller has the flexibility to switch from TOD period $i$ to TOD period $i+1$ at any time within $\textsc{SegmentWindow}(\tau_{i})$ and  the switch must occur within this window. The acceptable window is a design parameter, \emph{e.g.}, it may be chosen by a traffic engineer to meet other system requirements.
As an example, if the user-selected criterion allows for switching to a new TOD plan up to 45 minutes before or after the nominal segmentation time, we have $\textsc{SegmentWindow}(\tau)=\{\tau-3,\tau-2,\ldots,\tau+2,\tau+3\}$ when $\Delta=\text{15 minutes}$. 

 Now  suppose that the current TOD period is $i$, the currently active parameter vector is $\mu^*_i\in\mathbb{R}^{M}$, and that $t\in\textsc{SegmentWindow}(\tau_i)$ so that it is allowable to switch to the $(i+1)$-th TOD period at the current time $t$. To determine if switching is desirable, we use a prediction of future traffic flow to determine if it is possible to achieve a lower segment cost by choosing a different segmentation time. To this end, let
 \begin{align}
   \label{eq:37}
   \hat{y}=\begin{bmatrix}\hat{y}(t+1)&\hat{y}(t+2)&\ldots&\hat{y}(\tau_{i+1})\end{bmatrix}\in\mathbb{R}^{(\tau_{i+1}-t)M}
 \end{align}
be a prediction of traffic flow from time $t+1$ to time $\tau_{i+1}$, the nominal ending time for the next TOD period.  We wish to find the segmentation time which minimizes the predicted remaining segment cost for the $i$-th TOD period and the predicted segment cost for the $(i+1)$-th TOD period. We thus compute
\begin{align}
  \label{eq:38}
  t_\text{opt}=\argmin_{\substack{u\in\textsc{SegmentWindow}(\tau_i)\\\text{s.t. }u\geq t}} \left(F(t+1,u,\hat{y},\mu_i^*)+\min_\mu F(u+1,\tau_{i+1},\hat{y},\mu)\right).
\end{align}
We interpret $t_\text{opt}$ as the optimal time to switch from TOD period $i$ to TOD period $i+1$ (recall that for $u<t+1$, we have $F(t+1,u,\hat{y},\mu_i^*)=0$). If $t_\text{opt}=t$, then it is best to switch to the $(i+1)$-th TOD period immediately. In this case, we define
\begin{align}
  \label{eq:39}
  \tau^*_i&=t_\text{opt}\\
  \label{eq:39-2}
\mu^*_{i+1}&=\min_\mu  F(u+1,\tau_{i+1},\hat{y},\mu)
\end{align}
as the \emph{predictive} segmentation time for the $i$-th period and the predictive parameter vector for the $(i+1)$-th TOD period. We then repeat the process for the next TOD segmentation time.
However, if $t_\text{opt}>t$, then it is optimal to continue with the current TOD period, delaying the switch to the $(i+1)$-th period. Time advances to $t+1$ and the process repeats, where we update the measurement vector and recompute the predicted flow measurements. In particular, this recomputation implies that $t_\text{opt}$ updates based on the latest measurements. To initialize this approach, we define $\mu^*_1:=\mu_1$, that is, the parameter vector for the first TOD period is assumed to be the nominal parameter vector $\mu_1$ since no measurements are available yet that would alter this prediction.

Finally, we highlight one particularly important possible modification to the above procedure. In computing $t_\text{opt}$ in \eqref{eq:38}, we assumed that we are able to establish a new parameter vector for the $(i+1)$-th TOD period, however due to constraints on the traffic signaling hardware or the preference of practitioners, this may not be possible. For example, the traffic signal controller may allow variability in the segmentation times defining TOD periods but  not allow the timing plans themselves to be altered. In this case, we choose a new segmentation time assuming that the parameter vector remains fixed. We modify \eqref{eq:38} as 
\begin{align}
\tag{\ref{eq:38}b}
  \label{eq:40}
    t_\text{opt}=\argmin_{\substack{u\in\textsc{SegmentWindow}(\tau_i)\\\text{s.t. }u\geq t}} \left(F(t+1,u,\hat{y},\mu_i)+ F(u+1,\tau_{i+1},\hat{y},\mu_{i+1})\right).
\end{align}

Algorithm \ref{fig:algo1} summarizes this traffic predictive approach. We now prove a desirable consistency property of this algorithm. In particular, we show that, for the fit function given in \eqref{eq:46}, if the predicted traffic flow is equal to the nominal traffic flow, then the segmentation times and parameter vectors obtained via Algorithm \ref{fig:algo1} are equal to the nominal segmentation times and parameter vectors. Below, we assume for simplicity that minimizing arguments are unique.

\begin{prop}
\label{prop:consist}  
Consider times $\tau_a<\tau_b<\tau_c$  defining two TOD periods $\{\tau_a+1,\ldots,\tau_b\}$ and $\{\tau_b+1,\ldots,\tau_c\}$, and consider the fit function given in \eqref{eq:46}. Assume that $\tau_b$ is the optimal time to switch from the first TOD period to the second TOD period when the flow is $x\in\mathbb{R}^{TM}$, and suppose $\mu_1$ and $\mu_2$ are the optimal parameter vectors for these periods, that is,

\begin{align}
  \label{eq:43}
  (\tau_b,\mu_1,\mu_2)=\argmin_{\tilde{\tau}_b,\tilde{\mu}_1,\tilde{\mu}_2} F(\tau_a+1,\tilde{\tau}_b,x,\tilde{\mu}_1)+F(\tilde{\tau}_b+1,\tau_c,x,\tilde{\mu}_2).
\end{align}
Then, for all $t\in\{\tau_a,\ldots,\tau_b\}$, 
\begin{align}
  \label{eq:44}
  (\tau_b,\mu_2)=\argmin_{\tilde{\tau}_b,\tilde{\mu}_2} F(t+1,\tilde{\tau}_b,x,\mu_1)+F(\tilde{\tau}_b+1,\tau_c,x,\tilde{\mu}_2).
\end{align}
\end{prop}
Proposition \ref{prop:consist} states that, if the measured data coincides with $x$, $\tau_b$ remains the optimal time to switch TOD periods even if the segmentation time and parameter vector of the second TOD is recomputed at time $t\geq \tau_a$.
\begin{proof} Consider $t\in\{\tau_a,\ldots,\tau_b\}$.
Let 
\begin{align}
  \label{eq:45}
(\tau_b',\mu_2')=\argmin_{\tilde{\tau}_b,\tilde{\mu}_2} F(t+1,\tilde{\tau}_b,x,\mu_1)+F(\tilde{\tau}_b+1,\tau_c,x,\tilde{\mu}_2).   
\end{align}
Observe that
\begin{align}
  \label{eq:52}
F(t_1,t_2,x,\mu)=F(t_1,t_3,x,\mu)+F(t_3+1,t_2,x,\mu)\quad \text{ for all $t_1,t_2,t_3$}
\end{align}
for any  $x$, $\mu$. Thus, by \eqref{eq:43}, we have that
\begin{align}
  \label{eq:49}
F(\tau_a+1,\tau_b,x,\mu_1)+F(\tau_b+1,\tau_c,x,\mu_2)&=  F(\tau_a+1,t,x,\mu_1)+  F(t+1,\tau_b,x,\mu_1)+  F(\tau_b+1,\tau_c,x,\mu_2) \\
&\leq \min_{\tilde{\mu}_1,\tilde{\mu}_2} F(\tau_a+1,\tau_b',x,\tilde{\mu}_1)+F(\tau '_b+1,\tau_c,x,\tilde{\mu}_2).
\end{align}
It follows that
\begin{align}
  \label{eq:50}
F(t+1,\tau_b,x,\mu_1)+  F(\tau_b+1,&\tau_c,x,\mu_2)\\
&\leq \min_{\tilde{\mu}_1,\tilde{\mu}_2} F(\tau_a+1,\tau_b',x,\tilde{\mu}_1)+F(\tau '_b+1,\tau_c,x,\tilde{\mu}_2)-  F(\tau_a+1,t,x,\mu_1)\\
&=\min_{\tilde{\mu}_1} F(\tau_a+1,\tau_b',x,\tilde{\mu}_1) -  F(\tau_a+1,t,x,\mu_1)+\min_{\tilde{\mu}_2}F(\tau '_b+1,\tau_c,x,\tilde{\mu}_2)\\
  \label{eq:50-end}&\leq F(t+1,\tau_b',x,\mu_1)+\min_{\tilde{\mu}_2}F(\tau '_b+1,\tau_c,x,\tilde{\mu}_2)
\end{align}
where the last inequality follows because
\begin{align}
  \label{eq:51}
  \min_{\tilde{\mu}_1} F(\tau_a+1,\tau_b',x,\tilde{\mu}_1)\leq F(\tau_a+1,t,x,\mu_1)+F(t+1,\tau_b',x,\mu_1).
\end{align}
Then \eqref{eq:44} follows from \eqref{eq:45} and \eqref{eq:50}--\eqref{eq:50-end}.
\end{proof}

The form of the fit function given in \eqref{eq:46} ensures \eqref{eq:52} in the proof of Proposition \ref{prop:consist}. Thus, the proposition may fail to hold for alternative choices of fit functions.

\begin{cor}
  If $\hat{y}=\begin{bmatrix}x^s(t+1)& x^s(t+2)&\ldots&x^s(\tau_{i+1})\end{bmatrix}$, that is, the predicted traffic flow is equal to the measured traffic flow, then $\tau_i=\tau^*_i$ and $\mu_i=\mu^*_i$ for all $i=1,\ldots,S$, that is, the predictive segmentation times and predictive parameter vectors are equal to the nominal segmentation times and parameter vectors.
\end{cor}

From a computational perspective, Algorithm \ref{fig:algo1} consists of an online component which computes the optimal predictive segmentation time for each TOD period as well as an offline component which processes historical data to obtain the PLS components for the predictions required in line \ref{line:predict}. For the offline component, we compute PLS predictor/predicted components for each possible segmentation time. Thus, suppose $|\textsc{SegmentWindow}(\tau)|=W$ for all $\tau$ for some $W>0$, that is, there are $W$ allowable segmentation times around each nominal segmentation time. Then we execute the PLS algorithm a total of $SW$ times. While this process is potentially computationally taxing, it can be accomplished offline by processing historical data. To obtain $\hat{y}$ in line \ref{line:predict} of Algorithm \ref{fig:algo1} requires only matrix multiplication using the PLS components that are computed offline and stored. Next, computing $t_\text{opt}$ requires solving at most $W$ convex optimization problems if using \eqref{eq:38}, or $W$ evaluations of the fit function $F$ if using \eqref{eq:40}. Solving the convex optimization problem is typically fast, as are evaluations of $F$, thus $t_\text{opt}$ is easily computed within any practically sized time step $\Delta$, affording ample time to decide predictive segmentation times and parameter vectors online.

\begin{algorithm}
  \centering
\begin{minipage}{\linewidth}
\begin{algorithmic}[1]
\algblockdefx{InputS}{EndInputS}{\textbf{inputs: }}{}
\algtext*{EndInputS}
\algblockdefx{OutputS}{EndOutputS}{\textbf{outputs: }}{}
\algtext*{EndOutputS}
\Function{PredictiveTrafficControl}{$(\tau_1,\ldots,\tau_{S-1})$, $(\mu_1,\ldots,\mu_S)$, $x^s$}

\InputS \hspace{.08in}$(\tau_1,\ldots,\tau_{S-1})$, nominal segmentation times defining $S$ TOD periods
\State \pushcodeb $(\mu^1,\ldots,\mu^S)$, parameter vectors for the TOD periods
\State \pushcodeb $x^s\in\mathbb{R}^{TM}$, measured traffic flow, available in real-time
\EndInputS 
\OutputS $(\tau_1^*,\ldots,\tau_{S-1}^*)$, predictive segmentation times
\State \pushcodeb $(\mu_1^*,\ldots,\mu_S^*)$, predictive parameter vectors
\EndOutputS
\State $\mu^*_1:=\mu_1$ %
\State $i:=1$  \Comment Current TOD period
\For{$t=1,2,\ldots,T$}   \Comment Current time
\State $x^\text{meas}:=\begin{bmatrix}x^s(1)&x^s(2)&\ldots x^s(t)\end{bmatrix}\in\mathbb{R}^{tM}$  \Comment Flow measurements up to current time
\If{$t\in\textsc{SegmentWindow}(\tau_i)$}
\State $\hat{y}=\textsc{Predict}(x^\text{meas},t+1,\tau_{i+1})$ \Comment Predicted traffic flow up to end of next TOD period \label{line:predict}
\State $\displaystyle t_\text{opt}:=$ according to
 \eqref{eq:38} or \eqref{eq:40} \Comment Best time to switch according to current prediction

\If{$t_\text{opt}=t$}
\State $\tau^*_i:=t_\text{opt}$
\State $\displaystyle \mu^*_{i+1}:=\begin{cases}\argmin_\mu F(\tau^*_i+1,\tau_{i+1},\hat{y},\mu)&\text{ if using \eqref{eq:38}}\\ \mu_{i+1}&\text{ if using \eqref{eq:40}}\end{cases}$
\State $i:=i+1$
\EndIf
\EndIf
\EndFor
\State \Return $((\tau_1^*,\ldots,\tau_{S-1}^*), (\mu_1^*,\ldots,\mu_S^*))$
\EndFunction
\vspace{5pt}

\algblockdefx{InputS}{EndInputS}{\textbf{inputs: }}{}
\algtext*{EndInputS}
\algblockdefx{OutputS}{EndOutputS}{\textbf{output: }}{}
\algtext*{EndOutputS}
\algblockdefx{Summary}{EndSummary}{\textbf{summary: }}{}
\algtext*{EndSummary}
  \Function{Predict}{$x^\text{meas}$, $t_a$, $t_b$}
\InputS \hspace{.08in}$x^\text{meas}\in\mathbb{R}^{tM}$, measured traffic flow up to time $t$
\State \pushcodeb $t_a$ and $t_b$, start and end times for traffic flow prediction
\EndInputS
\OutputS $\hat{y}\in\mathbb{R}^{(t_b-t_a+1)M}$, traffic flow prediction from time $t_a$ to time $t_b$
\EndOutputS
\Summary \emph{Using historical data and the prediction scheme developed in Section \ref{sec:traff-pred-from}, predict traffic}
\State \pushcodeb \emph{flow from time $t_a$ to time $t_b$ using real-time measurements $x^\textnormal{meas}$.}
\EndSummary
\EndFunction
\end{algorithmic}
\end{minipage}
  \caption{Algorithm for determining predictive segmentation times and predictive parameter vectors given a set of nominal segmentation times and vectors as well as a vector of current flow measurements. The algorithm runs online as time progress from $t=1$ to $t=T$.}

\label{fig:algo1}
\end{algorithm}

\subsection{Case Study}
We assume nominal segmentation times for seven TOD plans as shown in Figure \ref{fig:seg}b. Now consider a traffic predictive controller as developed in Section \ref{sec:predictive-time-day} that is able to adjust in real-time the TOD periods so that the segmentation times are up to 45 mins earlier or later than the nominal segmentation times.

\begin{figure}
  \centering
  \includegraphics[width=.7\textwidth,clip=true, trim=0in .75in 0in 0in]{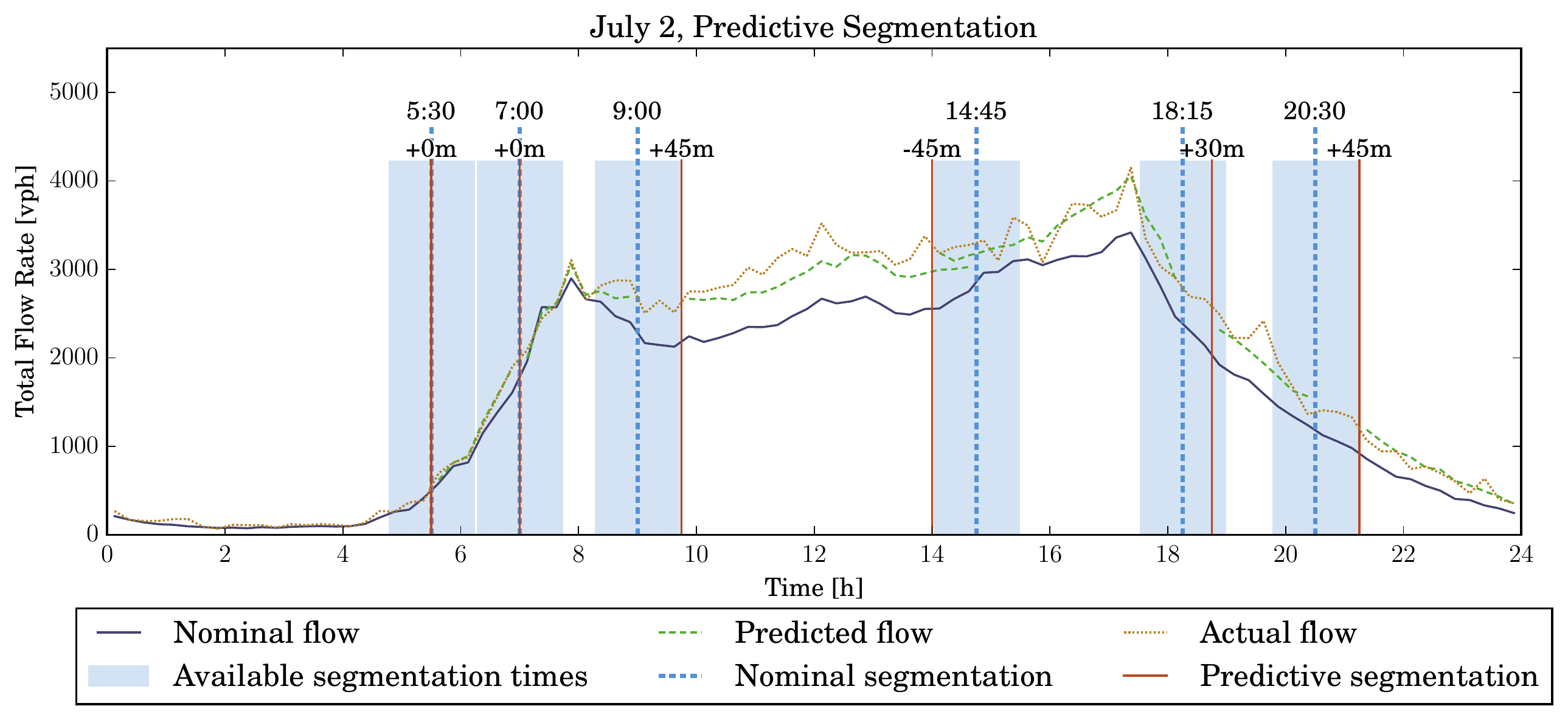}\\
  \includegraphics[width=.7\textwidth]{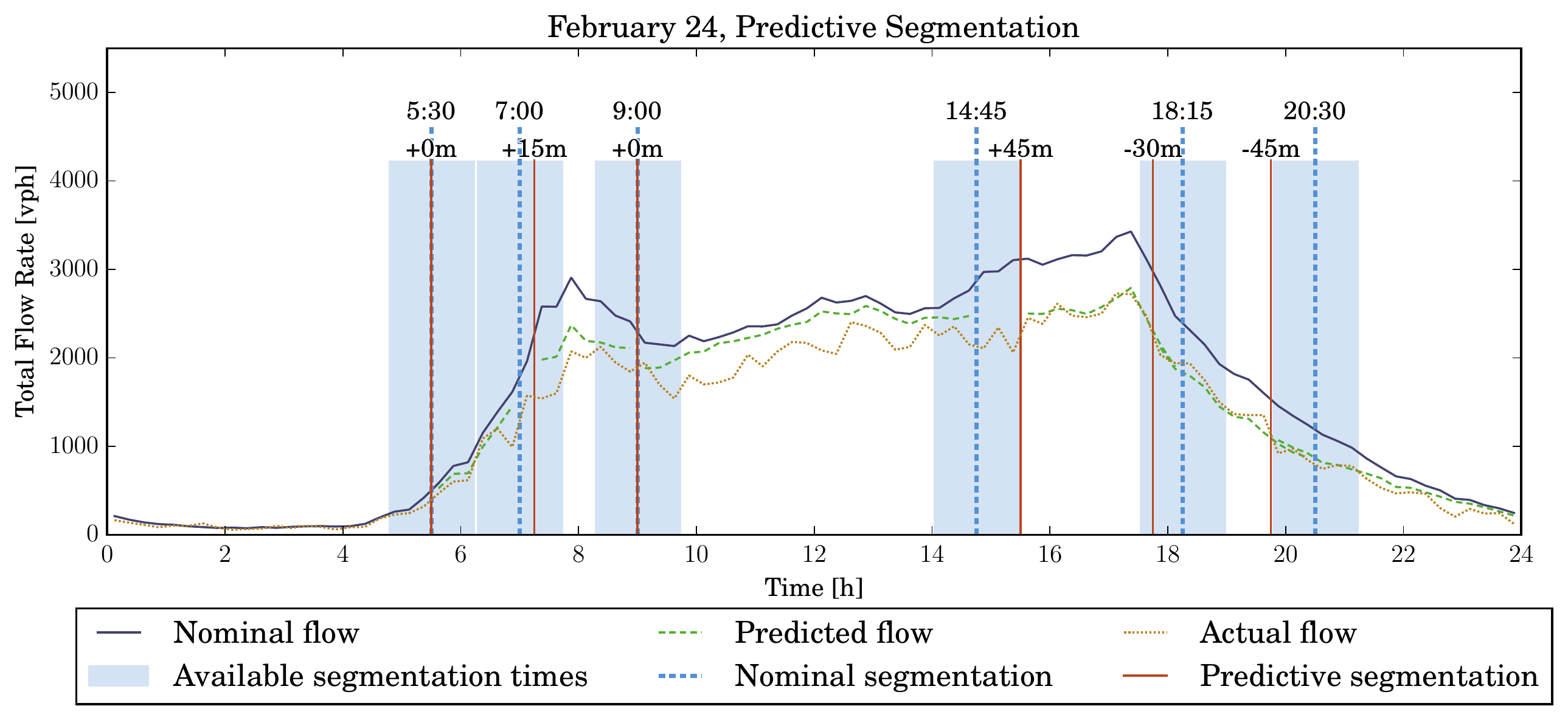}\\
  \caption{Example of traffic predictive control for two unusual days. The traffic predictive controller computes predictive segmentation times and adjusts the TOD periods to accommodate the predicted traffic flow, which deviates from the nominal average computed over the dataset. For clarity, the plots show total traffic flow through the intersection, summed over the twelve movements, however the prediction algorithm considers separate flow measurements for each movement.}
  \label{fig:pred_cont}
\end{figure}

 We first consider the case for which the traffic predictive controller is only able to establish predictive segmentation times and not predictive parameters.  Figure \ref{fig:pred_cont} plots the results of executing the traffic predictive controller on the unusual dates February 24 (below average traffic due to weather conditions) and July 2 (above average flow due to holiday). The solid line in the figure depicts the mean \emph{nominal} total traffic flow over the dataset (excluding the unusual date), summed over the 12 movements, in 15 minute intervals. Here, we plot total traffic flow only for clarity; the analysis considers separate flow measurements for all 12 movements as above. The dashed line is the total predicted traffic flow for the next TOD period as predicted at the start of each TOD period.  The vertical dashed blue lines are the nominal segmentation times, the lightly shaded bars represent the segmentation windows during which a change to a new TOD period is allowed, and the solid vertical lines are the actual segmentation times as determined using the traffic predictive controller.

From the figure, it is evident that the traffic predictive controller behaves in a reasonable way. For example, on July 2, the nominal TOD period between 7:00 and 9:00 corresponds to the morning peak period. Since traffic is predicted to be above average at this time, this TOD period is extended by 45 minutes. Likewise, since traffic in the afternoon peak period is predicted to be above average, the corresponding TOD period is extended, beginning 45 minutes earlier and ending 30 minutes later. On February 24, we see the reverse phenomenon due to the prediction that traffic will be below average; the TOD periods for the morning and afternoon peak are correspondingly shortened.

A key feature of the traffic predictive controller as developed in Section \ref{sec:predictive-time-day} is that the scheme is agnostic regarding the specific green split algorithm utilized, that is, the algorithm seeks to predict flows and segmentation times, not green splits directly. However, to estimate the gain in performance, we must choose a particular green split algorithm.  We employ a classical delay minimizing green split optimization algorithm as developed in \cite{Allsop:1971fh}, where we substitute modified delay formulas from the Highway Capacity Manual (HCM) \cite{Manual:2000qd}. These delay formulas account for queue buildup when the green splits are too short to accommodate queued vehicles. It is assumed that a fixed time control strategy is used within each TOD period calculated using the parameter vector, which is inflated by a factor to accommodate random fluctuations that would occur if vehicles arrive according to a Poisson process, a reasonable assumption.

\begin{figure}
  \centering
\begin{tabular}{@{}c@{} @{}c@{}}
\includegraphics[width=.5\textwidth,clip=true,trim=1.65in .75in 1.7in 0in]{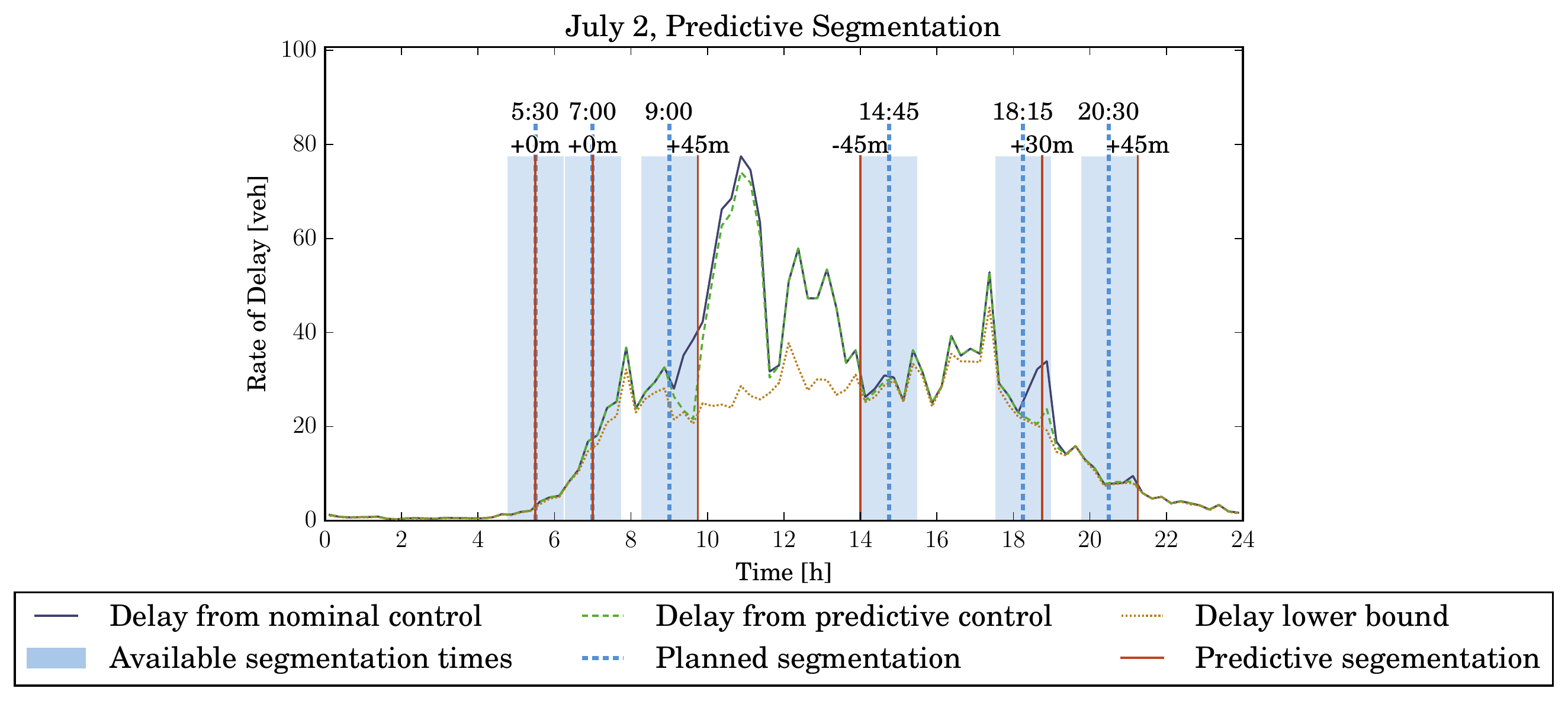}&
\includegraphics[width=.5\textwidth,clip=true,trim=1.65in .75in 1.7in 0in]{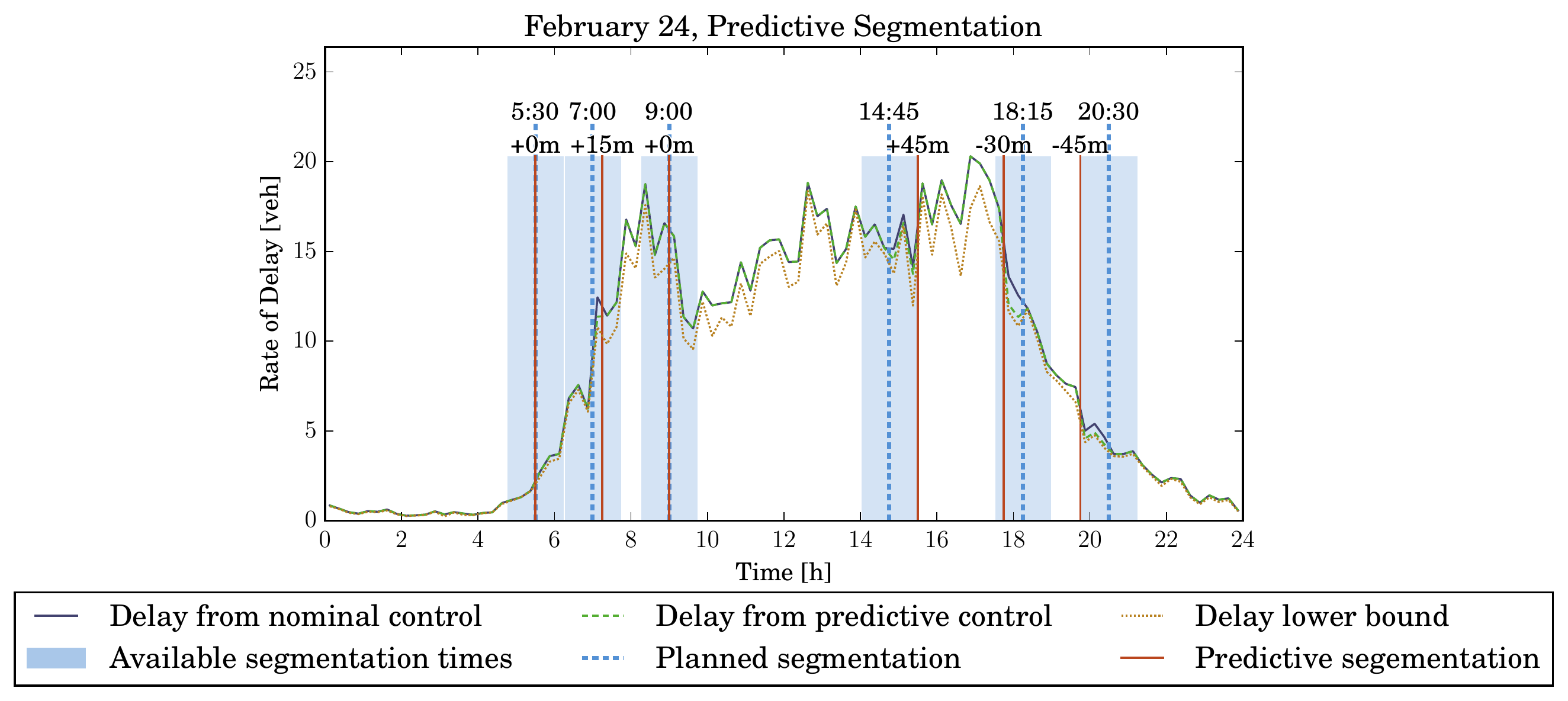}\\
\includegraphics[width=.5\textwidth,clip=true,trim=1.65in .75in 1.7in 0in]{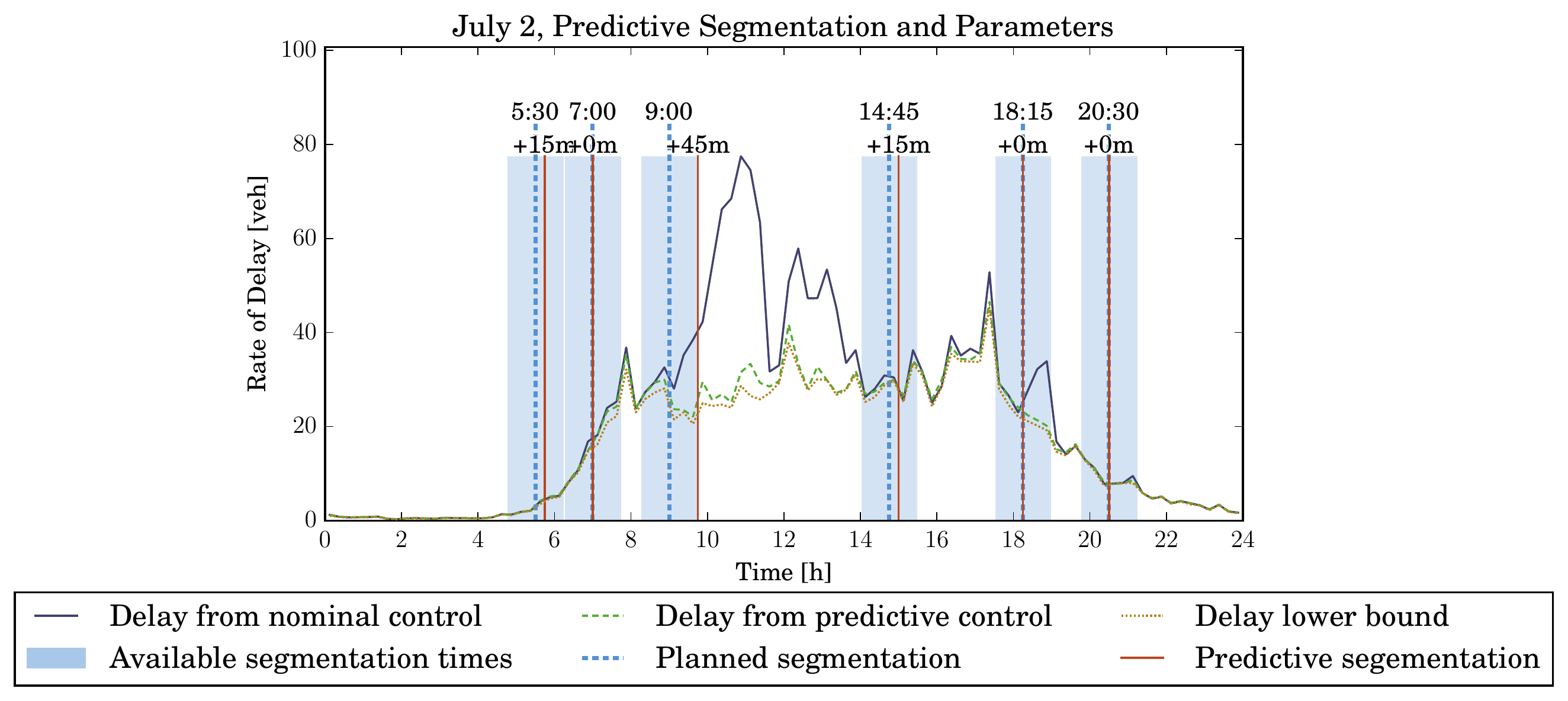}&
\includegraphics[width=.5\textwidth,clip=true,trim=1.65in .75in 1.7in 0in]{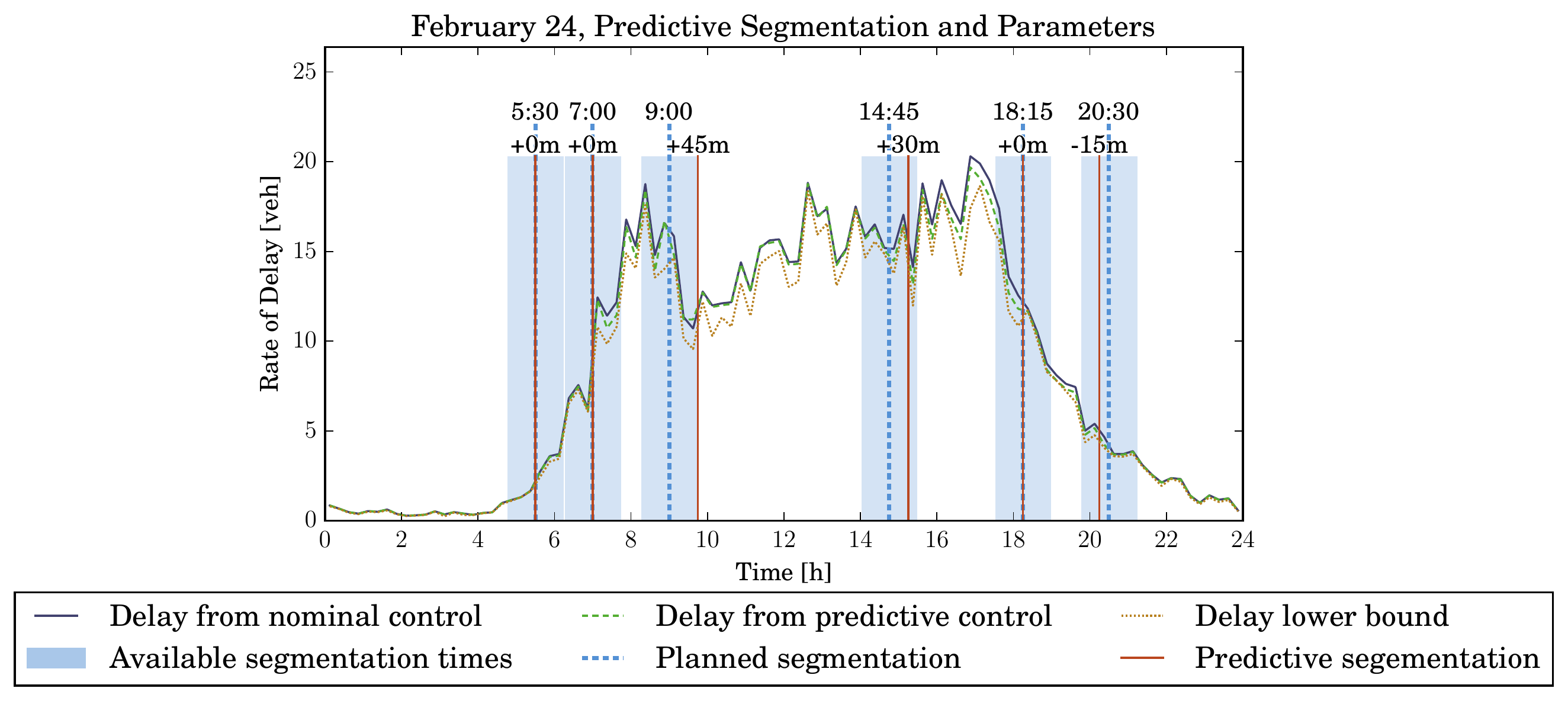}\\
\multicolumn{2}{c}{
\includegraphics[width=.7\textwidth,clip=true,trim=0in 0in 0in 3.95in]{img006}
}
\end{tabular}
  \caption{Rate of delay induced by the traffic predictive controller that adjusts TOD periods and TOD plans based on real-time traffic data as compared to the nominal controller that does not adjust TOD periods or plans. It is assumed that the green splits are computed to minimize delay using the parameter vectors for each TOD period. The top plots are the case when only predictive segmentation is considered while the bottom plots consider predictive segmentation and predictive parameter vectors. Traffic predictive control eliminates excessive delay caused by queued vehicles that require multiple cycles to clear the intersection and reduces wasted green time. Table \ref{tab:final} quantifies these delay improvements.}
\label{fig:delay}
\end{figure}

In Figure \ref{fig:delay}, we plot the result of using this delay minimizing green split optimization with our traffic predictive controller. In each subplot, the solid trace is the rate of delay for the intersection plotted over time for the case where the green splits are designed to minimize delay using the nominal segmentation times and nominal parameter vectors. The rate of delay is calculated using the analytical formulas found in the HCM. The dashed trace is the rate of delay obtained using the traffic predictive controller. The top plots consider a traffic predictive controller that uses predictive segmentation times but nominal parameter vectors, that is, the green splits are not allowed to differ from nominal during each TOD period, however, the duration and starting point of the TOD period is adjusted based on the predicted flow. The bottom plots consider the case when the traffic predictive controller uses predictive segmentation times and predictive parameter vectors, thereby adjusting the green splits based on predicted traffic flow. The dotted trace indicates a lower bound on the rate of delay that is computed assuming that the optimal green splits for each fifteen minute interval are applied at each time step.

Suboptimality of green splits occurs due to two reasons: either the splits are too short for some movements, resulting in queued vehicles that must wait more than one cycle to clear the intersection; or the splits are too long, resulting in wasted green time whereby vehicles at other movements must wait longer than necessary to receive a green light. For the nominal control case, the former condition occurs on July 2 and the latter condition occurs on February 24. Indeed, the large increase in the rate of delay between 10:00 and 11:00 on July 2 results from queued vehicles waiting multiple cycles to move through the intersection, so-called \emph{cycle failures}. These queued vehicles contribute substantially to the rate of delay at the intersection. Allowing predictive segmentation mitigates the issue somewhat as the traffic predictive controller delays by 45 minutes the change to a new TOD period that nominally occurs at 9:00, which in turn delays the onset of queued vehicles. However, since the nominal parameter vectors are used within each TOD period, the issue returns between 10:00 and 14:00. However, when predictive parameter vectors are considered as in the top right figure, leading to recomputed green splits for each TOD period, the rate of delay substantially decreases and is nearly equal to the lower bound. 

On February 24, cycle failures are not a concern due to the below average traffic, however the nominal green splits lead to wasted green time. The rate of delay can be modestly improved by using predicted parameter vectors. The two trends exhibited by the conditions on July 2 and February 24 are general: cycle failures lead to large increases in the rate of delay while wasted green time typically affects delay by a lesser degree. 

Total delay is computed by integrating rate of delay over the course of the day. Table \ref{tab:final} collects the total delay values for the four cases in Figure \ref{fig:delay} as well as mean values taken over the entire data set. The traffic predictive controller reduces total delay at the intersection by up to 113.3 veh$\cdot$hr on July 2 by mitigating excessive delay caused by cycle failures. This leads to a 22.1\% decrease in total delay, a marked improvement especially considering that a lower bound on the best achievable delay reduction is 26.4\%. Approximately 45\,400 vehicles transited the intersection on this day, resulting in average savings of 9.0s per vehicle. Over the 132 days in the dataset, the traffic predictive controller that uses predictive segmentation and predictive parameters results in a delay improvement of 7.8 veh$\cdot$hr on average, and the delay improvement exceeds 25 veh$\cdot$hr on 11 occasions.
\begin{table}
  \centering

\begin{tabular}{l | l l l }
&Feb 24 & July 2 & Dataset Mean\\
\hline \hline
    Delay, nominal control [veh$\cdot$hr]&209.1 & 512.1 & 305.7 \\
    Delay, predictive segmentation [veh$\cdot$hr]&207.5&490.5&303.1\\
    Delay, predictive segmentation and parameters [veh$\cdot$hr]&  204.2 &398.8 & 297.9 \\
    Delay lower bd. [veh$\cdot$hr]& 192.5&377.5&276.4\\
\hline
Delay improvement, predictive segmentation [veh$\cdot$hr]&1.7&21.6&2.6\\
Delay improvement, predictive seg. and param.  [veh$\cdot$hr] &5.0&113.3&7.8\\

  \end{tabular}

  \caption{Illustrative delay savings from using traffic predictive control to adjust TOD periods and plans. The traffic predictive controller reduces total delay at the intersection by 113.3 veh$\cdot$hr on July 2 by mitigating excessive delay caused by queued vehicles that wait multiple cycles to clear the intersection, so-called cycle failures. On average, the traffic predictive controller improves total delay by 7.8 veh$\cdot$hr per day by preventing cycle failures and reducing wasted green time.}
  \label{tab:final}
\end{table}

\section{Conclusions}
\label{sec:conclusions}

We have proposed a traffic predictive control scheme that identifies trends in historical traffic flow data and uses real-time measurements to predict future traffic flow. These trends manifest as low-rank structure in the data which are identified using decomposition techniques akin to principal component analysis and particularly suited for prediction. Using a rich dataset of traffic flow measurements over the course of eight months, we provide evidence that much of the day-to-day variation in traffic flow consists of these low-rank, latent structures.%

The traffic predictive control adjusts the time periods and parameters for time-of-day traffic signal scheduling based on predictions of traffic flow using real-time measurements. This scheme is particularly well-suited for implementation on existing traffic control hardware, which universally support time-of-day plans and often are capable of remote updating of signal timing parameters. Furthermore, this approach is well-aligned with standard signal timing practices, increasing the likelihood of successful adoption by practitioners. Additionally, the traffic predictive control requires minimal tuning and accommodates any green split optimization scheme that requires expected traffic flow as input.

The savings in delay for the case study intersection is found to be 7.8 veh$\cdot$hr per day. Valuing a driver's time at \$20 per hour and conservatively assuming that each vehicle carries only one occupant, this suggests annual savings of around \$57\,000. In addition, this metric is compared to a well-timed but pre-specified controller that does not account for real-time measurements; the savings are likely to be higher for intersections that are currently poorly timed.  If these calculations are even only approximately correct, the savings are likely to be well worth the relatively small implementation costs.

An important future direction of research is to consider the case of an arterial corridor. In this case, a straightforward extension of the proposed approach is to consider all movements along the corridor in aggregate. However, there is high correlation between the traffic flows for some sets of movements as vehicles progress along the corridor. This suggests an approach that explicitly accounts for these spatial correlations as is done in \cite{Min:2011kq}.

 Additionally, future research will investigate confidence bounds on the predictions. For example, can low-rank structure be used to predict the 90th or the 99th percentile traffic flow? Furthermore, traffic prediction from historical trends captures phenomena that have occurred previously in the historical data, and thus may not be well-suited for events that are truly one-off events such as lane closures due to construction. However, it may be that lane closures in one direction over an interval of time affects traffic in a similar manner as lane closures on a different leg at a different time. Learning these more universal trends is another future direction of research. Finally, these questions will be pursued alongside an implementation pilot planned in collaboration with Sensys Networks, Inc.

\section*{Acknowledgements}
The authors acknowledge Beaufort County, SC for use of the intersection data and thank Montasir Abbas for discussions regarding technical capabilities of common traffic control hardware and best practices for signal timing and plan selection.

\section*{References}

\bibliographystyle{ieeetr}

\end{document}